\documentclass[aps,nofootinbib, superscriptaddress, pra, twocolumn]{revtex4-1}
\usepackage{ORI_Group_style}
\usepackage{blindtext}
\usepackage[dvipsnames]{xcolor}
\usepackage{array}
\usepackage{subfigure}
\usepackage{float}
\usepackage[normalem]{ulem}

\usepackage{array}
\newcolumntype{P}[1]{>{\centering\arraybackslash}p{#1}}
\newcolumntype{M}[1]{>{\centering\arraybackslash}m{#1}}

\hypersetup{colorlinks,linkcolor=red,citecolor=blue,urlcolor=blue}

\begin{document}

\title{Theory of Quantum Acoustomagnonics and Acoustomechanics with a Micromagnet}


\author{C. Gonzalez-Ballestero}
\email{c.gonzalez-ballestero@uibk.ac.at}
\affiliation{Institute for Quantum Optics and Quantum Information of the Austrian Academy of Sciences, 6020 Innsbruck, Austria.}
\affiliation{Institute for Theoretical Physics, University of Innsbruck, A-6020 Innsbruck, Austria.}

\author{D. H\"ummer}
\affiliation{Institute for Quantum Optics and Quantum Information of the Austrian Academy of Sciences, 6020 Innsbruck, Austria.}
\affiliation{Institute for Theoretical Physics, University of Innsbruck, A-6020 Innsbruck, Austria.}

\author{J. Gieseler}
\affiliation{Department of Physics, Harvard University, 17 Oxford Street, Cambridge, MA 02138, USA}
\affiliation{ICFO-Institut de Ciencies Fotoniques, Mediterranean Technology Park, 08860 Castelldefels (Barcelona), Spain}

\author{O. Romero-Isart}
\affiliation{Institute for Quantum Optics and Quantum Information of the Austrian Academy of Sciences, 6020 Innsbruck, Austria.}
\affiliation{Institute for Theoretical Physics, University of Innsbruck, A-6020 Innsbruck, Austria.}

\begin{abstract}
Recently~\cite{GonzalezBallesteroarXiv2019}, we proposed a new way to engineer a flexible \emph{acoustomechanical} coupling between the center-of-mass motion of an isolated micromagnet and one of its internal acoustic phonons by using a magnon as a passive mediator. In our approach, the coupling is enabled by the strong magnetoelastic interaction between magnons and acoustic phonons which originates from the small particle size. Here, we substantially extend our previous work. First, we provide the full theory of the quantum acoustomagnonic interaction in small micromagnets and analytically calculate the magnon-phonon coupling rates. Second, we fully derive the acoustomechanical Hamiltonian presented in Ref.~\cite{GonzalezBallesteroarXiv2019}. Finally, we extend our previous results for the fundamental acoustic mode to higher order modes. Specifically, we show the cooling of the center-of-mass motion with a range of internal acoustic modes. Additionally, we derive the power spectral densities of the center-of-mass motion which allow to probe the same acoustic modes.
\end{abstract}


\maketitle


\section{Introduction}

Micromagnets are a powerful resource in nanotechnology, enabling applications such as magnetic resonance microscopy~\cite{DegenPNAS2009} and serving as mediators for quantum spin-mechanical interfaces~\cite{RablPRB2009,RablNatPhys2010,HongNanoLett2012,LongeneckerACSNano2012,KolkowitzScience2012,vanWezelPRSA2012,PigeauNatComm2015}. Recently, their elementary solid-state excitations, known as magnons, have been harnessed in the quantum regime to realize active quantum components for applications in quantum science and technology, even at the level of single quanta~\cite{LachanceQuirioneSciAdv2017}.
In this context, magnons are extensively studied due to their long coherence times~\cite{MaierFlaigPRB2017,BaiPRL2015,HueblPRL2013,TabuchiPRL2014,ZhangPRL2014,SoykalPRL2010} and the perspective of interfacing them with other quantum excitations, such as other magnons~\cite{LambertPRA2016}, spin qubits~\cite{TabuchiScience2015,LachanceQuirioneSciAdv2017,TabuchiCRP2016}, acoustic phonons~\cite{ZhangSciAdv2016,LiPRL2018,LiPRA2019,WangIEEE2019}, and optical~\cite{ZhangPRL2016,OsadaPRL2016,ViolaKusminskiyPRA2016,OsadaPRL2018,kusminskiy2019quantum} and microwave photons~\cite{ZhangPRL2014,GoryachevPRB2018,GoryachevPRApplied2014,ZhangNPJQ2015,TabuchiPRL2014,HueblPRL2013,TabuchiCRP2016,GloppePRApplied2019}. This versatility enables a variety of applications, ranging from fundamental physics~\cite{SikiviePRL2014,BarbieriPDU2017} to quantum technologies~\cite{GoryachevPRApplied2014,TabuchiPRL2014,YaoNatComm2017,ZhangNPJQ2015,ZhangPRL2014,ZhangNatComm2015,BaiPRL2015,LachanceQuirionJSAP2019} or microwave-to-optical conversion~\cite{ZhangPRL2016,TabuchiPRL2014,ZhangNatComm2015,BaiPRL2015}. 
A particularly promising prospect is to largely isolate single micromagnets from their environment, either by clamping them to high-Q microresonators~\cite{BurgessScience2013,VinanteNatComm2011,ShamsudhinSmall2016,DrugeNJP2014,FischerNJP2019,KolkowitzScience2012} or by levitating them, as theoretically studied~\cite{RusconiPRB2016,KimballPRL2016,RusconiPRL2017,RusconiPRB2017,PratCampsPRApp2017,GonzalezBallesteroarXiv2019} and experimentally implemented~\cite{DrugeNJP2014,HuilleryArxiv2019,WangPRApplied2019,Jantoappear,BarowskiJTP1993}. The large degree of isolation allows to explore rich internal mesoscopic quantum physics, such as the strong interaction between magnons and acoustic phonons inside the magnet, and the interplay between the internal and the external degrees of freedom, that is the center-of-mass motion and the rotation of the micromagnet.

Following this idea, we proposed in our recent work~\cite{GonzalezBallesteroarXiv2019} a way to couple the lowest energy acoustic phonon of an isolated micromagnet to its center-of-mass motion by using a magnetization wave (magnon) as a passive mediator. 
We showed that the resulting \emph{acoustomechanical} system is widely tunable and that it can be exploited to implement well known optomechanical protocols, where the acoustic phonon plays the role of a built-in internal acoustic ``cavity''. Specifically, we demonstrated the possibility of acoustically cooling the center-of-mass motion to its ground state and of experimentally probing the elusive acoustic mode by measuring the center-of-mass displacement in the strong coupling regime. In this article, we present in detail the theoretical derivation leading to the acoustomechanical Hamiltonian, namely Eq.~(1) of Ref.~\cite{GonzalezBallesteroarXiv2019}, and extend our previous results to higher order acoustic modes. 

\begin{figure}
	\centering
	\includegraphics[width=\linewidth]{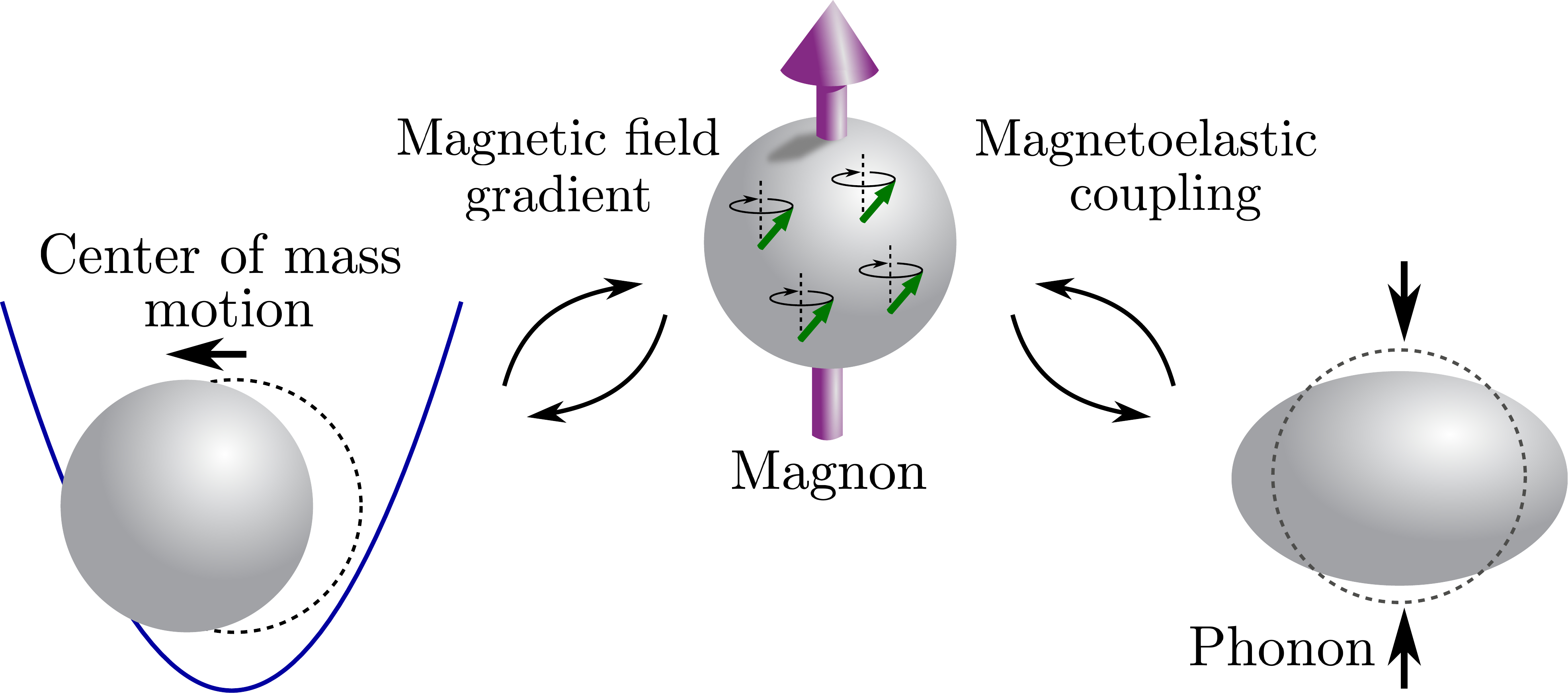}
	\caption{In this work, we theoretically study the interaction between the center-of-mass motion of a harmonically trapped micromagnet, the magnetization fluctuations about its fully magnetized state, namely magnons, and the quanta of elastic deformations in the magnet, i.e.\ acoustic phonons. }\label{figsystem}
\end{figure}

This article is organized as follows. First, in Sec.~\ref{SecFreeInternalHamiltonian}, we summarize the derivation of the magnon and acoustic phonon modes inside a spherical micromagnet. Subsequently, we derive the magnetoelastic interaction Hamiltonian in Sec.~\ref{SecMagnetoelastic} and analytically compute the coupling rates and the selection rules for the couplings between every acoustic phonon and selected magnon modes. In Sec.~\ref{SecAcoustomechanics}, we include the motion of the micromagnet and derive the acoustomechanical Hamiltonian. We also extend the results of Ref.~\cite{GonzalezBallesteroarXiv2019} to higher order acoustic phonons. Finally, we draw our conclusions in Sec.~\ref{Secconclusions}. 
The four appendices of this article provide further details about the calculation of the magnon and the phonon modes (appendices \ref{appendixMAGNONS} and \ref{appendixPHONONS}, respectively), the magnetoelastic coupling rates (appendix \ref{appendixCOUPLINGS}), and the internal heating of the micromagnet (appendix \ref{appendixFurtherDerivations}).

 \section{Free magnon and phonon Hamiltonians of a micromagnet}\label{SecFreeInternalHamiltonian}
 
 The system under consideration, schematically depicted in \figref{figsystem}, is a spherical micromagnet with radius $R \approx 10\text{\,nm}-10\,\mu\text{m}$, which we consider to be well-isolated from its environment, for example due to levitation in high vacuum. 
The Hamiltonian describing the micromagnet is given by
 \begin{equation}\label{totalH}
     \hat{H} = \hat{H}_{\text{ex}} + \hat{H}_{\rm in} + \hat{V}.
 \end{equation}
 Here, the first and third contributions describe the external degrees of freedom of the micromagnet, i.e.\ translation and rotation, and their interaction with the internal degrees of freedom, respectively. These terms will be discussed in Sec.~\ref{SecAcoustomechanics}. The second term in Eq.~\eqref{totalH}, namely $\hat{H}_{\rm in}$, describes the relevant internal degrees of freedom of the micromagnet. In the absence of optical fields these degrees of freedom are acoustic and magnetic, i.e., phonons and magnons, and we write
 \begin{equation}\label{Hinternal}
     \hat{H}_{\rm in} = \hat{H}_{\rm p} + \hat{H}_{\rm m} + \hat{H}_{\rm m-p},
 \end{equation}
as the sum of three contributions describing the phonons, the magnons, and the phonon-magnon interaction, respectively. The term $\hat{H}_{\rm m-p}$ will be discussed in Sec.~\ref{SecMagnetoelastic}. In the following section, we summarize the derivation of the free Hamiltonians $\hat{H}_{\rm p}$ and $\hat{H}_{\rm m}$, while more details are provided in Appendices~\ref{appendixPHONONS} and \ref{appendixMAGNONS}.

 \subsection{Acoustic Phonon Hamiltonian}

 We begin with the acoustic degrees of freedom of the micromagnet. They are described by a continuous displacement field $\mathbf{u}(\mathbf{r},t)$, which we determine with the theory of linear elastodynamics~\cite{Gurtinbook,Eringenbook,Achenbachbook}.
 As detailed in Appendix~\ref{appendixPHONONS}, we first derive the classical acoustic eigenmodes of a spherical, homogeneous, and isotropic body under free-stress boundary conditions, which is a very good approximation for a well-isolated body \cite{ZhangSciAdv2016}. After canonical quantization, the Hamiltonian of linear elastodynamics takes the form
 \begin{equation}\label{Hp}
     \hat{H}_{\rm p} = \hbar\sum_\alpha \omega_\alpha\hat{a}_{\alpha}^\dagger\hat{a}_\alpha,
 \end{equation}
  \begin{table}[t]
	\centering
	\begin{tabular}{ p{4.5cm} | p{3.5cm} }
		\hline
		\hline
		Parameter & Value \\
		\hline
		Longitudinal sound velocity\cite{ZhangSciAdv2016}
		 & $c_L = 7118$ m s$^{-1}$ \\
		Transverse sound velocity\cite{ZhangSciAdv2016} & $c_T = 3871$ m s$^{-1}$ \\
		Mass density\cite{ZhangSciAdv2016,StancilBook2009} & $\rho = 5170$ kg m$^{-3}$ \\
		\hline
		Gyromagnetic ratio\cite{ZhangSciAdv2016,StancilBook2009} &  $\gamma=-1.76\times 10^{11}$T$^{-1}$ s$^{-1}$ \\
		Saturation magnetization & $M_S=5.87\times 10^5$A m$^{-1}$ \\
		\hline
		Magnetoelastic constant $B_1$\cite{ZhangSciAdv2016} &
	     $B_1=3.48\times 10^5$J m$^{-3}$  \\
		Magnetoelastic constant $B_2$\cite{ZhangSciAdv2016} & $B_2=6.4\times 10^5$J m$^{-3}$ \\
		\hline
		\hline
	\end{tabular}
	\caption{Material parameters for YIG.} \label{tablePARAMS}
\end{table}
\begin{figure} 
	\centering
	\includegraphics[width=\linewidth]{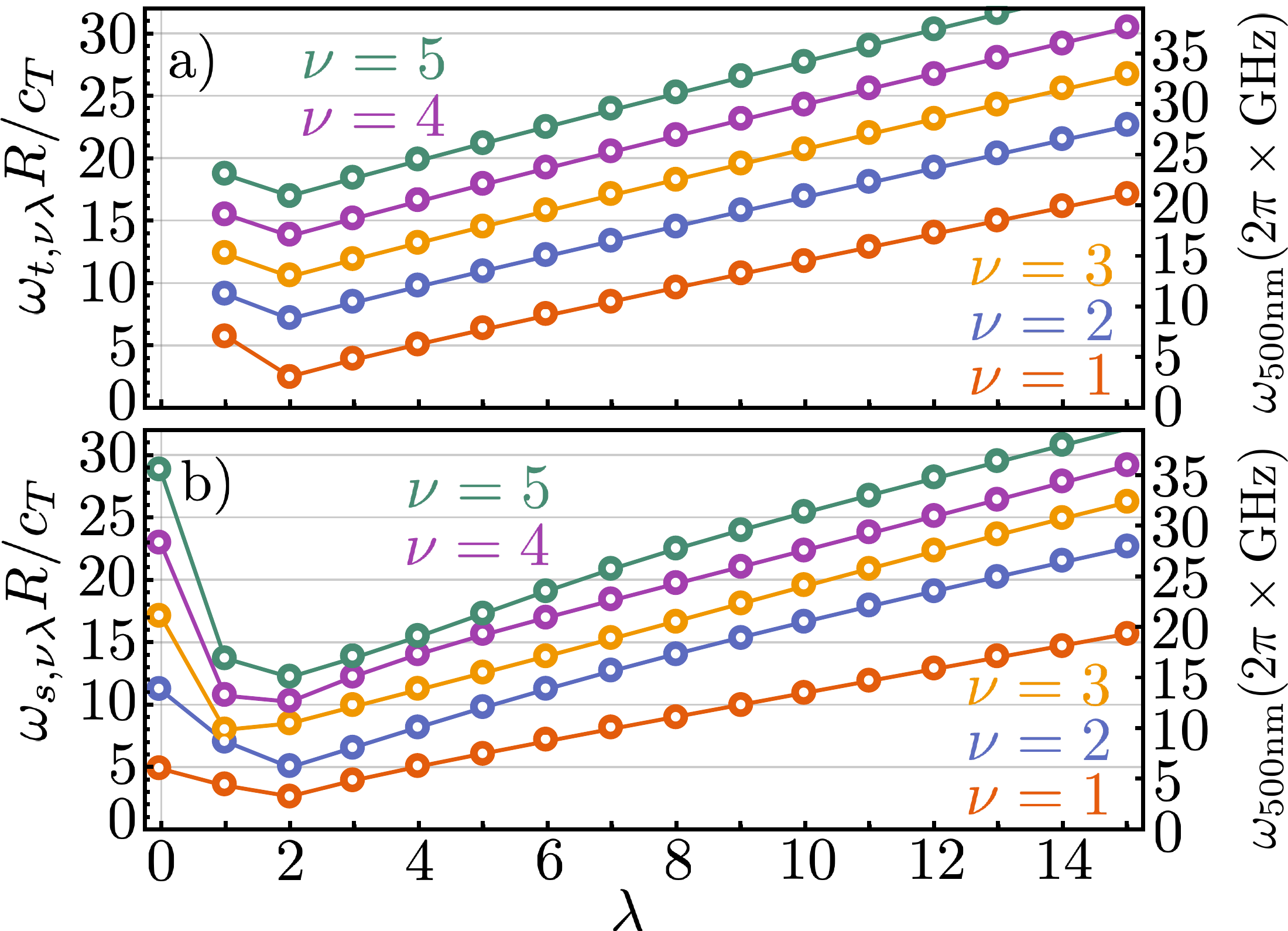}
	\caption{ Lowest order torsional (a) and spheroidal (b) acoustic mode eigenfrequencies of a YIG sphere, versus mode index $\lambda$. Different colors correspond to different values of the mode index $\nu$. The joining lines are a guide to the eye. The right vertical axis shows the eigenfrequencies for a YIG sphere of $R=500$\,nm.}\label{figphononfreqs}
\end{figure}
in terms of phononic operators, that satisfy bosonic commutation relations, $[\hat{a}_\alpha,\hat{a}_{\alpha'}^\dagger]=\delta_{\alpha\alpha'}$.
 The above sum runs over the compound index $\alpha\equiv\{\sigma,\nu,\lambda,\mu\}$, which labels the different acoustic modes of a free sphere, also known in the literature as Lamb modes \cite{LambPLMS1881,Eringenbook}. The first mode index $\sigma \in \{s,t\}$ is a polarization index that divides the eigenmodes into two families, namely the purely shear \emph{torsional} modes ($T_{\nu\lambda\mu}$), and the hybrid shear-compression \emph{spheroidal} modes ($S_{\nu\lambda\mu}$). The remaining mode indices take values $\nu = 1,2,3,...$, $\lambda = 0,1,2,3,...$, and $\mu\in[-\lambda,\lambda]$, and determine the azimuthal ($\mu$), polar ($\lambda$), and radial ($\nu$) geometry of the displacement mode functions. 
The corresponding acoustic eigenfrequencies, namely  $\omega_\alpha$ in Eq.~\eqref{Hp}, are independent of the mode index $\mu$, and are exclusively determined by the sphere size $R$, to which they are inversely proportional ($\omega_\alpha \propto R^{-1}$), and by
 the longitudinal and transverse sound velocities of the material, namely $c_L$ and $c_T$. In \figref{figphononfreqs}, we show the acoustic mode eigenfrequencies $\omega_\alpha$ for the torsional (panel a) and spheroidal (panel b) mode families, using the material parameters for Yttrium-Iron-Garnet (YIG), listed in Table~\ref{tablePARAMS}. The left axes of both panels show the size-independent parameter $\omega_{\sigma,\nu\lambda}R/c_T$, whereas the right axes show the corresponding eigenfrequency for $R=500$\,nm, which reaches values above $2\pi\times1$\,GHz even for the lowest energy mode $S_{12\mu}$.  
 Note that, by definition, the frequency increases at larger values of $\nu$, and that no torsional mode exists with $\lambda=0$. Finally, the quantum displacement field operator in the Schr\"odinger picture reads
\begin{equation}\label{Uquantum}
    \hat{\mathbf{u}}(\mathbf{r}) = \sum_{\alpha}\mathcal{U}_{0\alpha}\left[\mathbf{f}_\alpha(\mathbf{r})\hat{a}_\alpha + \text{H.c.}\right],
\end{equation}
where the zero-point displacement of mode $\alpha$ is given by
\begin{equation}
    \mathcal{U}_{0\alpha} \equiv \sqrt{\frac{\hbar}{2\rho\omega_\alpha\mathcal{N}_\alpha}},
\end{equation}
with $\rho$ being the mass density of the micromagnet. Both the classical mode functions $\mathbf{f}_\alpha(\mathbf{r})$ and their normalization constants $\mathcal{N}_\alpha \equiv \int dV \vert \mathbf{f}_\alpha(\mathbf{r})\vert^2$ are given in Appendix~\ref{appendixPHONONS}.

 \subsection{Magnon Hamiltonian}
 
We now focus on the magnetization waves, or spin waves, supported by the spherical micromagnet. These waves are described by a continuous magnetization field $\mathbf{M}(\mathbf{r},t)$ and its associated electromagnetic fields $\mathbf{E}(\mathbf{r},t)$ and $\mathbf{H}(\mathbf{r},t)$, which obey both Maxwell's equations and the phenomenological nonlinear Landau-Lifshitz equation
\cite{aharoni2000introduction,StancilBook2009}
 \begin{equation}\label{LLE}
     \frac{d}{dt}\mathbf{M}(\mathbf{r},t) = -\vert\gamma\vert \mu_0\mathbf{M}(\mathbf{r},t)\times\mathbf{H}_{\rm eff}(\mathbf{M},\mathbf{r},t).
 \end{equation}
 Here, $\gamma$ is the gyromagnetic ratio, $\mu_0$ is the vacuum permeability, and $\mathbf{H}_{\rm eff}(\mathbf{M},\mathbf{r},t) = \mathbf{H}(\mathbf{r},t) + \Delta \mathbf{H}(\mathbf{M},\mathbf{r},t)$ is the \emph{effective field}, composed of the Maxwell field $\mathbf{H}(\mathbf{r},t)$ and an additional contribution that accounts for the solid-state interactions in the magnetic material~\cite{StancilBook2009}. As detailed in Appendix~\ref{appendixMAGNONS}, we obtain the magnetization eigenmodes under the following approximations: 
 \begin{enumerate}
     \item We assume that a large external homogeneous field $H_0\mathbf{e}_z$ is applied to fully magnetize the micromagnet to its saturation magnetization $M_S$. This allows us to describe small deviations of the fields from their equilibrium as
        \begin{equation}\label{Hlinearization}
            \mathbf{H} (\mathbf{r},t) = H_0\mathbf{e}_z + \mathbf{h}(\mathbf{r},t),
        \end{equation}
        \vspace{-0.7cm}
        \begin{equation}\label{Mlinearization}
            \mathbf{M} (\mathbf{r},t) = M_S\mathbf{e}_z + \mathbf{m}(\mathbf{r},t),
        \end{equation}
     where $\mathbf{m} \ll M_S$ and $\mathbf{h} \ll H_0$ are the dynamical variables, whose eigenmodes we calculate and quantize later. The above approximation, known in the literature as the spin wave approximation or spin wave limit~\cite{StancilBook2009}, allows to linearize the Landau-Lifshitz equation by keeping first order terms in the small variables $\mathbf{m}/M_S$ and $\mathbf{h}/H_0$ (see Appendix~\ref{appendixMAGNONS}).
     \item We assume that the micromagnet is larger than the usual domain wall length~\cite{Fletcher59,StancilBook2009} ($2R\gtrsim 10$\,nm), and that it has cubic internal symmetry, as it is the case for YIG. This allows to largely simplify the effective field to $\mathbf{H}_{\rm eff}\approx H_I\mathbf{e}_z$, where $H_I\equiv H_0-M_S/3$ is known as the internal field~\cite{StancilBook2009} (see Appendix~\ref{appendixMAGNONS}).
     \item We undertake the magnetostatic approximation\cite{StancilBook2009}
     \begin{equation}
         \nabla\times\mathbf{h}(\mathbf{r},t)\approx 0,
     \end{equation}
     which is valid for micromagnet sizes much smaller than the wavelength of the electromagnetic component of the spin wave, i.e., for $R\ll 2\pi c/\omega_{\rm sw}$, where $c$ is the vacuum speed of light and $\omega_{\rm sw}$ is the frequency of the spin wave. For the energies considered in this paper, this implies $R\lesssim 1$cm. This approximation simplifies the problem by uncoupling the electric field from the system of equations and by allowing to define a magnetostatic potential; see Appendix~\ref{appendixMAGNONS} for details.
 \end{enumerate}
 
  \begin{figure} 
	\centering
	\includegraphics[width=\linewidth]{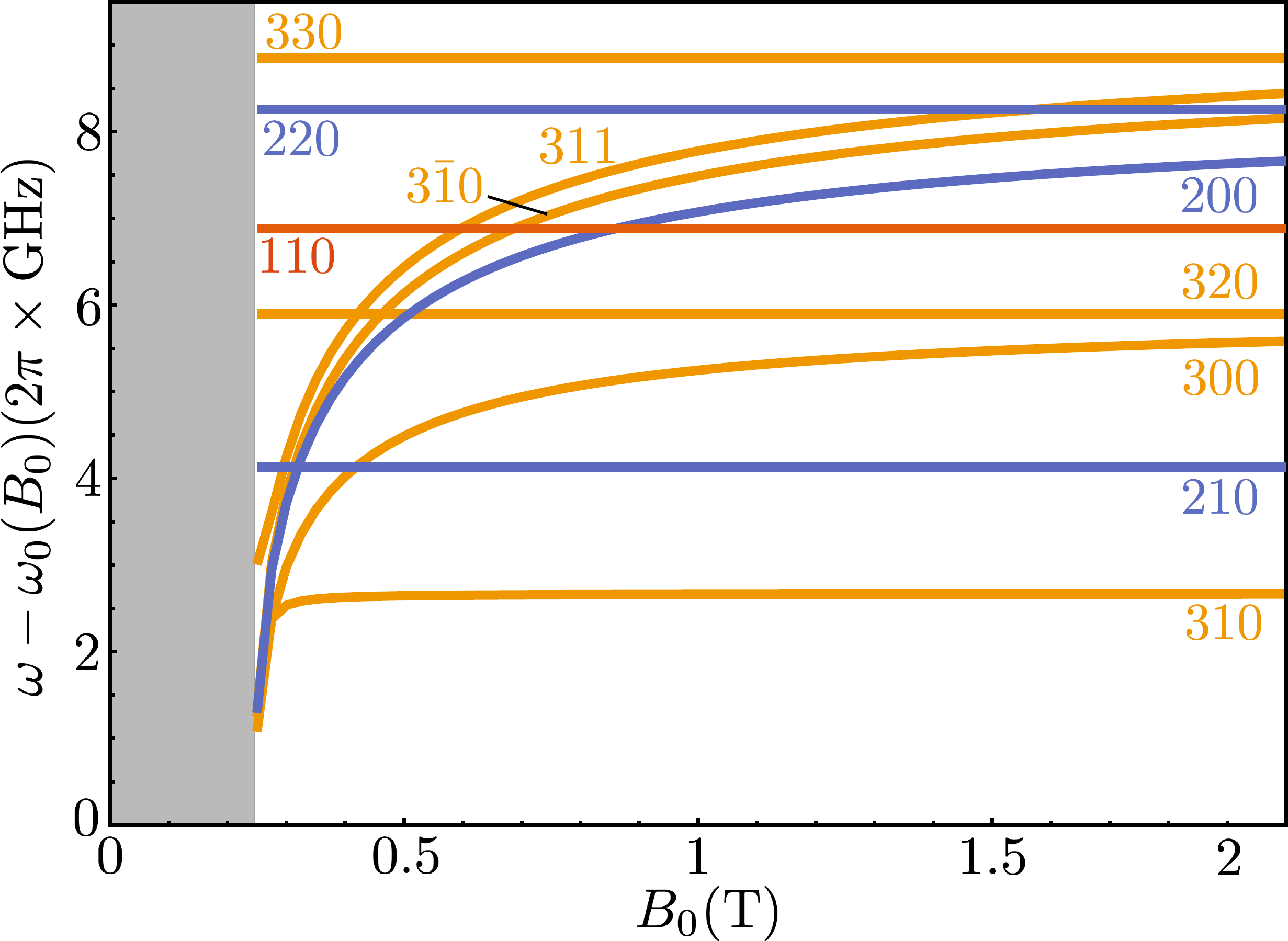}
	\caption{Magnonic eigenfrequencies of a YIG sphere for $l=1$ (red), $l=2$ (blue), and $l=3$ (orange), as a function of external static field $B_0$. The frequency $\omega_0$ (see main text) has been substracted on the vertical axis for a better visualization. The shaded area shows the unstable region when $B_0<\mu_0 M_S/3 =246$\,mT for YIG.}\label{figmagnonfreqs}
\end{figure}

The above approximations allow to obtain magnetization eigenmodes, known in the literature as Magnetostatic Dipolar Spin Waves, or Walker modes \cite{WalkerPhysRev1957,Roschmann,Fletcher59}. The Walker eigenmodes are labeled by a compound multi-index $\beta \equiv\{lmn\}$, with $l=1,2,3,...$, $m\in [-l,l]$, and $n=0,1,2,...,n_{\rm max}(l)$. These three indices determine the shape of the magnetization profile (with $m$ being an azimuthal wave number) through complicated mode functions for which there is no simple general expression\cite{Roschmann,Fletcher59} (see Appendix~\ref{appendixMAGNONS}). The resulting mode eigenfrequencies $\omega_\beta\equiv\omega_{lmn}$ depend on the external field $B_0\equiv \mu_0H_0$, but are independent of the micromagnet size $R$ for the sizes consistent with our approximations ($10\text{\,nm}\lesssim R \lesssim 1$\,cm). In \figref{figmagnonfreqs}, we show these eigenfrequencies for a YIG micromagnet (see Table~\ref{tablePARAMS}) as a function of $B_0$ for all the magnon modes with $l\le 3$. Note that not all the possible triplets $\{lmn\}$ correspond to a physical mode; for instance, there is no $\{3\bar{2}0\}$ mode (here, the bar above an index denotes a negative value). Note also that, in the vertical axis of \figref{figmagnonfreqs}, we have subtracted $\omega_0 (B_0)\equiv \vert\gamma\vert\mu_0H_I$ so that horizontal lines in the figure reflect linear dependencies with the external field. 
  The shaded area in \figref{figmagnonfreqs} corresponds to a region of negative internal field $H_I$, where some of the solutions become imaginary, i.e., unstable \cite{Mills2006}. In this paper we  consider solutions without this instability. 
  Finally, note that the energy does not monotonically depend on the mode indices, and that the lowest-order mode $\{110\}$ (the celebrated Kittel mode~\cite{ZhangPRL2014,GoryachevPRB2018,GoryachevPRApplied2014,ZhangNPJQ2015,TabuchiPRL2014,HueblPRL2013,TabuchiCRP2016,TabuchiScience2015,LachanceQuirioneSciAdv2017,LambertPRA2016,WangPRB2016,kusminskiy2019quantum,ViolaKusminskiyPRA2016}) does not have the lowest energy. Within the magnetostatic dipolar spin wave approximation, the modes with the lowest energy are the $\{l10\}$ modes, whose frequency tends to the absolute lower bound $\omega_{l10}\to\omega_0$ as $l$ tends to infinity \cite{Roschmann}\footnote{Note that in practice, the description in terms of Walker modes breaks down for modes with sufficiently large $l$, as
  their mode functions display short-wavelength spatial oscillations, and thus do not fulfill some underlying asumptions of the theory. }.

As detailed in Ref.~\cite{Mills2006}, the Walker modes can be quantized by using, on the one hand, the micromagnetic energy functional generating the Landau-Lifshitz equations, 
\begin{equation}\label{micromagneticenergy}
\begin{split}
      E_m(&\{\mathbf{m}\},\{\mathbf{h}\}) =
      \\
      &=\frac{\mu_0}{2}\int dV \mathbf{m}(\mathbf{r},t)
      \cdot\left[\frac{H_I}{M_S}\mathbf{m}(\mathbf{r},t)-\mathbf{h}(\mathbf{r},t)\right],
\end{split}
\end{equation}
and, on the other hand, the orthogonality relations between different magnonic eigenmodes, the so-called Walker identities~\cite{WalkerPhysRev1957,brown1962book,Mills2006} (see Appendix~\ref{appendixMAGNONS} for details).
With the bosonic ladder operators, $[\hat{s}_\beta,\hat{s}_{\beta'}^\dagger]=\delta_{\beta\beta'}$, the resulting magnon Hamiltonian is given by
\begin{equation}\label{Hm}
    \hat{H}_{\rm m} = \hbar\sum_{\beta}\omega_\beta\hat{s}^\dagger_\beta\hat{s}_\beta,
\end{equation}
and the spin-wave magnetization operator in the Schr\"odinger picture can be written as
\begin{equation}\label{mquantum0}
    \hat{\mathbf{m}}(\mathbf{r})=\sum_\beta\mathcal{M}_{0\beta}\left[\tilde{\mathbf{m}}_\beta(\mathbf{r})\hat{s}_\beta + \text{H.c.}\right].
\end{equation}
Here, the zero-point magnetization is given by
\begin{equation}\label{zeropointM}
\begin{split}
    \mathcal{M}_{0\beta}&\equiv\sqrt{\frac{\hbar\vert\gamma\vert M_S}{\tilde\Lambda_\beta}},
\end{split}
\end{equation}
$\tilde{\mathbf{m}}_\beta(\mathbf{r})$ is the mode function of the spin wave and $\tilde{\Lambda}_\beta \equiv 2\text{Im}\int dV \tilde{m}_x^*(\mathbf{r})\tilde{m}_y(\mathbf{r})$ its normalization constant (see Appendix~\ref{appendixMAGNONS}).

 \section{Magnetoelastic interaction}\label{SecMagnetoelastic}
 
 In this section we derive the magnetoelastic interaction Hamiltonian, namely $\hat{H}_{\rm m-p}$ in Eq.~\eqref{Hinternal}, starting with the exact form of this Hamiltonian up to second-order in magnon operators. Then, we focus on the small particle limit, and we analytically compute the coupling rates and selection rules between every acoustic phonon mode and selected magnon modes.
 
 \subsection{Magnetoelastic Hamiltonian}
 
 The magnetic energy density depends microscopically on the positions and orientations of individual spins inside the material. As a consequence, elastic deformations of the lattice modify the energy density yielding the magnetoelastic interaction between magnons and phonons. The general expression for the magnetoelastic energy density is well known \cite{KittelRMP1949} and can be easily derived assuming only time-reversal invariance \cite{Landaubook}. For a cubic material such as YIG, and neglecting exchange effects in the spirit of the Walker mode approximations (see Appendix~\ref{appendixMAGNONS}), the lowest-order contribution to the energy density takes on the form
  \cite{Gurevich1996magnetization,ZhangSciAdv2016}
 \begin{equation}\label{LandauUme}
 \begin{split}
     U_\text{me}(\mathbf{r},t) =&\sum_{i,j=x,y,z}\frac{B_2+\delta_{ij}(B_1-B_2)}{M_S^2}
     \\
     &
     \times M_i(\mathbf{r},t)M_j(\mathbf{r},t)\overline{\varepsilon}_{ij}(\mathbf{r},t).
\end{split}
 \end{equation}
 Here, $M_i(\mathbf{r},t)\equiv\mathbf{e}_i\cdot\mathbf{M}(\mathbf{r},t)$ are the Cartesian components of the total magnetization field, $B_1$ and $B_2$ are the magnetoelastic coefficients of the material (see Table~\ref{tablePARAMS}), and we define the adimensional strain tensor
 \begin{equation}\label{straintensor}
     \overline{\varepsilon}_{ij}(\mathbf{r},t)\equiv \frac{1}{2}\left[\frac{\partial u_i(\mathbf{r},t)}{\partial r_j} + \frac{\partial u_j(\mathbf{r},t)}{\partial r_i}\right],
 \end{equation}
 which is symmetric in $i$ and $j$. From Eq.~\eqref{LandauUme}, one can derive all the magnetoelastic energy terms to first and second order in ($\mathbf{m}/M_S$). Although the relevant physics in this work will stem only from the former, it is insightful to compute both of them in order to compare with previous works. We first write the magnetization in the spin wave limit, Eq.~\eqref{Mlinearization}, as
 \begin{equation}\label{Mquadratic}
\begin{split}
    \mathbf{M}(\mathbf{r},t) &= m_x(\mathbf{r},t)\mathbf{e}_x + m_y(\mathbf{r},t)\mathbf{e}_y+
    \\
    &
    +\sqrt{M_S^2-m_x^2(\mathbf{r},t)-m_y^2(\mathbf{r},t)}\mathbf{e}_z,
\end{split}
 \end{equation}
 where the inclusion of second-order terms in the $z-$component is necessary to correctly account for all the second-order magnetoelastic contributions~\cite{ZhangSciAdv2016}. Combining Eqs.~\eqref{LandauUme} and \eqref{Mquadratic}, and neglecting all terms of order $(m_j/M_S)^3$ or higher, we rewrite the magnetoelastic energy density in terms of our dynamical variables,  $\mathbf{m}(\mathbf{r},t)$ and $\mathbf{u}(\mathbf{r},t)$, as
 \begin{equation}\label{Umespinwave}
\begin{split}
     U_\text{me}&(\mathbf{r},t)\approx\frac{2B_2}{M_S}\sum_{i=x,y}m_i(\mathbf{r},t)\overline{\varepsilon}_{iz}(\mathbf{r},t) +
     \\
     &
     +\sum_{i,j=x,y}\frac{B_2+\delta_{ij}(B_1-B_2)}{M_S^2}m_i(\mathbf{r},t)m_j(\mathbf{r},t)
     \\
     &
     \times\left[\overline{\varepsilon}_{ij}(\mathbf{r},t)-\overline{\varepsilon}_{zz}(\mathbf{r},t)\delta_{ij}\right].
\end{split}
 \end{equation}
 The first term in the above equation describes the first-order contribution to the magnetoelastic energy density, whereas the remaining two lines describe the second order contribution.

 \begin{table*}[t]
\centering
\begin{tabular}{|M{1.2cm}|M{2.6cm}|M{11cm}|}
\hline
Magnon&\centering Allowed phonons (selection rules)& \vspace{0.23cm} $g_{\alpha\beta}/g_{\alpha\beta}^0$\\[2ex]\hline\hline
Kittel&$\displaystyle S_{\nu21}$&\vspace{0.2cm}$\displaystyle\frac{6}{5}\left[\tilde\eta z_{2\nu}j_1(\tilde\eta z_{2\nu})-6z_{2\nu}j_1(z_{2\nu})\frac{j_2(\tilde\eta z_{2\nu})-\tilde\eta z_{2\nu} j_3(\tilde\eta z_{2\nu})}{(6-z_{2\nu}^2)j_2(\tilde\eta z_{2\nu})+2z_{2\nu}j_3(z_{2\nu})}\right]$
\\[3.4ex]\hline
$\{210\}$&$\displaystyle T_{\nu21}$&\vspace{0.2cm}$\displaystyle \frac{3i}{5}j_2(z_{2\nu}^{(t)})$
\\[3.4ex]\cline{2-3}
&$\displaystyle S_{\nu11}$&\vspace{0.2cm}$\displaystyle \frac{3}{5}\left[\frac{2}{3}\tilde\eta z_{1\nu}j_2(\tilde\eta z_{1\nu})-\frac{1}{2}\frac{j_1(\tilde\eta z_{1\nu})}{j_1(z_{1\nu})}z_{1\nu}j_2(z_{1\nu})\right]$
\\[3.4ex]\cline{2-3}
&$\displaystyle S_{\nu31}$&\vspace{0.2cm}$\displaystyle -\frac{3}{5}\left[\frac{8}{7}\tilde\eta z_{3\nu} j_2(\tilde\eta z_{3\nu})-\frac{64}{7}\frac{2j_3(\tilde\eta z_{3\nu})-\tilde\eta z_{3\nu}j_4(\tilde\eta z_{3\nu})}{\left[16-(z_{3\nu})^2\right]j_3(z_{3\nu})+2 z_{3\nu} j_4(z_{3\nu})}z_{3\nu} j_2(z_{3\nu})\right]$
\\[3.4ex]\hline
\end{tabular}
\caption{Summary of the magnetoelastic couplings for the Kittel and the $\{210\}$ magnon modes. The second column shows the phonons to which the corresponding magnon can couple according to the selection rules. In the third column, we define $z_{\lambda\nu}\equiv\omega_{s,\lambda\nu}R/c_T$, $z_{\lambda\nu}^{(t)}\equiv\omega_{t,\lambda\nu}R/c_T$, and $\tilde\eta \equiv c_T/c_L$.}\label{TableCouplings}
\end{table*} 
 
 In order to obtain the quantum magnon-phonon interaction Hamiltonian in the Schr\"odinger picture, $\hat{H}_{\rm m-p}$, we substitute in Eq.~\eqref{Umespinwave} the corresponding quantum operators, namely $\hat{\mathbf{u}}(\mathbf{r})$ in Eq.~\eqref{Uquantum}, and $\hat{\mathbf{m}}(\mathbf{r})$ in Eq.~\eqref{mquantum0}, and integrate over the micromagnet volume, $V$. The resulting Hamiltonian can be split into two contributions,
 \begin{equation}\label{Hmp2terms}
     \hat{H}_{\rm m-p} = \hat{H}_{\rm m-p}^{(1)}+\hat{H}_{\rm m-p}^{(2)}.
 \end{equation}
 The first term, originating from the first line in Eq.~\eqref{Umespinwave}, contains only single-magnon operators,
 \begin{equation}\label{Hmp1}
     \hat{H}_{\rm m-p}^{(1)} = \hbar\sum_{\alpha\beta}\hat{s}_\beta\left(\tilde g_{\alpha\beta}\hat{a}_\alpha+g_{\alpha\beta}\hat{a}^\dagger_\alpha\right)+\text{H.c.}
 \end{equation}
 Here, the single-magnon coupling rates are given by
 \begin{equation}\label{singlemagnoncouplings}
\begin{split}
    \left[
     \begin{array}{c}
          \tilde g_{\alpha\beta}  \\
          g_{\alpha\beta} 
     \end{array}
     \right]&=g_{\alpha\beta}^0\frac{1}{V}\sum_{i}\int dV\tilde{m}_{\beta i}(\mathbf{r})\left[\begin{array}{c}
          \tilde\varepsilon_{iz}^{(\alpha)}(\mathbf{r})  \\
          \tilde\varepsilon_{iz}^{(\alpha)*}(\mathbf{r}) 
     \end{array}\right]
     ,
\end{split}
 \end{equation}
and for convenience we introduce the bare coupling rate
 \begin{equation}
     g_{\alpha\beta}^0 \equiv 2\frac{B_2 V}{\hbar}\frac{\mathcal{U}_{0\alpha}}{R}\frac{\mathcal{M}_{0\beta}}{M_S},
 \end{equation}
 and the adimensional strain tensor for acoustic mode $\alpha$,
 \begin{equation}
     \tilde\varepsilon^{(\alpha)}_{ij} \equiv  \frac{R}{2}\left[\frac{\partial f_{\alpha,i}(\mathbf{r})}{\partial r_j} + \frac{\partial f_{\alpha,j}(\mathbf{r})}{\partial r_i}\right].
 \end{equation}
  The second term in Eq.~\eqref{Hmp2terms} stems from the second and third lines of Eq.~\eqref{Umespinwave}, and contains two magnon operators:
 \begin{equation}\label{Hmp2}
\begin{split}
    \frac{\hat{H}_{\rm m-p}^{(2)}}{\hbar}& = \sum_{\alpha\beta\beta'}\left[\hat{s}_\beta\hat{s}_{\beta'}\left(\Psi_{\beta\beta'}^\alpha\hat{a}_\alpha+\tilde{\Psi}_{\beta\beta'}^\alpha\hat{a}^\dagger_\alpha\right) + \text{H.c.}\right]+
    \\
    &
    + \sum_{\alpha\beta\beta'}\left[\hat{s}^\dagger_\beta\hat{s}_{\beta'}\left(\Omega_{\beta\beta'}^\alpha\hat{a}_\alpha+\tilde{\Omega}_{\beta\beta'}^\alpha\hat{a}^\dagger_\alpha\right) + \text{H.c.}\right].
\end{split}
 \end{equation}
 The two-magnon coupling rates $\Psi^{\alpha}_{\beta\beta'}$ can be written compactly as
 \begin{equation}\label{psiMP}
\begin{split}
    \Psi^{\alpha}_{\beta\beta'} &=
     \frac{g_{\alpha\beta}^0}{2}\frac{\mathcal{M}_{0\beta'}}{M_S}
     \\
     &
     \times\frac{1}{V}\int dV \tilde{\mathbf{m}}_\beta(\mathbf{r})^T \tilde{A}_{\varepsilon}^{(\alpha)}(\mathbf{r}) \tilde{\mathbf{m}}_{\beta'}(\mathbf{r}),
\end{split}
 \end{equation}
 where we defined the adimensional matrix
 \begin{equation}
 \begin{split}
     \left[\tilde{A}_\varepsilon^{(\alpha)}(\mathbf{r})\right]_{ij}  \equiv & \left[1+\delta_{ij}\left(\frac{B_1}{B_2}-1\right)\right]
     \\
     &
     \times\left[\tilde{\varepsilon}_{ij}^{(\alpha)}(\mathbf{r})-\delta_{ij}\tilde{\varepsilon}_{zz}^{(\alpha)}(\mathbf{r})\right].
 \end{split}
 \end{equation}
 The coupling rate $\Omega_{\beta\beta'}^\alpha$ has a similar expression as Eq.~\eqref{psiMP}, with $\tilde{A}_\varepsilon^{(\alpha)} \to \tilde{A}_\varepsilon^{(\alpha)*}$.
 Finally, the corresponding tilded couplings, $\tilde{\Psi}^{\alpha}_{\beta\beta'}$ and $\tilde \Omega^{\alpha}_{\beta\beta'}$, are given by the same expressions as $\Psi^{\alpha}_{\beta\beta'}$ and $ \Omega^{\alpha}_{\beta\beta'}$ under the substitution $\tilde{\mathbf{m}}_\beta \to \tilde{\mathbf{m}}^*_\beta$.
 The expressions derived above are so far exact up to second order in the spin wave magnetization $\mathbf{m}/M_S$.
 
 \subsection{Acoustomagnonics in the small particle limit}

 Once the magnetoelastic interaction has been computed, we can write the total internal Hamiltonian, Eq.~\eqref{Hinternal} in terms of bosonic operators, by adding up Eqs.~\eqref{Hp}, \eqref{Hm}, and \eqref{Hmp2terms}.
The internal Hamiltonian is usually simplified by a rotating wave approximation. However, this approximation depends on the particle size and differs radically between small and large particles. For very large particles ($R\gtrsim 100-1000\,\mu$m), the acoustic mode frequencies $\omega_\alpha \propto R^{-1}$ are negligibly small compared to the magnon frequencies $\omega_\beta$ (see Fig.~\ref{figphononfreqs}). Consequently, one can neglect the first order magnetoelastic coupling $H_{\rm m-p}^{(1)}$ as $\vert\omega_\beta \pm\omega_\alpha\vert \approx \omega_\beta \gg g_{\alpha\beta},\tilde g_{\alpha\beta}$. Neglecting also the largely energy non-conserving terms in the second-order contribution $\hat{H}_{\rm m-p}^{(2)}$, one obtains
\begin{equation}\label{dispersiveMPInt}
\begin{split}
    \hat{H}_{\rm in}&\big\vert_{\text{RWA, large }R} \approx \hat{H}_{\rm p} + \hat{H}_{\rm m} +
    \\
    &
    + \hbar\sum_{\alpha\beta} \left[\hat{s}^\dagger_\beta\hat{s}_{\beta}\left(\Omega_{\beta\beta}^\alpha\hat{a}_\alpha+\tilde{\Omega}_{\beta\beta}^\alpha\hat{a}^\dagger_\alpha\right) + \text{H.c.}\right].
\end{split}
\end{equation}
This is the dispersive magnon-phonon interaction employed so far in acoustomagnonics  \cite{ZhangSciAdv2016,LiPRL2018,LiPRA2019,WangIEEE2019}. Note that, precisely because it stems from a second-order contribution, the corresponding interaction rates are small, usually in the $\sim 10$mHz range for $R\sim 100\,\mu$m \cite{ZhangSciAdv2016}.

In this paper, we focus on the opposite limit, namely the small micromagnet limit $R\approx 10$\,nm$-10\,\mu$m where the acoustic eigenfrequencies lie in the $\gtrsim 1$\,GHz range, as evidenced by \figref{figphononfreqs}. We will assume that the frequency of a specific magnon $\hat{s}_{\beta_0}$ is brought close to resonance with one of the acoustic phonons $\hat{a}_{\alpha_0}$ through the external magnetic field $B_0$ (see \figref{figmagnonfreqs}). In this situation, we can disregard any contribution to the first-order interaction $\hat{H}_{\text{m-p}}^{(1)}$ except for the quasi-resonant contribution $\sim \hat{s}_{\beta_0}^\dagger\hat{a}_{\alpha_0}$, since, for any other magnon-phonon pair, $\vert \omega_\alpha \pm \omega_\beta \vert \gtrsim \text{\,GHz}\gg g_{\alpha\beta},\tilde g_{\alpha\beta}$. The entire second-order contribution $\hat{H}_{\text{m-p}}^{(2)}$ can also be neglected under a rotating wave approximation, as $\vert \pm \omega_\beta-\omega_{\beta'} \pm \omega_\alpha\vert \gtrsim \text{\,GHz} \gg \Omega_{\beta\beta'}^\alpha,\Psi_{\beta\beta'}^\alpha,\tilde\Omega_{\beta\beta'}^\alpha,\tilde\Psi_{\beta\beta'}^\alpha$.
The resulting small-particle Hamiltonian in the RWA then reads
\begin{equation}\label{beampslitterMPInt}
\begin{split}
    \hat{H}_{\rm in}&\big\vert_{\text{RWA, small }R} \approx \hat{H}_{\rm p} + \hat{H}_{\rm m} +
    \\
    &
    + \hbar\left( g_{\alpha_0\beta_0}\hat{s}_{\beta_0}\hat{a}_{\alpha_0}^\dagger + \text{H.c.}\right) .
\end{split}
\end{equation}
This beam-splitter interaction is the key component of our work. As we will see below, its magnon-phonon coupling rates are much stronger than the coupling rates in the large particle limit Eq.~\eqref{dispersiveMPInt} because they stem from the first-order magnetoelastic term $\hat{H}_{\rm m-p}^{(1)}$. 

\begin{figure*} 
 	\centering
 	\includegraphics[width=\linewidth]{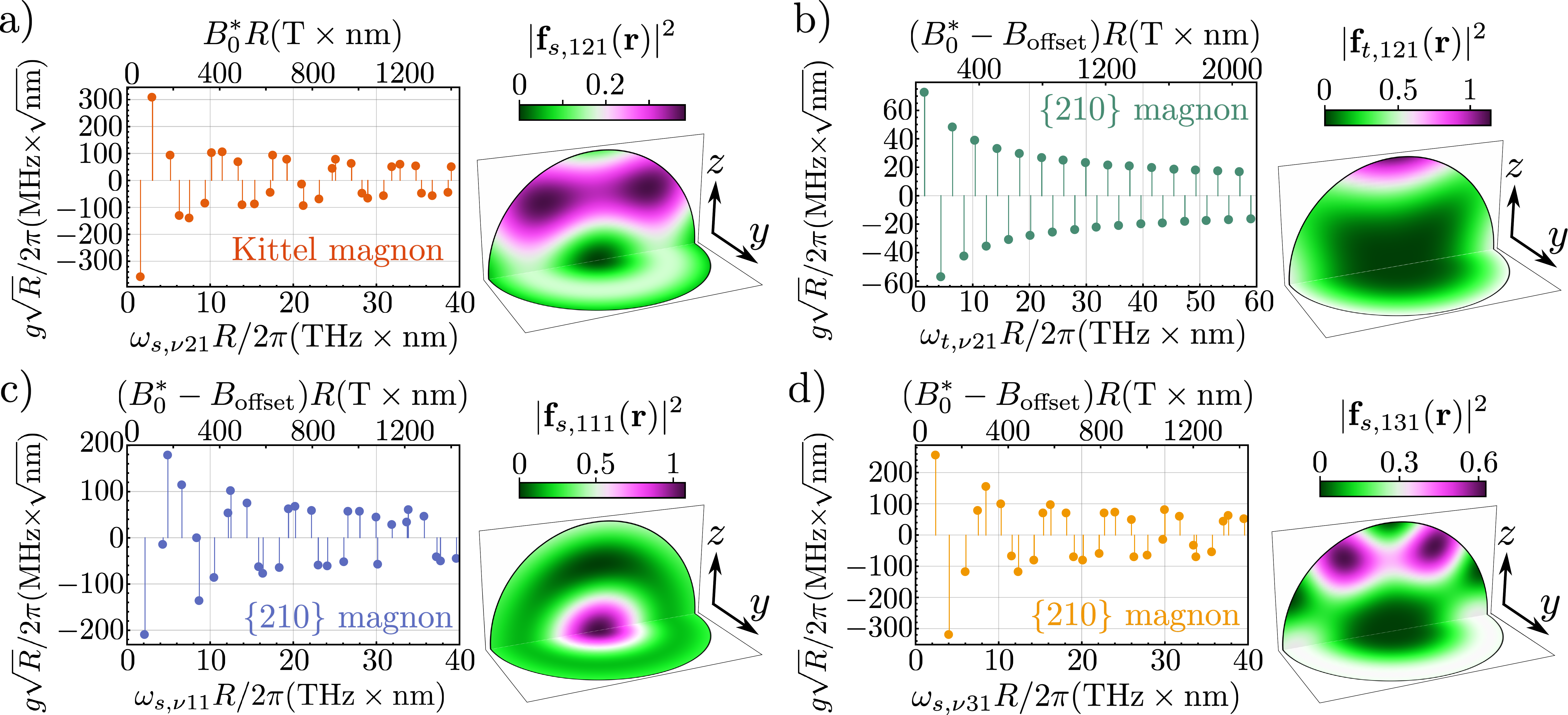}
 	\caption{Magnetoelastic couplings $g_{\alpha\beta}$ for the Kittel magnon-$S_{\nu21}$ phonon (a), the $\{210\}$ magnon-$T_{\nu21}$ phonon (b), the $\{210\}$ magnon-$S_{\nu11}$ phonon (c), and the $\{210\}$ magnon-$S_{\nu31}$ phonon pairs (d), versus acoustic mode frequency and for YIG parameters (see Table \ref{tablePARAMS}). The discrete acoustic modes $\nu=1,2,3,...$ are represented by solid circles. The upper axes show the external field $B_0$ required to tune the magnon in resonance with the corresponding acoustic phonon [see Eq.~\eqref{magnonfreqDEFFS}]. All  axes are normalized to be size-independent. The contour plots show the normalized acoustic mode intensities corresponding to $\nu=1$, i.e.\ to the lowest-frequency acoustic mode in each panel. }\label{figmagnetoelastic}
\end{figure*}

Let us now characterize the two-mode interaction Eq.~\eqref{beampslitterMPInt} in greater detail. More specifically, let us determine the coupling rates $g_{\alpha\beta}$ and the magnon-phonon selection rules. In general, it is possible to show that the rates scale as $g_{\alpha\beta}\propto R^{-1/2}$ and fulfill the azimuthal selection rule $g_{\alpha\beta} \propto \delta_{\mu m}$, i.e.\ the coupling is nonzero only if the acoustic azimuthal mode index $\mu$ is equal to the magnonic azimuthal mode index $m$
\footnote{This selection rule is derived by writing $g_{\alpha\beta}$ in terms of cylindrical mode functions, $\tilde{m}_{\beta\pm}(\mathbf{r}) \equiv \tilde{m}_{\beta x}(\mathbf{r}) \pm i\tilde{m}_{\beta y}(\mathbf{r})$ and $\tilde{\varepsilon}^{(\alpha)}_{\pm}(\mathbf{r})\equiv \tilde{\varepsilon}^{(\alpha)}_{xz}(\mathbf{r}) \pm i\tilde{\varepsilon}^{(\alpha)}_{xz}(\mathbf{r})$. One then easily demonstrates that $\tilde{m}_{\beta\pm}(\mathbf{r})=\tilde{m}_{\beta\pm}^\perp(r,\theta)e^{i\phi(m\pm 1)}$ and
$\tilde{\varepsilon}^{(\alpha)}_{\pm}(\mathbf{r})=\tilde{\varepsilon}^{(\alpha)}_{\pm\perp}(r,\theta)e^{i\phi(\mu\pm 1)}$. The selection rule $\delta_{\mu m}$ follows directly from the integration over the azimuthal angle $\phi$.}.
Note that precisely this selection rule allows us to write the interaction Eq.~\eqref{beampslitterMPInt} as a two-mode coupling, instead of as a sum over all the degenerate acoustic modes $\alpha_0 = \{\sigma_0,\nu_0,\lambda_0,-\lambda_0\},...,\{\sigma_0,\nu_0,\lambda_0,+\lambda_0\}$. Aside from these two, it is not possible to derive more general properties analytically due to the lack of a general analytical expression for the magnon mode functions $\tilde{\mathbf{m}}_\beta (\mathbf{r})$. In order to continue, one must compute the coupling rates separately for each magnon mode of interest. 

To provide an example, we will focus on two particular magnonic modes in the following: first, the relevant Kittel mode $\beta_0 \equiv \{110\} \equiv K$, which is widely used in magnonics and characterized by an homogeneously magnetized mode function
\begin{equation}\label{mtildeKittel}
     \tilde{\mathbf{m}}_{\rm K}(\mathbf{r}) = \mathbf{e}_x + i\mathbf{e}_y;
 \end{equation}
 and second, the $\beta_0 \equiv \{210\}$ mode, whose mode function is given by
 \begin{equation}\label{mtilde210}
     \tilde{\mathbf{m}}_{210}(\mathbf{r}) = \frac{z}{R}\left[\mathbf{e}_x + i\mathbf{e}_y\right],
 \end{equation}
 as an example of a non-homogeneous magnetization wave. Their respective zero-point magnetizations are
 \begin{equation}\label{zeropointMmagnons}
     \mathcal{M}_K = \frac{1}{\sqrt{5}}\mathcal{M}_{210}=\sqrt{\frac{\hbar\vert\gamma\vert M_S}{2V}},
 \end{equation}
 and their respective eigenfrequencies are
 \begin{equation}\label{magnonfreqDEFFS}
     \omega_K = \vert\gamma\vert B_0 \hspace{0.3cm};\hspace{0.3cm} \omega_{210} = \vert\gamma\vert (B_0-B_{\rm offset}),
 \end{equation}
 where $B_{\rm offset} \equiv 2\mu_0 M_S/15 =49.2$\,mT for YIG. Note that for sufficiently low external fields $B_0$ the frequency of the $\{210\}$ mode becomes negative, resulting in potential dynamical instabilities \cite{KusturaPRA2019}. Here we will consider $B_0$ large enough to avoid these instabilities.
 We have analytically calculated the magnetoelastic couplings $g_{\alpha\beta}$ for both of the above magnons as detailed in Appendix~\ref{appendixCOUPLINGS}.
  The magnetoelastic couplings and selection rules for the Kittel and $\{210\}$ magnons are compiled in Table~\ref{TableCouplings}. As evidenced by these results, the Kittel mode is subject to strict selection rules, and only couples to the $S_{\nu21}$ acoustic phonons. In contrast, the selection rules for the more complex $\{210\}$ magnon are less restrictive. Specifically, it interacts with the $T_{\nu21}$, the $S_{\nu11}$, and the $S_{\nu31}$ phonon families. The coupling rates $g_{\alpha\beta}$ versus acoustic mode frequency are plotted in \figref{figmagnetoelastic} for the Kittel magnon-$S_{\nu21}$ interaction (a), the $\{210\}$ magnon-$T_{\nu21}$ interaction (b), the $\{210\}$ magnon-$S_{\nu11}$ interaction (c), and the $\{210\}$ magnon-$S_{\nu31}$ interaction (d), for the first $30$ acoustic modes ($\nu=1,...30$) of each family. In the upper axis of each panel we show the external magnetic field $B_0^*$ required to tune the involved magnon in resonance with the corresponding acoustic phonon. Note that the horizontal and vertical axes are multiplied by $R$ and $\sqrt{R}$ respectively, such that the information displayed in the figure is size-independent. In \figref{figmagnetoelastic}, we observe that the coupling rates can reach very large values regardless of the chosen magnon, especially for spheroidal acoustic modes. For instance, for $R=100$\,nm we find $\vert g\vert\approx 2\pi\times 36$\,MHz for the Kittel-magnon-to-$S_{121}$ phonon coupling and  $\vert g\vert\approx 2\pi\times 32$\,MHz for the $\{210\}$-magnon-to-$S_{231}$ phonon coupling, eight orders of magnitude larger than the dispersive couplings reported for magnets with $R=250\,\mu$m~\cite{ZhangSciAdv2016}. Moreover, the couplings decrease slowly with the acoustic energies, allowing to couple each magnon efficiently to several acoustic modes within experimentally feasible requirements on the external magnetic fields. As an example, for $R=100$\,nm and $B_0\lesssim 5$T, the magnon-phonon resonance condition can be met for the Kittel magnon and acoustic modes up to $S_{10,21}$, and for the $\{210\}$ magnon and spheroidal modes up to $S_{831}$ and $S_{911}$. We conclude that strong magnon-phonon interaction can be reached for multiple magnon-phonon pairs.

 The results in Table~\ref{TableCouplings} and \figref{figmagnetoelastic} show that the micromagnet represents a flexible acoustomagnonic system where two modes, one magnonic and one acoustic, can be tuned into resonance and can interact coherently. In order to characterize the coherent interaction, we estimate the acousto-magnonic cooperativity 
 \begin{equation}\label{cooperativityAM}
     C_{\alpha\beta}\equiv\frac{4g_{\alpha\beta}^2}{\gamma_\alpha\gamma_\beta} \equiv\frac{4g_{\alpha\beta}^2}{\omega_\alpha\gamma_\beta}Q_\alpha,
 \end{equation}
 where $\gamma_\alpha $ and $\gamma_\beta $ are, respectively, the decoherence rates of the phonon and the magnon, and $Q_\alpha$ represents the quality factor of the acoustic mode $\alpha$. Regarding the magnon, linewidths $\gamma_\beta \approx 2\pi\times 1$\,MHz have been reported at cryogenic temperatures and magnon frequencies of $\sim 10$\,GHz for the Kittel mode~\cite{TabuchiPRL2014,KlinglerAPL2017,MaierFlaigPRB2017}, and even lower for inhomogeneous modes such as the $\{210\}$ magnon \cite{MaierFlaigPRB2017,KlinglerAPL2017,GloppePRApplied2019}. 
 Regarding the acoustic quality factors, no measurements have been performed in  micromagnets, although unusually high values ($Q_p \approx 10^5-10^7$) have been reported for larger samples \cite{ZhangSciAdv2016,LeCrawPRL1961}. Moreover, for sufficiently isolated micromechanical resonators, $Q_\alpha$ is known to be limited by indirect (i.e.\ environment-mediated) interactions with other acoustic modes, and reportedly reaches values up to $Q_\alpha \gtrsim  5\times10^{10}$ when consecutive acoustic modes are, as in the present case, largely detuned ($\gtrsim $\,GHz)\cite{MaccabearXiv2019}. Even for moderate values $Q_\alpha \approx 10^6-10^7$, the high-cooperativity regime $C_{\alpha\beta}>1$ can be reached between either the Kittel or the $\{210\}$ magnons and multiple acoustic modes. Moreover, for $Q_\alpha \sim 10^9$, several magnon-phonon pairs reach cooperativities above $100$. These results show that the acoustomagnonic system introduced in this paper can reach the strong coupling regime, where the experimentally elusive acoustic phonons could be probed and coherently manipulated through the magnonic degrees of freedom, which are experimentally accessible, for example through cavities \cite{ZhangSciAdv2016,BaiPRL2015,TabuchiPRL2014,ZhangPRL2014,GoryachevPRApplied2014,YaoNatComm2017,ZhangNPJQ2015,ZhangNatComm2015,GoryachevPRB2018,TabuchiCRP2016,TabuchiScience2015,LachanceQuirioneSciAdv2017,WangPRB2016}, waveguides~\cite{ZhangPRL2016,MaierFlaigPRB2017}, or  near-field magnetic probes \cite{GloppePRApplied2019,Jantoappear,HuilleryArxiv2019}. Additionally, as we will see in the next section, and originally proposed in Ref.~\cite{GonzalezBallesteroarXiv2019}, the acoustic modes could also be probed by the much narrower center-of-mass degrees of freedom, since the center-of-mass motion can be coupled to the Kittel mode through an oscillating inhomegeneous magnetic field (see \figref{figSetupAcMec}).
 
\section{Quantum acoustomechanics}\label{SecAcoustomechanics}

In this section, we consider the interaction between the external and the internal degrees of freedom. First, in Sec.~\ref{secCMmagnonGeneral}, we derive the coupling between the center-of-mass motion of the micromagnet and its magnonic modes, induced by an inhomogeneous magnetic driving field. In Sec.~\ref{secCMmagnonCasestudy} we provide a case study for a specific driving field. In Sec.~\ref{SecACOUSTOMECHANICShigherorder} we derive the acoustomechanical Hamiltonian of Ref.~\cite{GonzalezBallesteroarXiv2019}, and extend our previous result to higher order acoustic phonons. These modes are particularly attractive, because, at a given temperature, the entropy of the high frequency acoustic phonons is lower than the entropy of the fundamental mode, which improves the acoustic cooling of the center-of-mass motion.

\subsection{Magnon-Motion Coupling Through Inhomogeneous Magnetic Driving Field}\label{secCMmagnonGeneral}

\begin{figure} 
	\centering
	\includegraphics[width=\linewidth]{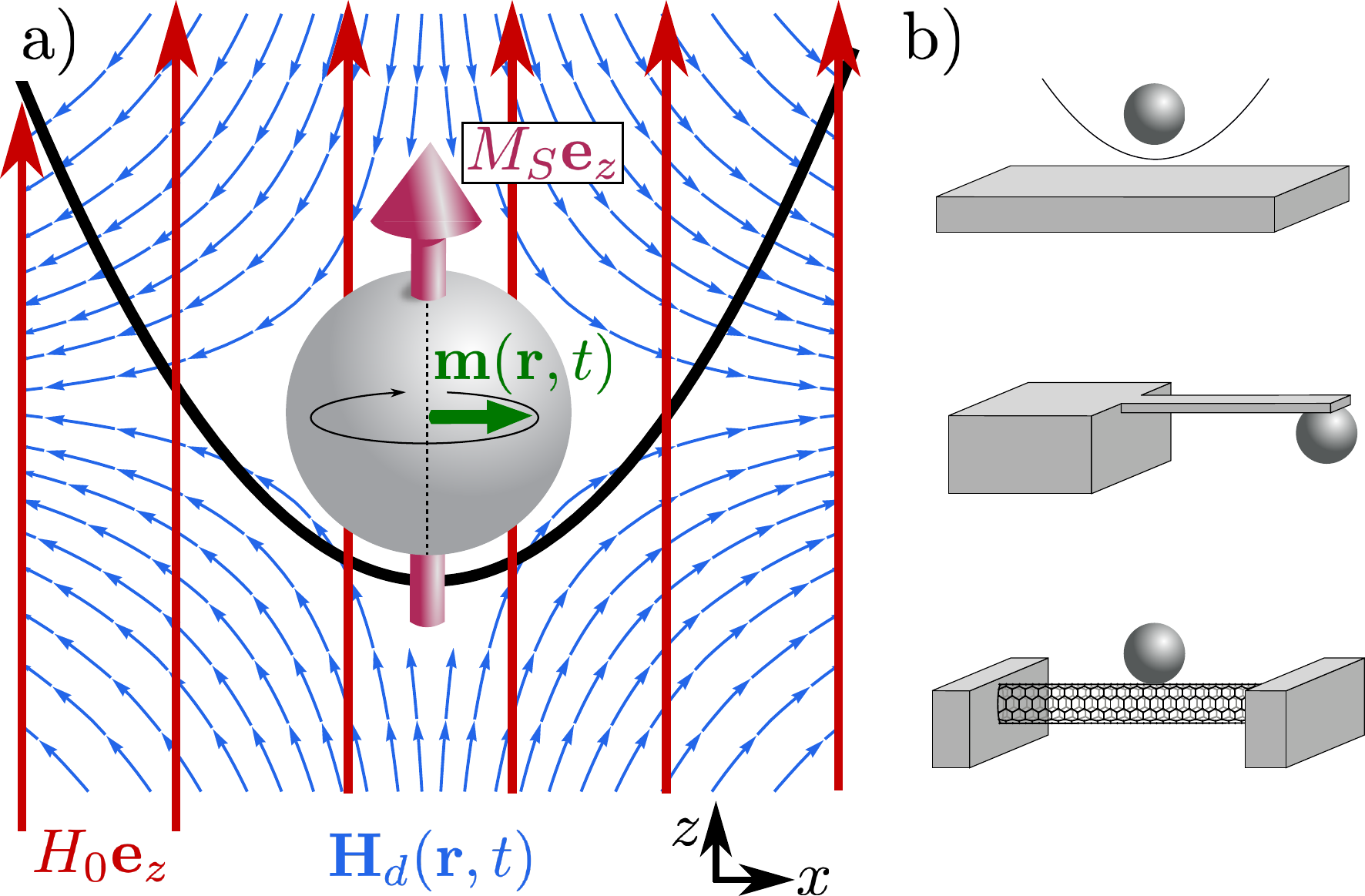}
	\caption{a) A weak inhomogeneous magnetic field drive $\mathbf{H}_d(\mathbf{r},t)$ (blue), superposed with the homogeneous bias field $H_0\mathbf{e}_z$ (red), results in an interaction between the center-of-mass motion of a harmonically trapped micromagnet and its internal magnetization field $\mathbf{m}(\mathbf{r},t)$ (green), i.e.\ its magnonic modes. b) Some potential setups for non-magnetic trapping of the micromagnet: quasielectrostatic levitation (top), tethering to a cantilever (middle), and deposition on a carbon nanotube (bottom).}\label{figSetupAcMec}
\end{figure}

Let us derive the coupling between the center-of-mass motion of the micromagnet and its magnonic modes.
 We assume that the micromagnet is trapped in a three-dimensional harmonic potential with frequencies $\omega_{tx}$, $\omega_{ty}$, and $\omega_{tz}$, either by levitation~\cite{PratCampsPRApp2017,RusconiPRL2017,HuilleryArxiv2019,BarowskiJTP1993} or by weak clamping to a high-Q micromechanical oscillator~\cite{BurgessScience2013,VinanteNatComm2011,ShamsudhinSmall2016,DrugeNJP2014,FischerNJP2019,KolkowitzScience2012} (see \figref{figSetupAcMec}b). In the former case, we assume for simplicity the levitation to be non-magnetic, for instance by optical levitation or by levitation in a Paul trap. We remark, however, that our results could also apply to magnetic levitation~\cite{PratCampsPRApp2017,RusconiPRL2017,HuilleryArxiv2019,BarowskiJTP1993,DrugeNJP2014,Jantoappear}, by appropriately including any additional magnetic fields in the Landau-Lifshitz equations. In the case of a clamped micromagnet, the weak clamping condition amounts to assuming that the expressions for acoustic modes computed under zero-stress boundary conditions remain valid \cite{ZhangSciAdv2016}. 
 The external Hamiltonian in Eq.~\eqref{totalH} is thus given by
\begin{equation}\label{Hcm}
    \hat{H}_{\rm ex}= \hbar\sum_{j=x,y,z}\omega_{tj}\hat{b}^\dagger_j\hat{b}_j,
\end{equation}
in terms of bosonic ladder operators, $[\hat{b}_j,\hat{b}_{j'}^\dagger]=\delta_{jj'}$, which describe
 annihilation and creation of a motional quantum along the direction $j$. Note that, throughout this work, we will not refer to the motional quanta as phonons to avoid confusion with the acoustic phonons described by the operators $\hat{a}_\alpha$. Note also that the micromagnet rotation can be neglected within our approximations \footnote{Indeed, under the assumption of a cubic material undertaken in Sec.~\ref{SecFreeInternalHamiltonian}, the magnetocrystalline anisotropy can be neglected (see Appendix~\ref{appendixMAGNONS}), and with it the main mechanism enabling the interaction between the micromagnet rotation and its internal degrees of freedom~\cite{RusconiPRL2017,RusconiPRB2017}}.

In order to couple the center-of-mass motion of the micromagnet to its magnetic degrees of freedom, we apply an external driving field $\mathbf{H}_d(\mathbf{r},t)$ which is spatially inhomogeneous and time-dependent. This method has been used to couple different phononic reservoirs through a spin qubit \cite{KepesidisPRB2016}. 
We consider the driving field $\mathbf{H}_{d}(\mathbf{r},t)$ as a classical degree of freedom, and assume that it fulfills the weak-driving and the small-curl conditions
\begin{equation}\label{weakdrivingcondition}
    \vert\mathbf{H}_{d}(\mathbf{r},t)\vert \ll H_0 \hspace{0.6cm} \forall \hspace{0.1cm}\mathbf{r}_{\rm magnet},t,
\end{equation}
\begin{equation}\label{curlcondition}
    \vert \nabla\times \mathbf{H}_d(\mathbf{r},t)\vert \vert \mathbf{r}\vert \ll \vert \mathbf{H}_d(\mathbf{r},t) \vert \hspace{0.6cm} \forall \hspace{0.1cm}\mathbf{r}_{\rm magnet},t,
\end{equation}
where the sub-index ``magnet'' above indicates that these conditions must be fulfilled for all spatial positions occupied by the micromagnet during its dynamical evolution.
We include the magnetic driving in the theory by adding it to
the total magnetic field, i.e.\ by redefining Eq.~\eqref{Hlinearization} as $
    \mathbf{H}(\mathbf{r},t) = H_0\mathbf{e}_z  + \mathbf{h}(\mathbf{r},t) + \mathbf{H}_{d}(\mathbf{r},t).
$ 

The addition of the driving magnetic field $\mathbf{H}_d(\mathbf{r},t)$ increases the total energy by
\begin{equation}\label{DeltaEtotal}
    \Delta E = \Delta E_{\rm static} + \Delta E_{\rm magnons},
\end{equation}
where the first and second terms describe the interaction between the driving field and, respectively, the static magnetization, $M_S\mathbf{e}_z$, and the magnetization of the spin waves, $\mathbf{m}(\mathbf{r},t)$. The former contribution is given by the well-known electrodynamical expression \cite{Jackson1975classical}
\begin{equation}\label{DeltaEstatic}
    \Delta E_{\rm static}(t) = -\frac{\mu_0 M_S}{2}\int_{\mathcal{V}(\mathbf{R}_{\rm CM})} dV \mathbf{e}_z\cdot \mathbf{H}_d(\mathbf{r},t),
\end{equation}
where $\mathcal{V}(\mathbf{R}_{\rm CM})$ is a spherical volume of radius $R$ centered at the center-of-mass position of the micromagnet, $\mathbf{R}_{\rm CM}(t)$.
The calculation of the second contribution to Eq.~\eqref{DeltaEtotal} is less straightforward, because the micromagnetic energy functional for the magnons, Eq.~\eqref{micromagneticenergy}, is phenomenological, like the Landau-Lifshitz equations generated by it (see Appendix~\ref{appendixMAGNONS}). Hence, it cannot be generalized directly to arbitrary driving fields. However, for driving fields satisfying the two conditions Eqs.~\eqref{weakdrivingcondition} and \eqref{curlcondition}, it is possible to show that\footnote{This can be proven by following an analogous reasoning as in Appendix~\ref{appendixMAGNONS}: first, we  linearize the Landau-Lifshitz equations also in the variable $\mathbf{H}_d/H_0$, which is small by virtue of the assumption Eq.~\eqref{weakdrivingcondition}. The derivation is then analogous as in Appendix~\ref{appendixMagnetostaticEnergy}, with the substitution $\mathbf{h}\to \mathbf{h}'\equiv\mathbf{h} + \mathbf{H}_d$, as all the assumptions undertaken in the derivation of the magnons still hold, including $\nabla\times\mathbf{h}'\approx0$ due to Eq.~\eqref{curlcondition}. It is then possible to show that the micromagnetic energy under driving is modified to $E_m(\{\mathbf{m}\},\{\mathbf{h}+\mathbf{H}_d\})$.} 
\begin{equation}\label{DeltaEmagnons}
\begin{split}
    \Delta E&_{\rm magnons}(t) =
    \\
    &
    =E_m\left(\{\mathbf{m}\},\{\mathbf{h}+\mathbf{H}_d\}\right)-E_m\left(\{\mathbf{m}\},\{\mathbf{h}\}\right)=
    \\
    &
    =-\frac{\mu_0}{2}\int_{\mathcal{V}(\mathbf{R}_{\rm CM})} dV \mathbf{m}(\mathbf{r}-\mathbf{R}_{\rm CM},t)\cdot\mathbf{H}_d(\mathbf{r},t).
\end{split}
\end{equation} 
The derivation from the Landau-Lifshitz equations coincides in this case with the purely electrodynamical expression Eq.~\eqref{DeltaEstatic}. The total variation in magnetic energy can thus be written as
\begin{equation}
\begin{split}
    &\Delta E (t)  =
    \\
    &
    =
    -\frac{\mu_0}{2}\int_{\mathcal{V}(0)} \!\!\! dV\left[M_S\mathbf{e}_z + \mathbf{m}(\mathbf{r},t)\right]\!\cdot\!\mathbf{H}_d(\mathbf{r}+\mathbf{R}_{\rm CM},t),
\end{split}
\end{equation}
where we have changed the integration variable from $\mathbf{r}$ to $\mathbf{r}-\mathbf{R}_{\rm CM}$. Substituting in the equation above the dynamical variables by their corresponding quantum operators in the Schr\"odinger Picture, we obtain the quantum Hamiltonian describing the interaction between the center-of-mass motion $\hat{\mathbf{R}}_{\rm CM}\equiv(\hat{X},\hat{Y},\hat{Z})$ and the spin wave magnetization,
\begin{equation}\label{potentialHd}
    \hat{V}(t) = -\frac{\mu_0}{2}\int dV\left[M_S\mathbf{e}_z + \hat{\mathbf{m}}(\mathbf{r})\right]\cdot\mathbf{H}_d(\mathbf{r}+\hat{\mathbf{R}}_{\rm CM},t),
\end{equation}
where from now on we omit the explicit specification of the integration domain, namely a spherical volume with radius $R$.

\subsection{Case Study}\label{secCMmagnonCasestudy}

As a specific example, we now discuss the efficient coupling of the center-of-mass motion to a particular magnon mode, namely the Kittel mode. We also assume the following specific form for the driving field,
\begin{equation}\label{Hdparticular}
    \mathbf{H}_d(\mathbf{r},t) = \frac{b}{\mu_0}\left[-x\mathbf{e}_x + z\mathbf{e}_z\right]\cos(\omega_d t),
\end{equation}
i.e., a harmonic oscillation at frequency $\omega_d$ that is parametrized by the field gradient $b$ (dimensions [T/m]). The above spatial profile can be realized, for instance, close to the center of a zero-bias Ioffe-Pritchard trap, or, if the center-of-mass is highly confined along the $Y$ axis, a quadrupole trap \cite{PerezRiosAJP2013,foot2005book,bookReichelVuletic}. 
Conveniently, the above field has exactly zero curl, i.e., $\nabla\times\mathbf{H}_d(\mathbf{r},t) = 0$, thus automatically satisfying the condition Eq.~\eqref{curlcondition}. Furthermore, for bias field $B_0\gtrsim 0.1$T (see \figref{figmagnonfreqs}), the weak driving assumption Eq.~\eqref{weakdrivingcondition} is also fulfilled even for large field gradients $b\lesssim 10^5$T/m,
as the average displacement of the center of mass at room temperature, namely $\langle\hat{X}^2\rangle^{1/2}\approx (k_B T/\rho V \omega_{tj}^2)^{1/2}$, remains small for usual parameters, e.g. it remains below $200$\,nm for $R \ge 10$\,nm and $\omega_{tj} \gtrsim 2\pi\times50$\,kHz.

Using the field Eq.~\eqref{Hdparticular} and the selection rule $\int dV \tilde{\mathbf{m}}_\beta (\mathbf{r}) \propto \delta_{\beta,\{110\}}$, derived in Appendix~\ref{appendixFurtherDerivations}, we write the interaction Eq.~\eqref{potentialHd} as
\begin{equation}\label{Vsecondquantization}
\begin{split}
    \hat{V}&(t)=\frac{bV}{2}\cos(\omega_dt)\bigg[ \mathcal{M}_{K}\hat{X}(\hat{s}_K + \hat{s}^\dagger_K)+
    \\
    &
    -M_S\hat{Z}+
    \sum_{l\text{ even}}\sum_{m=0,\pm2}\sum_n\mathcal{M}_{0\beta}\left(L_\beta\hat{s}_\beta + \text{H.c.}\right)
    \bigg],
\end{split}
\end{equation}
where the selection rules in the last term are easily obtained from the coupling integral $L_\beta \sim V^{-1} \mathbf{e}_x \int dV x\tilde{\mathbf{m}}_\beta(\mathbf{r})$ 
by following a similar procedure as in Appendix~\ref{appendixSELECTION}. According to Eq.~\eqref{Vsecondquantization}, three distinct terms arise from the driving: first, a quadratic coupling between the Kittel magnon and the center-of-mass motion along the $x-$axis; second, a coherent driving of the center of mass along the $z-$axis; third, a coherent driving of a subset of magnons, from which the Kittel mode $(l=1)$ is excluded. Let us remark that these interactions are by no means general, as one can choose a driving field configurations that couples the CM motion to other magnons than the Kittel mode, albeit usually with weaker coupling rates.

We are finally in a position to write explicitly the whole system Hamiltonian, Eq.~\ref{totalH}, by combining the free center-of-mass Hamiltonian, Eq.~\eqref{Hcm}, the internal Hamiltonian describing magnons and phonons, Eq.~\eqref{beampslitterMPInt}, and the interaction with the driving, Eq.~\eqref{Vsecondquantization}. As discussed above, we will assume a bias field $H_0$ such that the Kittel magnon is close to resonance with a single acoustic phonon of the family $S_{\nu21}$, namely $\hat{a}_{\alpha_0}$, while the remaining magnon-phonon pairs are far detuned. We emphasize that, within the $S_{\nu21}$ family, the choice of acoustic phonon is free, and in the following we will characterize the system for different acoustic modes.
In the scenario described above, the total Hamiltonian of the system is
\begin{equation}\label{Htotalprov}
\begin{split}
    \frac{\hat{H}(t)}{\hbar}&=
    \sum_{j}\omega_{tj}\hat{b}^\dagger_j\hat{b}_j + \omega_{\alpha_0}\hat{a}^\dagger_{\alpha_0}\hat{a}_{\alpha_0} + \sum_\beta\omega_\beta\hat{s}^\dagger_\beta\hat{s}_\beta+
    \\
    &+ 
    (g_{\alpha_0,K}\hat{s}_K\hat{a}^\dagger_{\alpha_0} + \text{H.c.})
    \\
    &
    +\frac{bV}{2\hbar}\cos(\omega_dt)\bigg[  \mathcal{M}_{K}\hat{X}(\hat{s}_K + \hat{s}^\dagger_K)+
    \\
    &
    -\! M_S\hat{Z} + \!\!
    \sum_{l\text{ even}}\sum_{m=0,\pm2}\!\sum_n\mathcal{M}_{0\beta}\!\left(L_\beta\hat{s}_\beta \!+ \!\text{H.c.}\right)\!
    \bigg]\!.
\end{split}
\end{equation}
We focus on the regime $\omega_d\sim\omega_K$, where the contribution $\propto \hat{X}(\hat{s}_K + \text{H.c.})\cos(\omega_dt)$ oscillates slowly in the interaction picture, thus maximizing the interaction between the center of mass and the Kittel mode. In this regime, the rapidly oscillating terms $\propto \cos(\omega_d t)\hat{Z}$ and $\propto \cos(\omega_d t)\hat{s}_\beta$ can be neglected under a rotating wave approximation, since $\omega_d\gtrsim2\pi \times 1$\,GHz$ \gg \omega_{tx}$ for typical trapping frequencies, and since consecutive magnon modes are largely detuned, $\vert\omega_d\pm\omega_\beta\vert \sim\omega_K$ (see \figref{figmagnonfreqs})\footnote{Note that at some particular values of the external field $B_0$ some magnons become degenerate. In this work we assume $B_0$ does not take any of these critical values, such that the Kittel mode is sufficiently detuned from any other magnonic mode. Note also that the $\{430\}$ magnon, which is always degenerate with the Kittel mode~\cite{WalkerPhysRev1957,Fletcher59,Roschmann}, is not included in the last term of Eq.~\eqref{Htotalprov} and can thus also be ignored.}.
 Under this approximation, Eq.~\eqref{Htotalprov} reduces to a three-mode Hamiltonian involving the selected acoustic phonon $\hat{a}_{\alpha_0}$, the Kittel magnon $\hat{s}_K$, and the motion along the $x-$axis, 
\begin{equation}\label{H3mode}
 \begin{split}
    \frac{\hat{H}(t)}{\hbar}&=
    \omega_{tx}\hat{b}^\dagger\hat{b} + \omega_{p}\hat{a}^\dagger\hat{a} + \omega_m\hat{s}^\dagger\hat{s}+ 
    (g\hat{s}\hat{a}^\dagger + \text{H.c.})+
    \\
    &
    +G_x\cos(\omega_dt) (\hat{b} + \hat{b}^\dagger)(\hat{s} + \hat{s}^\dagger).
\end{split}   
\end{equation}
This is the starting Hamiltonian in Ref.~\cite{GonzalezBallesteroarXiv2019}.
Here and in the following, we drop the indices in the operators and, for simplicity, relabel the magnon and phonon frequencies and the magnon-phonon coupling as $\omega_m$, $\omega_p$, and $g$, respectively. We also define the magnon-to-center-of-mass coupling rate
\begin{equation}\label{Gx}
    G_x \equiv \frac{bx_0\mathcal{M}_{K}V}{2\hbar} = \frac{b}{4}\sqrt{\frac{\vert\gamma\vert M_S}{\rho \omega_{tx}}},
\end{equation}
with $x_0\equiv(2\rho V\omega_{tx}/\hbar)^{-1/2}$. Equation~\eqref{Gx} has a straightforward interpretation as the magnetic dipole moment associated with the Kittel mode, namely $\mathcal{M}_K V$, times the average magnetic field felt by the micromagnet on its trajectory along the $x-$axis, namely $bx_0/2$. Note that this coupling does not depend on the size of the micromagnet.

As detailed in Ref.~\cite{GonzalezBallesteroarXiv2019}, the density matrix $\hat{\rho}$ of the three-mode system described above obeys the dynamical equation
\begin{equation}\label{Mequation3mode}
    \frac{d}{dt}\hat{\rho} = \frac{1}{i\hbar}\left[\hat{H}(t),\hat{\rho}\right]+\gamma_m\mathcal{L}_m[\hat \rho]+\gamma_p\mathcal{L}_p[\hat \rho]+\gamma_x\mathcal{L}_{x}[\hat \rho],
\end{equation}
where $\hat{H}(t)$ is given by Eq.~\eqref{H3mode}, and the remaining three terms represent the dissipation of the magnon, the phonon, and the CM motion, respectively, through contact with thermal reservoirs at a common temperature $T_e$~\cite{GonzalezBallesteroarXiv2019}, i.e., $\mathcal{L}_j[\hat \rho]=(\bar{n}_j+1)L_{\hat{o}_j}[\hat \rho]+\bar{n}_jL_{\hat{o}^\dagger_j}[\hat \rho]$ and $L_{\hat{o}_j}[\hat \rho]\equiv\hat{o}\hat \rho\hat{o}^\dagger-\{\hat{o}^\dagger\hat{o},\hat \rho\}/2$, where $\{\hat{o}_m,\hat{o}_p,\hat{o}_x\}\equiv\{\hat{s},\hat{a},\hat{b}\}$ and $\bar{n}_j\equiv (\exp\left[\hbar\omega_j/k_B T_{e,j}\right]-1)^{-1}$ is the Bose-Einstein occupation factor. Regarding the corresponding dissipation rates the Kittel magnon linewidth is on the order $\gamma_m \approx 2\pi\times 1$\,MHz~\cite{TabuchiPRL2014,KlinglerAPL2017,MaierFlaigPRB2017} for YIG as discussed in Sec.~\ref{SecMagnetoelastic}, and we will use this value from now on. 
To describe the dissipation of the acoustic phonon and the center-of-mass mode, we introduce their respective quality factors, $Q_p=\omega_p/\gamma_p$ and $Q_x=\omega_{tx}/\gamma_x$. Expected values for the former have been discussed in Sec.~\ref{SecMagnetoelastic}. Although the dissipation of the center-of-mass mode greatly depends on the trapping mechanism, experimental measurements of $Q_x \gtrsim 10^8$ have been reported both in nanofabricated resonators 
\cite{NortePRL2016,ReinhardtPRX0216,GhadimiScience2018,MasonNatPhys2019}
and levitated systems \cite{GieselerNatPhys2013,GieselerPRL2012}.
The master equation Eq.~\eqref{Mequation3mode} is quadratic, which allows us to solve it exactly in the following.

\subsection{Acoustomechanics with higher order acoustic phonons}\label{SecACOUSTOMECHANICShigherorder}

As detailed in Ref.~\cite{GonzalezBallesteroarXiv2019}, the parameters of the three-mode Hamiltonian Eq.~\eqref{H3mode} can be adjusted to efficiently couple an acoustic phonon to the center-of-mass motion of the micromagnet. Here, we briefly summarize the derivation and extend our previous results to higher order $S_{\nu21}$ acoustic modes.

We start by diagonalizing the internal Hamiltonian, i.e., by 
writing
\begin{equation}
    \omega_m\hat{s}^\dagger\hat{s} + \omega_p\hat{a}^\dagger\hat{a} + \left(g\hat{s}^\dagger\hat{a} + \text{H.c.}\right) = \sum_{q=1,2}\omega_q\hat{c}_q^\dagger\hat{c}_q,
\end{equation}
in terms of hybrid magnon-phonon modes described by the bosonic operators $\hat{c}_q$ and $\hat{c}_q^\dagger$, given by~\cite{KusturaPRA2019}
\begin{equation}\label{Bogoliubovtrafo}
    \left[
    \begin{array}{c}
         \hat{c}_1  \\
         \hat{c}_2 
    \end{array}
    \right]=\frac{-1}{\sqrt{1+\chi^2}}\left[
    \begin{array}{cc}
        \chi & -1 \\
        1 & \chi
    \end{array}
    \right]\left[
    \begin{array}{c}
         \hat{a}  \\
         \hat{s}
    \end{array}
    \right].
\end{equation}
Here, the hybridization parameter is $\chi = -2g[\Delta-(\Delta^2+4g^2)^{1/2}]^{-1}$,
with $\Delta \equiv \omega_m-\omega_p$, and the corresponding eigenfrequencies are $\omega_1=\omega_p-g/\chi$ and $\omega_2=\omega_m+g/\chi$, respectively. Both the hybridization parameter and the normal mode eigenfrequencies are tunable through the magnon frequency, i.e., through the external bias field $H_0$.

\begin{figure} 
	\centering
	\includegraphics[width=\linewidth]{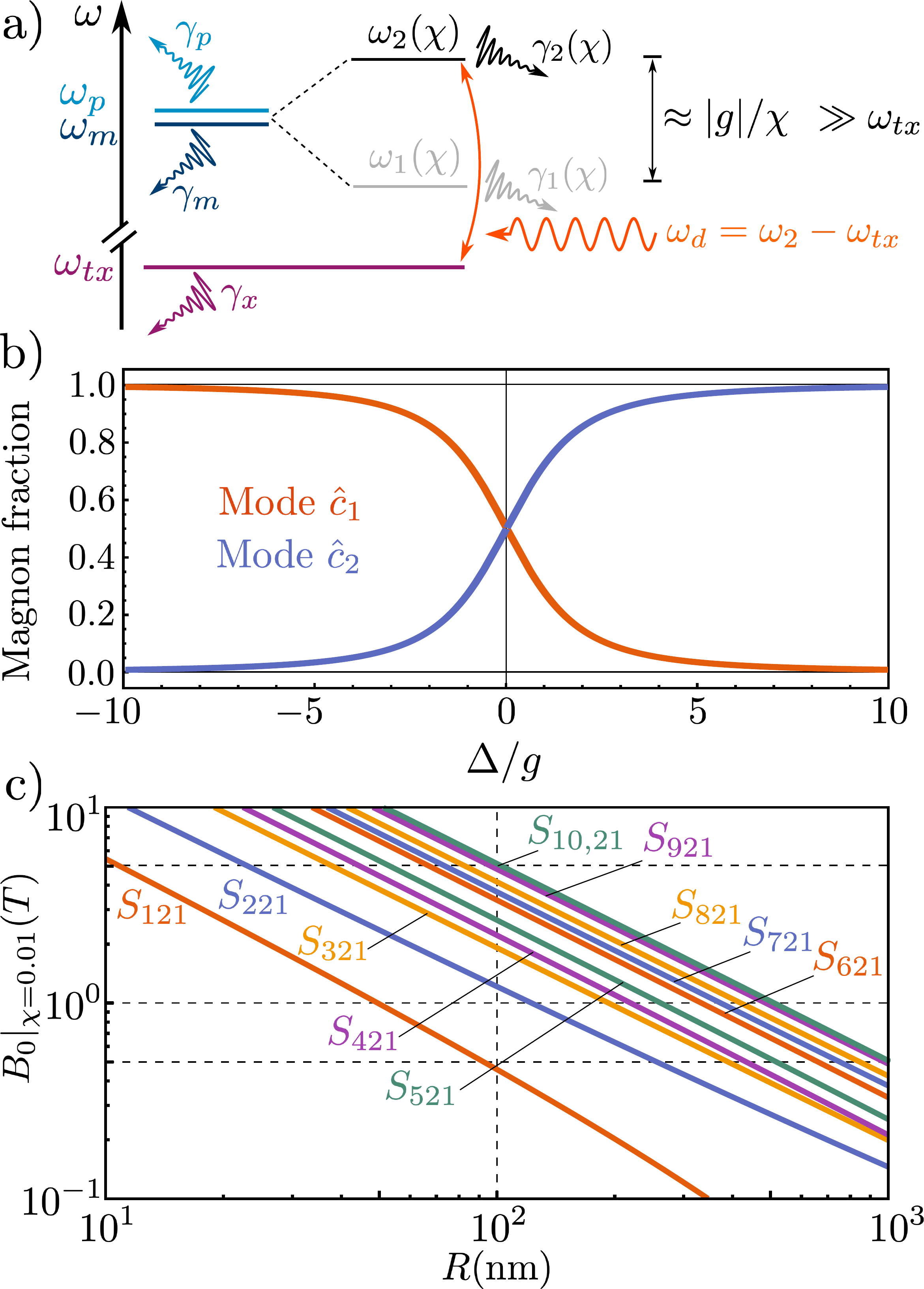}
	\caption{a) Energy level diagram of the proposed acoustomechanical system. b) Magnon fraction of each hybrid magnon-phonon normal mode as a function of the adimensional parameter $g/\Delta$. c) External bias field $B_0=\mu_0H_0$ required to achieve $\chi=\chi_0=10^{-2}$ versus micromagnet radius $R$, for the $10$ lowest energy acoustic modes, that are able to interact with the Kittel magnon. 
	}\label{figBogoliubovDecomposition}
\end{figure}

We now invert the transformation Eq.~\eqref{Bogoliubovtrafo}, introduce it into the Hamiltonian Eq.~\eqref{H3mode}, apply the unitary transformation $U(t)=\exp[i\omega_d t\sum_q\hat{c}^\dagger_q\hat{c}_q]$, and neglect the rapidly oscillating terms under the rotating wave approximation $2\omega_d \gg \vert G_{x}\vert$. This leads to the time-independent Hamiltonian
\begin{equation}\label{HbogoliubovRWA}
    \begin{split}
    \frac{\hat{H}}{\hbar}&\approx
    \omega_{tx}\hat{b}^\dagger\hat{b} +\sum_{q=1,2}\Delta_q\hat{c}_q^\dagger\hat{c}_q
    +
    \\
    &
    + (\hat{b} + \hat{b}^\dagger)\sum_{q=1,2}G_{xq}(\hat{c}_q + \hat{c}_q^\dagger),
\end{split}
\end{equation}
where we defined $\Delta_q \equiv \omega_q-\omega_d$ and $G_{x1}=-G_{x2}/\chi=G_x(2\sqrt{1+\chi^2})^{-1}$.
This simplified Hamiltonian allows to tailor a step-by-step recipe for devising an acoustomechanical system, which couples the center-of-mass motion of the micromagnet to a hybrid, predominantly acoustic, normal mode (see\figref{figBogoliubovDecomposition}a):
First, we set the hybridization parameter to a small value, say $\chi = \chi_0 = 10^{-2}$ for definiteness, by tuning the external magnetic bias $B_0$ to fulfill $\Delta \equiv \vert\gamma\vert B_0-\omega_p= g(\chi_0-\chi_0^{-1})$.
In this case, the mode $\hat{c}_2$ is predominantly acoustic, with only a small magnon fraction $\chi_0^2/(1+\chi_0^2) \approx \chi_0^2 = 10^{-4}$, while the mode $\hat{c}_1$ is mainly magnonic\footnote{Note that this magnon fraction can be attained by other choices of $\chi$ such as $\chi=-10^{-2}$, or $ \chi=\pm10^2$, the latter of which would result in the mode $\hat{c}_1$ becoming mainly acoustic instead of $\hat{c}_2$. All these routes are experimentally feasible albeit slightly more resource-demanding, as the required external fields $B_0$ are larger.}
. As shown in \figref{figBogoliubovDecomposition}c, the external field $B_0$ required to reach $\chi_0=10^{-2}$ lies within experimentally achievable values $B_0 <10$T for a wide range of particle sizes and numerous acoustic modes.
Second, we tune the mainly acoustic mode $\hat{c}_2$ in resonance with the center-of-mass motion by tuning the driving frequency $\omega_d$, i.e., by setting $\Delta_2 \equiv \omega_2 - \omega_d = \omega_{tx}$.
The third and final step is to increase the coupling rate $G_{x2}$ [Eq.~\eqref{Gx}] between the center-of-mass motion and the mode $\hat{c}_2$ by increasing the magnetic field gradient $b$.
Since $b$ is still a free parameter, it allows us to compensate for the decrease of $G_{x2}$ with $\chi$.
In this way, we engineer an effective two-mode system where the center of mass is coupled to the largely ($99.99\%$) acoustic mode $\hat{c}_2$. The remaining, mainly magnonic mode $\hat{c}_1$ is detuned with respect to these two modes by $\vert\delta_{12}\vert \equiv \vert\Delta_1-\Delta_2\vert = \sqrt{\Delta^2 + 4g^2}\approx \vert g \vert/\chi_0 \approx 100\vert g \vert \gg G_{x1}$ for all values of $R$ and $b$ consistent with our approximations, and thus plays a negligible role in the dynamics.

The dissipators in Eq.~\eqref{Mequation3mode} are transformed in a similar fashion as
in the derivation of Eq.~\eqref{HbogoliubovRWA}, namely we express them in terms of the hybrid normal modes $\hat{c}_q$, and apply the same unitary transformation to obtain
\begin{equation}\label{DissipatorsBogoliubov}
\begin{split}
    \gamma_m&\mathcal{L}_m[\hat{\rho}]+\gamma_p\mathcal{L}_p[\hat{\rho}] = 
    \\
    &
    =\sum_q \left(\gamma_{q +}L_{\hat{c}^\dagger_q}[\hat{\rho}]+\gamma_{q -}L_{\hat{c}_q}[\hat{\rho}] \right)+
    \\
    &
    + \Gamma_{12}\left(e^{-i\delta_{12}t}L_{\hat{c}_1\hat{c}_2}[\hat{\rho}]+e^{i\delta_{12}t}L_{\hat{c}_2\hat{c}_1}[\hat{\rho}]\right)
    \\
    &
    + (\Gamma_{12}+\delta\Gamma)\left(e^{i\delta_{12}t}L_{\hat{c}_1^\dagger\hat{c}_2^\dagger}[\hat{\rho}]+e^{-i\delta_{12}t}L_{\hat{c}_2^\dagger\hat{c}_1^\dagger}[\hat{\rho}]\right).
\end{split}
\end{equation}
Here, the second line corresponds to the individual decoherence of each hybrid mode, with emission and absorption rates $\gamma_{q+}=(\bar{n}_m\gamma_m[\delta_{q1}+\chi^2\delta_{q2}]+\bar{n}_p\gamma_p[\delta_{q2}+\chi^2\delta_{q1}])/(1+\chi^2)$ and $\gamma_{q-}=([\bar{n}_m+1]\gamma_m[\delta_{q1}+\chi^2\delta_{q2}]+[\bar{n}_p+1]\gamma_p[\delta_{q2}+\chi^2\delta_{q1}])/(1+\chi^2)$, respectively. As can be readily checked from the equations of motion for $\langle \hat{c}_q\rangle$ generated by Eq.~\eqref{Mequation3mode}, the differences  $\gamma_q\equiv\gamma_{q-}-\gamma_{q+}=(\gamma_m[\delta_{q1}+\chi^2\delta_{q2}]+\gamma_p[\delta_{q2}+\chi^2\delta_{q1}])/(1+\chi^2)$ correspond to the linewidths of the normal modes,
which are also hybridized linewidths composed by a magnonic and a phononic contribution. 
The third and fourth lines in Eq.~\eqref{DissipatorsBogoliubov} correspond to an incoherent interaction between the two normal modes, described by the generalized Lindblad dissipator $L_{\hat{a}\hat{b}}\left[\hat{\rho}\right] \equiv \hat{a}\hat{\rho}\hat{b}^\dagger -\{\hat{b}^\dagger\hat{a},\hat{\rho}\}/2$, and the rates $\Gamma_{12}=(\gamma_p[1+\bar{n}_p]-\gamma_m[1+\bar{n}_m])\chi/(1+\chi^2)$ and $\delta\Gamma=\chi(\gamma_m-\gamma_p)/(1+\chi^2)$. According to the equations of motion for $\langle \hat{c}_q\rangle$ generated by Eq.~\eqref{Mequation3mode}, 
these dissipators induce an incoherent coupling between the two normal modes $\hat{c}_1$ and $\hat{c}_2$ characterized by a time-dependent rate $(\delta\Gamma/2)\exp(\pm i\delta_{12}t)$. This allows us to neglect the dissipative interaction terms in Eq.~\eqref{DissipatorsBogoliubov} under a rotating wave approximation, as $\vert\delta\Gamma/\delta_{12}\vert\approx\chi^2\vert(\gamma_m-\gamma_p)/g\vert \ll 1$ for $Q_p\gtrsim 10^2$, $R\lesssim 10\,\mu$m and for all acoustic modes $S_{\nu21}$ up to at least $\nu=20$. The final Master Equation of the acoustomechanical system thus reduces to
\begin{equation}\label{finalMasterEquation}
\begin{split}
    \frac{d}{dt}\hat{\rho} \approx & \frac{1}{i\hbar}\left[\hat{H},\hat{\rho}\right]+\gamma_x\mathcal{L}_{x}[\hat \rho]+
    \\
    &
    +\sum_q \left(\gamma_{q +}L_{\hat{c}^\dagger_q}[\hat{\rho}]+\gamma_{q -}L_{\hat{c}_q}[\hat{\rho}] \right),
\end{split}
\end{equation}
with $\hat{H}$ given by Eq.~\eqref{HbogoliubovRWA}.

\begin{figure} 
	\centering
	\includegraphics[width=\linewidth]{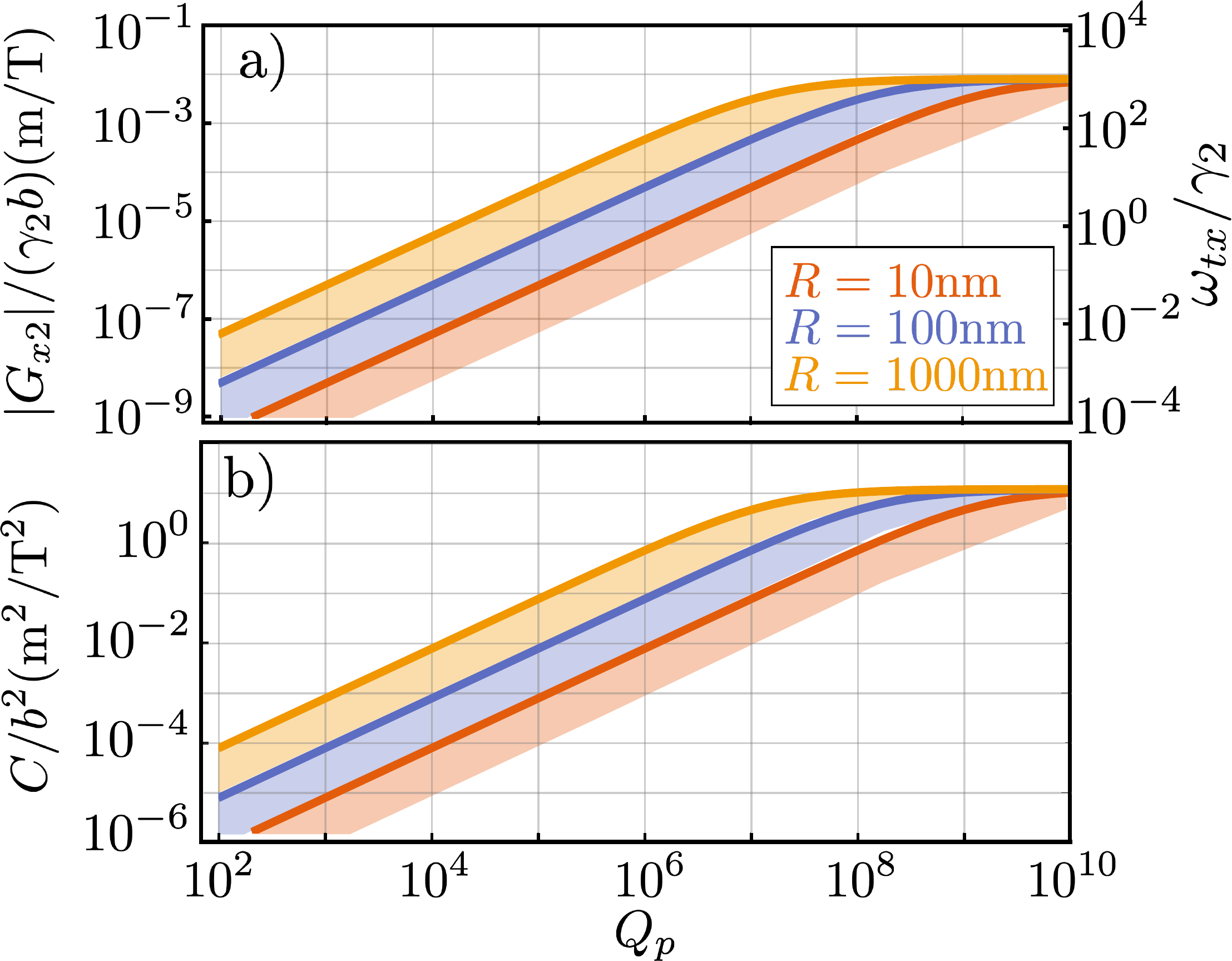}
	\caption{Acoustomechanical figures of merit versus acoustic quality factor $Q_p$, for $Q_x = 10^8$ and $\omega_{tx}=2\pi\times200$\,kHz. a) Coupling between center of mass and mode $\hat{c}_2$ normalized to magnetic field gradient $b$ and linewidth $\gamma_2$ (left), and ratio of center-of-mass frequency to linewidth $\gamma_2$ (right). b) Cooperativity $C=4G_{x2}^2/(\gamma_x\gamma_m)$ normalized to square of the field gradient $b^2$. The solid lines for each radius correspond to the $S_{121}$ acoustic mode,
	whereas the lower ends of the shaded areas correspond to the $S_{10,21}$ acoustic mode. The curves for all the acoustic modes with $1\le\nu\le10$ lie, in descending order, within the shaded area. 
	}\label{figAMparameters}
\end{figure}

As detailed in Ref. \cite{GonzalezBallesteroarXiv2019}, we can draw an analogy between our acoustomechanical system and a linearized optomechanical system where a low-frequency mechanical mode, in our case the center-of-mass motion of the micromagnet, is linearly coupled to a high-frequency optical mode, in our case the hybrid mode $\hat{c}_2$, which for simplicity we will refer to as acoustic mode from now on. The three most common optomechanical figures of merit~\cite{AspelmeyerRMP2014}, namely the resolved sideband parameter $\omega_{tx}/\gamma_2$, the normalized coupling rate $\vert G_{x2}\vert/\gamma_2$, and the cooperativity $C\equiv 4\vert G_{x2}\vert^2/(\gamma_2\gamma_x)$ were shown in Ref.~\cite{GonzalezBallesteroarXiv2019} to be high and tunable when the chosen acoustic phonon for the acoustomechanical system is the $S_{121}$ mode. In \figref{figAMparameters}, 
we show the same figures of merit for the first $10$ acoustic $S_{\nu21}$ modes and typical parameters in a levitation setup, namely $\omega_{tx}=2\pi\times200$\,kHz and $Q_x = 10^8$. As evidenced by \figref{figAMparameters}, all the figures of merit decrease for higher frequency phonons, because the acoustic linewidth $\gamma_p$, and thus the linewidth $\gamma_2$, increases for a given quality factor $Q_p$. Aside from this effect, the strong performance and tunability of the acoustomechanical system reported in Ref.~\cite{GonzalezBallesteroarXiv2019} clearly extends to higher order phononic modes. Specifically, both the resolved sideband regime ($\omega_{tx}>\gamma_2$), the weak and strong coupling regimes ($\vert G_{x2}\vert/\max[\gamma_x,\gamma_2]=\vert G_{x2}\vert/\gamma_2<1$ and $\vert G_{x2}\vert/\gamma_2>1$, respectively), and the high cooperativity regime ($C>1$) can be attained for all $S_{\nu21}$ acoustic modes up to $\nu=10$, even for moderate values $Q_p\sim 10^6-10^7$, $R\ge 100$\,nm, and feasible magnetic gradients~\cite{ReichelPRL1999,MachlufNatComm2013} $b\lesssim 10^3$T/m. Similarly, the strong quantum cooperativity regime $C>\bar{n}_p\bar{n}_x$ can be achieved at cryogenic temperatures ($\bar{n}_p\bar{n}_x<10^4$ at $T_e=100$mK) with slightly more demanding, but still feasible~\cite{MachlufNatComm2013} field gradients on the order of $b\approx 10^3$ T/m. 

\begin{figure} 
	\centering
	\includegraphics[width=\linewidth]{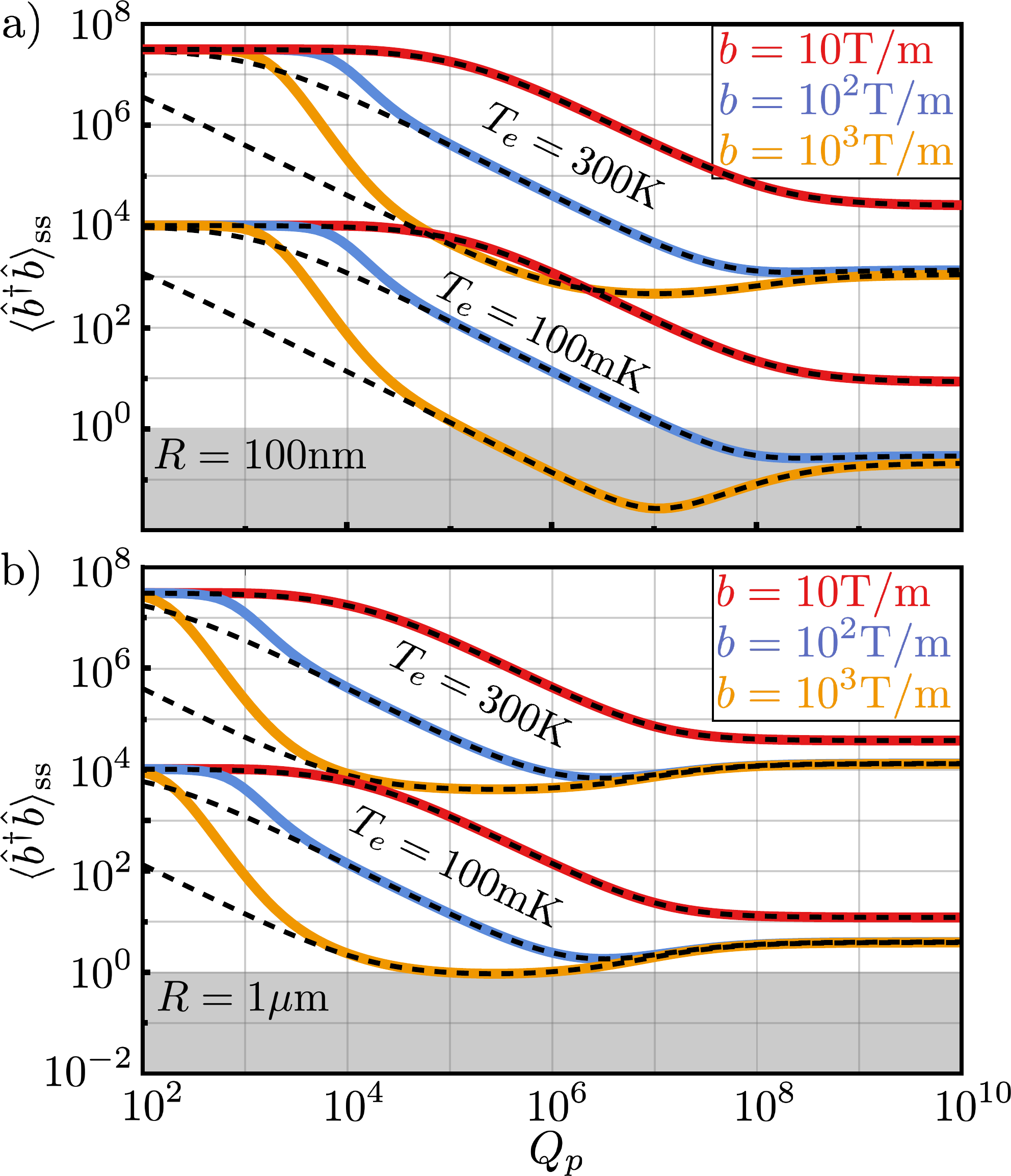}
	\caption{Steady-state center-of-mass occupation versus acoustic quality factor, for the $S_{121}$ acoustic phonon and three different values of field gradient $b$. The parameters are $Q_x=10^8$, $\omega_{tx} = 2\pi\times 200$\,kHz, and $R=100$\,nm (panel a) or $R=1\,\mu$m (panel b). The three upper (lower) curves in each panel correspond to environments at $T_e=300$K ($T_e=100$mK). The dashed lines depict the approximation Eq.~\eqref{approxOCCUPATION}. The shaded area corresponds to ground-state cooling, $\langle\hat{b}^\dagger\hat{b}\rangle_{\rm ss}<1$.
	}\label{figCMcoolingS121}
\end{figure}

\begin{figure} 
	\centering
	\includegraphics[width=\linewidth]{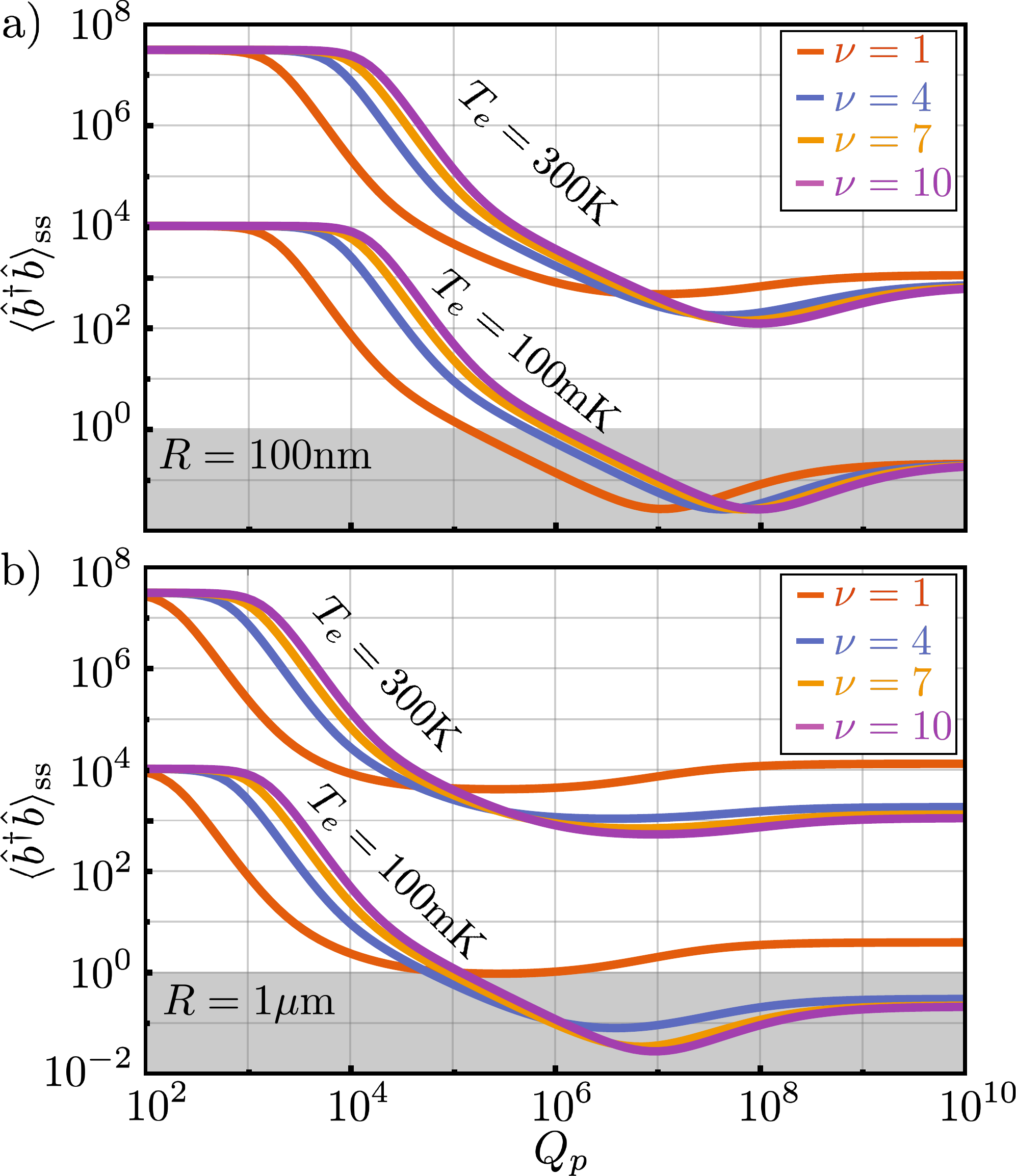}
	\caption{Steady-state center-of-mass occupation versus acoustic quality factor, for different acoustic $S_{\nu21}$ modes, and for $R=100$\,nm (panel a) or $R=1\,\mu$m (panel b). We take the same parameters as in \figref{figCMcoolingS121}, and  $b=10^3$T/m. The four upper (lower) curves in each panel correspond to environments at $T_e=300$K ($T_e=100$mK). The shaded area corresponds to ground-state cooling, $\langle\hat{b}^\dagger\hat{b}\rangle_{\rm ss}<1$.
	}\label{figCMcoolingNU}
\end{figure}

\begin{figure} 
	\centering
	\includegraphics[width=\linewidth]{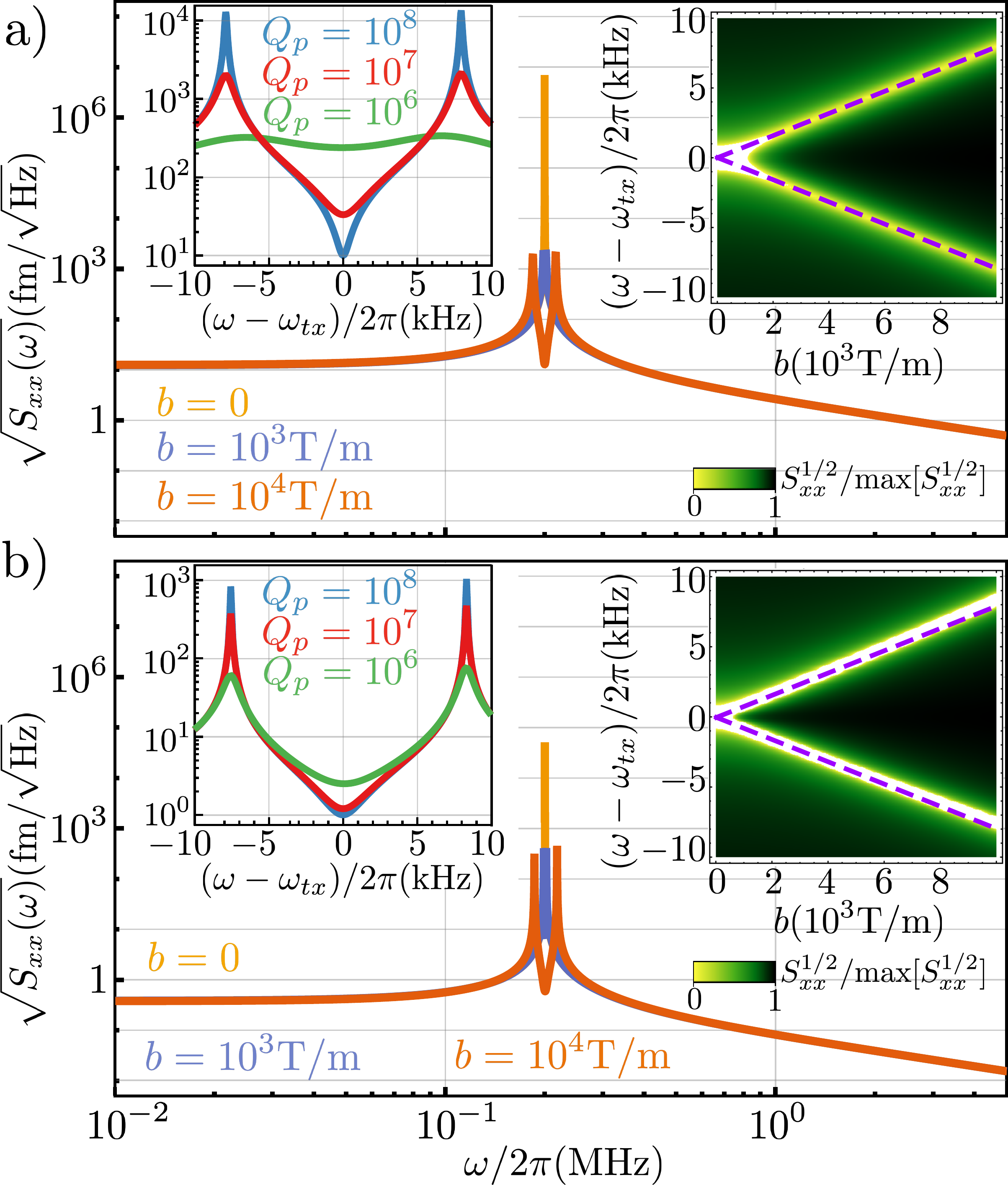}
	\caption{Position power spectral density for a micromagnet with $R=100$\,nm (panel a) or $R=1\,\mu$m (panel b), for $T_e=300$K, $Q_p=10^7$, $Q_x=10^5$, and coupling to the $S_{121}$ acoustic mode. In the main panels, different curves correspond to $b=0$ (orange), $b=10^3$T/m (blue), and $b=10^4$T/m (red). The left insets show the power spectral density near $\omega = \omega_{tx}$ at $b=10^4$T/m, for different values of the acoustic quality factor $Q_p$. The right insets show the power spectral density in normalized units as a function of frequency $\omega$ and magnetic field gradient $b$. The dashed curves correspond to $\pm\vert G_{x2}\vert$ as a function of $b$.
	}\label{figPSD}
\end{figure}

\begin{figure} 
	\centering
	\includegraphics[width=\linewidth]{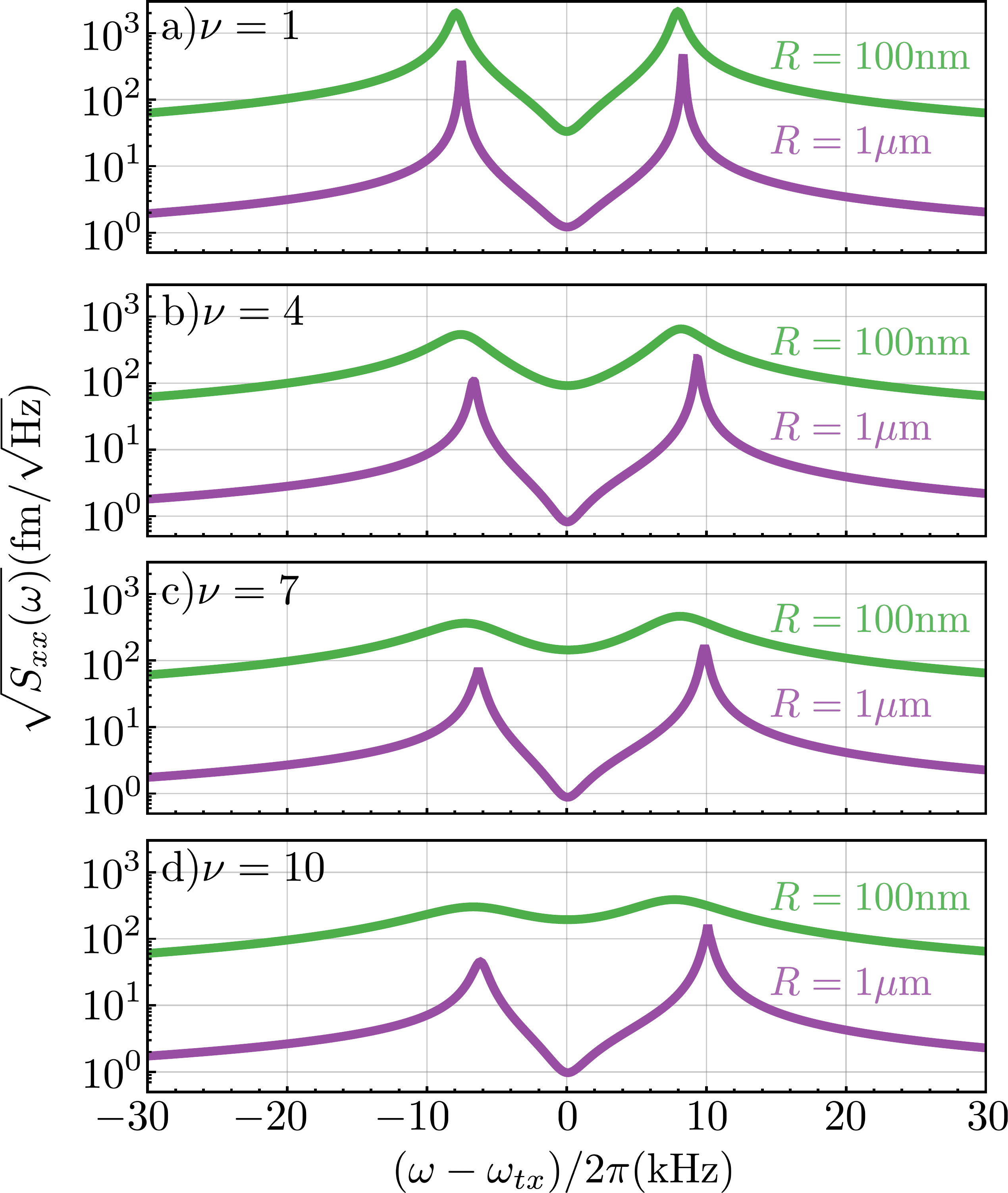}
	\caption{Position power spectral density for $Q_p=10^7$, $b=10^4$T/m, and two different micromagnet sizes, $R=100$\,nm (green curves) and $R=1\,\mu$m (purple curves). The different panels (a) to (d) correspond to different acoustic phonons $S_{\nu21}$ coupled to the center-of-mass motion.
	}\label{figPSDvariousNU}
\end{figure}

Since the large acoustomechanical coupling extends to the higher-order acoustic modes, they can be used for the same applications as the ones we envisioned for the fundamental mode $S_{121}$ in Ref.~\cite{GonzalezBallesteroarXiv2019}, namely acoustic cooling of the center-of-mass motion and probing the acoustic modes through displacement sensing. We begin our discussion with the acoustic cooling of the micromagnet, which is efficient in the resolved-sideband, high-cooperativity, and weak-coupling regime~\cite{GenesPRA2008,MarquardtPRL2007,WilsonRaePRL2007}. First we briefly revisit the steady-state occupation of the motional mode when it is coupled to the $S_{121}$ acoustic phonon, shown in  \figref{figCMcoolingS121}. As discussed in Ref.~\cite{GonzalezBallesteroarXiv2019}, lower center-of-mass  occupations are reached in cryogenic environments ($T_e=100$mK) and smaller micromagnet sizes due to the lower occupation $\bar{n}_p$ of the cooling mode, namely the acoustic phonon. Larger acoustomechanical couplings, enabled by larger field gradients $b$, also enhance the motional cooling. For $R=100$\,nm, ground state cooling is attained at $b=10^3$T/m and $T=100$mK. The dashed lines in \figref{figCMcoolingS121} represent the approximate expression
\begin{equation}\label{approxOCCUPATION}
    \langle\hat{b}^\dagger\hat{b}\rangle_{\rm ss} \approx \frac{1}{C+1}\bar{n}_x + \frac{C}{C+1}\left(\frac{\gamma_{2+} + \gamma_x\bar{n}_x}{\gamma_2}\right),
\end{equation}
obtained by neglecting the far-detuned mode $\hat{c}_1$ and the counter-rotating terms $\propto \hat{b}\hat{c}_2$ in the Master Equation, and in the limit $\gamma_x \ll\gamma_2,\gamma_{2+}$. 
This approximation is very accurate (see \figref{figCMcoolingS121}) 
except at very low values of $Q_p$ (very large acoustic linewidths), where the counter-rotating terms $\propto \hat{b}\hat{c}_2$ become relevant. Thus, Eq.~\eqref{approxOCCUPATION} allows us to identify the factors that limit the lowest possible occupation. At room temperature or considering large sizes $R$ (i.e.\ lower acoustic frequencies), and for sufficiently high cooperativity, the second term in Eq.~\eqref{approxOCCUPATION} dominates, and $\langle\hat{b}^\dagger\hat{b}\rangle_{\rm ss} \approx \gamma_{2+}/\gamma_{2}$, indicating that the limiting factor is the thermal ocupation $\bar{n}_p$ of the environment. In contrast, for $R=100$\,nm and lower temperatures, e.g. $T_e=100$mK, the higher acoustic frequencies result in a negligible acoustic occupation ($\bar{n}_p\approx 10^{-3}$), and the remaining two terms in  Eq.~\eqref{approxOCCUPATION} dominate. In this case, the minimum occupation is very well approximated by $\bar{n}_x(C^{-1}+\gamma_x/\gamma_2) =  \bar{n}_x(\gamma_x/\gamma_2)[1+(\gamma_2/2\vert G_{x2}\vert)^2]\approx 0.026$ for $b=10^3$T/m and $Q_p=10^7$. This corresponds to a cooperativity-limited minimum occupation, i.e., the occupation is limited by the decoherence rates of each component of the acoustomechanical system.

Since most of the curves in \figref{figCMcoolingS121} are limited by the thermal occupation of the acoustic mode, the cooling improves by choosing higher-frequency acoustic modes $S_{\nu21}$, which are less occupied. \figref{figCMcoolingNU} shows the steady-state center-of-mass occupation for $b=10^3$T/m and four different acoustic modes. Indeed, all the occupations that were thermally limited for the $S_{121}$ mode are significantly reduced for the higher frequency acoustic modes. For example, for $R=1\,\mu$m and $T=300$K the occupation remains thermally limited even beyond $\nu=20$, due to both the high-temperature environment and the low acoustic mode frequencies.
However, once the thermal occupation of the acoustic mode becomes negligible we enter the cooperativity limited regime and coupling to higher frequency acoustic modes is not advantageous anymore.
The transition into the cooperativity-limited regime is evidenced in  \figref{figCMcoolingNU} by a saturation of the curve minima, which become independent of the acoustic mode for $\nu\approx 7-10$, for example in the case  $(R,T_e)=(100$\,nm$,300\text{K})$ and $(R,T_e)=(1\,\mu$m$,100\text{mK})$.
Finally, note that at cryogenic temperatures, the improved cooling through higher order phonons allows for ground-state cooling even for relatively large magnets. 
The larger bias fields needed to work with these higher order modes are still well within experimental capabilities, e.g. for coupling to the $S_{10,21}$ phonon one needs $B_0\approx 5$T for $R=100$\,nm and $B_0\approx0.5$T for $R=1\,\mu$m, see \figref{figBogoliubovDecomposition}c).

The second acoustomechanical application studied in Ref.~\cite{GonzalezBallesteroarXiv2019} relies on the strong hybridization between the center-of-mass motion and the acoustic phonons in the strong coupling regime. 
For the $S_{\nu21}$ mode, this results in a drastic modification of the center-of-mass dynamics, specifically a peak splitting in the position power spectral density $S_{xx}(\omega)\equiv(2\pi)^{-1}\int_{-\infty}^\infty d\tau e^{i\omega\tau}\langle\hat{X}(0)\hat{X}(\tau)\rangle_{\rm ss}$. This is shown in \figref{figPSD}, where we display the power spectral density at $T_e=300$K, $Q_p=10^7$, and a moderate value $Q_x=10^5$, reachable in most experimental platforms \cite{GieselerNatPhys2013,BraakmanNanotech2019,TaoNatComm2014,BrawleyNatCom2016,WeberNatCom2016,HuttelNanoLett2009}. Upon increasing the magnetic field gradient $b$, the power spectral density transitions from a single peak, corresponding to a single mechanical oscillator, at zero coupling, i.e.\ at $b=0$ (orange curves), to a widening of the peak at moderate values $b=10^3$T/m (blue curves), which indicates motional cooling, to a doubly-peaked shape at $b=10^4$T/m (red curves), evidencing the hybridization of the center-of-mass motion and the acoustic mode~\cite{GroblacherNature2009}. As shown by the right insets of \figref{figPSD}, the frequency difference between these two peaks is given by $2\vert G_{x2}\vert$, which confirms the strong coupling between the center-of-mass motion and the mainly acoustic $\hat{c}_2$ mode.

The peak splitting offers a way to probe the acoustic phonons of the micromagnet by measuring the two peaks in its position power spectral density. According to the left inset of \figref{figPSD}, this measurement is experimentally feasible, especially for acoustic quality factors $Q_p>10^7$, as the two peaks reach values $\sim1$pm$/$Hz$^{1/2}$, even when the center-of-mass motion is only driven by thermal noise. This is within the
sensitivity regime of most state-of-the-art ultra-sensitive displacement sensors \cite{RossiNanoLett2019,BraakmanNanotech2019,JainPRL2016,GieselerNatPhys2013,BrawleyNatCom2016,MoserNatNano2013,WeberNatCom2016,MasonNatPhys2019,WollmanScience2015,LecocqPRX2015} and the signal-to-noise ratio in the power spectral density can be largely increased by resonant excitation of the center-of-mass mode. Remarkably, probing the internal modes remains feasible for higher-order acoustic phonons, as evidenced by \figref{figPSDvariousNU}. As the quality factor $Q_p=10^7$ is kept constant across all panels of the figure, the linewidth $\gamma_p$, and thus the linewidth $\gamma_2$, increases for higher order acoustic modes, resulting in a widening of the two peaks and a smoothing out of the curves as the ratio $2\vert G_{x2}\vert/\gamma_2$ becomes closer to $1$, i.e.\ as the system moves closer to the weak coupling regime. This effect is especially obvious for $R=100$\,nm where the linewidth $\gamma_2$ is entirely dominated by the large acoustic phonon linewidths. The  dependence of the power spectral density on the acoustic linewidth $\gamma_p$, combined with the possibility of addressing many distinct internal acoustic modes, represents a novel experimental approach to shed light on the hitherto unknown acoustic relaxation rates of extremely isolated mesoscopic bodies~\cite{RubioLopezPRB2018}.

\section{Conclusion}\label{Secconclusions} 

In this work, we derived in detail the acoustomechanical Hamiltonian reported in Ref.~\cite{GonzalezBallesteroarXiv2019}. The acoustomechanical interaction is enabled by the strong magnetoelastic coupling between magnons and acoustic phonons in small isolated micromagnets.
We uncovered a magnon-phonon interaction that is not only qualitatively different but also significantly stronger than in millimeter-sized 
samples~\cite{ZhangSciAdv2016}. In future research, this could be exploited together with a magnonic Kerr nolinearity~\cite{WangPRB2016} to devise a cavity QED analog with a magnonic ``qubit'' and an acoustic cavity with ultra-narrow linewidth \cite{MaccabearXiv2019}. A second research direction enabled by this new strong acoustomechanical coupling is to probe the acoustic modes using  magnetometry, for instance via NV centres as recently reported for levitated micromagnets \cite{Jantoappear,HuilleryArxiv2019}.

In addition, we extended our previous results~\cite{GonzalezBallesteroarXiv2019} on engineering a tunable acoustomechanical interaction between the fundamental acoustic phonon and the center-of-mass motion of a micromagnet to higher order acoustic phonon modes. Furthermore, we demonstrated how cooling through higher-frequency acoustic phonons, which have a lower thermal occupation, results in a more efficient acoustic cooling of the center-of-mass motion.  Finally, we showed how several acoustic modes of a microparticle can be probed by measuring their impact on its mechanical motion within experimentally feasible displacement sensitivities. 

\begin{acknowledgments}
C.~G.~B. and J.~G. acknowledge funding from the EU Horizon 2020 program under the Marie Sk\l{}odowska-Curie grant agreements no.~796725 and no.~655369, respectively.
We thank C. C. Rusconi and M. J. A. Schuetz for insightful discussions.
\end{acknowledgments}

 \appendix

 \section{Magnonic eigenmodes of a sphere}\label{appendixMAGNONS}
 
 In this appendix, we summarize the derivation and quantization of the magnon eigenmodes in a spherical magnet. First, in Sec.~\ref{appendixMagnoneqs}, we detail the simplification of the general Landau-Lifshitz equation into the magnetostatic dipolar spin wave equations. Then, we compute the spin wave eigenmodes in Sec.~\ref{appendixWalkermodes}. We then prove the adequacy of the magnetostatic energy functional in Sec.~\ref{appendixMagnetostaticEnergy}, which allows us to quantize the spin waves into magnon modes in Sec.~\ref{appendixMagnonquantization}.
 
 \subsection{Magnetostatic dipolar spin wave equations}\label{appendixMagnoneqs}
 
 As detailed in the main text, the spin waves supported by a magnet are described by a continuous magnetization field $\mathbf{M}(\mathbf{r},t)$ and its associated electromagnetic fields $\mathbf{E}(\mathbf{r},t)$ and $\mathbf{H}(\mathbf{r},t)$. To compute these fields, we combine Maxwell equations with the phenomenological nonlinear Landau-Lifshitz equation
\cite{aharoni2000introduction,StancilBook2009},
 \begin{equation}\label{LLEapp}
     \frac{d}{dt}\mathbf{M}(\mathbf{r},t) = -\vert\gamma\vert \mu_0\mathbf{M}(\mathbf{r},t)\times\mathbf{H}_{\rm eff}(\mathbf{M},\mathbf{r},t).
 \end{equation}
 Here, the effective field $\mathbf{H}_{\rm eff}(\mathbf{M},\mathbf{r},t) = \mathbf{H}(\mathbf{r},t) + \Delta \mathbf{H}(\mathbf{M},\mathbf{r},t)$ is composed of the Maxwell field $\mathbf{H}(\mathbf{r},t)$ and an extra contribution given by~\cite{StancilBook2009}
 \begin{equation}\label{DeltaH}
 \begin{split}
     \Delta&\mathbf{H}(\mathbf{M},\mathbf{r},t) =
     \\
     &=\mathbf{H}_{\rm x}(\mathbf{M},\mathbf{r},t) + \mathbf{H}_{\rm an}(\mathbf{M},\mathbf{r},t)+\mathbf{H}_{\rm dm}(\mathbf{M},\mathbf{r},t),
 \end{split}
 \end{equation}
 where the three terms represent the exchange field, the magnetocrystalline anisotropy field, and the demagnetizing field arising from the magnetic dipole-dipole interaction, respectively. In general, all the above contributions depend on the magnetization field $\mathbf{M}(\mathbf{r},t)$, rendering the Landau-Lifshitz equation Eq.~\eqref{LLEapp} nonlinear. The nonlinearity hinders the straightforward definition of eigenmodes and consequently their quantization. This section is devoted to simplifying the problem using the approximations detailed in Sec.~\ref{SecFreeInternalHamiltonian}, which result in the well-known magnetostatic dipolar spin wave equations.

To simplify the magnetization wave equations, we undertake the  following approximations:
\begin{enumerate}
    \item First, we take the spin wave approximation, i.e.\ we assume that a large field $H_0$ is applied along the $z-$ axis, which results in the magnet being fully magnetized. We then consider only small fluctuations around the fully magnetized state,
    \begin{equation}\label{Hlinearizationapp}
            \mathbf{H} (\mathbf{r},t) = H_0\mathbf{e}_z + \mathbf{h}(\mathbf{r},t),
        \end{equation}
        \vspace{-0.7cm}
        \begin{equation}\label{Mlinearizationapp}
            \mathbf{M} (\mathbf{r},t) = M_S\mathbf{e}_z + \mathbf{m}(\mathbf{r},t),
        \end{equation}
     where $\mathbf{m} \ll M_S$ and $\mathbf{h} \ll H_0$ are the dynamical variables whose eigenmodes we will calculate and quantize. Using the above expressions, we write the Landau-Lifshitz equation Eq.~\ref{LLEapp} as
    \begin{equation}\label{LLEprov0}
    \begin{split}
        &\frac{\dot{\mathbf{m}}(\mathbf{r},t)}{\vert\gamma\vert\mu_0M_SH_0}
        -\mathbf{e}_z\times\left[\frac{\mathbf{m}(\mathbf{r},t)}{M_S} -\frac{\mathbf{h}(\mathbf{r},t)}{H_0}\right] 
         \\
        &
        +\left[\mathbf{e}_z+\frac{\mathbf{m}(\mathbf{r},t)}{M_S}\right]\times\frac{\Delta\mathbf{H}(\mathbf{m},\mathbf{r},t)}{H_0}=
        \\
        &
        =
        \frac{\mathbf{h}(\mathbf{r},t)}{H_0}\times\frac{\mathbf{m}(\mathbf{r},t)}{M_S}.
    \end{split}
    \end{equation}
    The terms on the first line of Eq.~\eqref{LLEprov0} are of first order in the small variables $\mathbf{m}/M_S$ and $\mathbf{h}/H_0$, whereas the right hand side is of second order and can be neglected. In the following we will fully linearize the Landau-Lifshitz Equations by also keeping only the first order terms in the second line of Eq.~\eqref{LLEprov0}. Note that, to first order in $(\mathbf{m}/M_S)$, the dipolar spin wave magnetization fulfills $\mathbf{m}(\mathbf{r},t)\cdot\mathbf{e}_z=0$.
    
    \item We now simplify the contribution $\Delta\mathbf{H}$ through further approximations.    First, the exchange field can be neglected, $\mathbf{H}_{\rm x}\approx0$, for micromagnet sizes $2R$ larger than the usual domain wall length ($\sim 10$\,nm), since the magnetization waves are dominated by dipole-dipole interaction~\cite{aharoni2000introduction,StancilBook2009}. Second, for a cubic material, the magnetocrystalline anisotropy can also be neglected\footnote{Note that including the lowest-order contribution in the magnetocrystalline anisotropy, i.e.\ considering the next higher order in the spin wave expansion Eq.~\eqref{Mlinearization}, leads to the well-known magnon Kerr nonlinearity in the Hamiltonian \cite{StancilBook2009,WangPRB2016}}, $\mathbf{H}_{\rm an}\approx0$, as its lowest-order contribution to the Landau-Lifshitz equations is quadratic in $(\mathbf{m}/M_S)$  \cite{aharoni2000introduction,StancilBook2009,MaierFlaigPRB2017}. Third, we note that, within these assumptions, the demagnetizing field for a spherical magnet takes the simple form $\mathbf{H}_{\rm dm} = -(M_S/3)\mathbf{e}_z$~\cite{Jackson1975classical,aharoni2000introduction}. The above approximations allow us to write the fully linearized Landau-Lifshitz equations as~\cite{StancilBook2009}
    \begin{equation}\label{LLElinearized}
        \hspace{0.5cm}\left[
        \begin{array}{c}
            \dot{m}_x(\mathbf{r},t)    \\
            \dot{m}_y(\mathbf{r},t)   
        \end{array}
        \right]
        =
        \left[
        \begin{array}{c}
            -\omega_0m_y(\mathbf{r},t)  +\omega_M h_y(\mathbf{r},t)\\
            \omega_0m_x(\mathbf{r},t) -\omega_M h_x(\mathbf{r},t)
        \end{array}
        \right]
        .
    \end{equation}
    Here, we have defined the two relevant system frequencies
    \begin{equation}\label{omegaM0def}
        \omega_M \equiv \vert\gamma\vert\mu_0M_S \hspace{0.4cm};\hspace{0.4cm} \omega_0 \equiv \vert\gamma\vert\mu_0H_I,
    \end{equation}
    where the internal field is defined as $H_I \equiv H_0-M_S/3$.
    
    \item Finally, we apply the magnetostatic approximation $\nabla\times\mathbf{h}(\mathbf{r},t)\approx 0$. The simplifications stemming from this long-wavelength approximation are twofold: on the one hand, the electric field of the spin wave is eliminated as a variable, as it is uncoupled from $\mathbf{h}$ in the Maxwell equations~\cite{StancilBook2009}. On the other hand, this approximation allows to define a magnetostatic potential through $\mathbf{h}(\mathbf{r},t)=-\nabla\psi(\mathbf{r},t)$. 
    The problem thus reduces to solving for three coupled scalar fields, $m_x(\mathbf{r},t)$, $m_y(\mathbf{r},t)$, and $\psi(\mathbf{r},t)$, whose equations are given by the linearized Landau-Lifshitz equations Eq.~\eqref{LLElinearized} and by the zero-divergence condition for the magnetic field of the spin wave, $\mathbf{b}(\mathbf{r},t) \equiv \mu_0[\mathbf{h}(\mathbf{r},t) + \mathbf{m}(\mathbf{r},t)]$, or equivalently~\cite{StancilBook2009},
    \begin{equation}\label{eqforpsi}
        \nabla^2 \psi (\mathbf{r},t) = \partial_xm_x(\mathbf{r},t) + \partial_ym_y(\mathbf{r},t),
    \end{equation}
    which naturally implies $\nabla^2\psi=0$ outside of the micromagnet. These equations are complemented by the boundary conditions at the surface of the magnet, namely the continuity of $\mathbf{h}\times\mathbf{e}_n$ and $\mathbf{b}\cdot\mathbf{e}_n$, where $\mathbf{e}_n$ is the unit vector normal to the surface of the micromagnet.
\end{enumerate}
 
 The simplified equations obtained above are the starting point for the calculation of the spin wave eigenmodes. These eigenmodes are known in the literature as magnetostatic dipolar spin waves or Walker modes~\cite{StancilBook2009,aharoni2000introduction}.
 
 \subsection{Walker modes}\label{appendixWalkermodes}
 
 Let us calculate the Walker modes for a spherical magnet. We start by expressing the corresponding fields in terms of eigenmodes,
 \begin{equation}\label{meigenmodes}
     \mathbf{m}(\mathbf{r},t)=\sum_\beta \left[s_\beta\mathbf{m}_{\beta}(\mathbf{r})e^{-i\omega_\beta t}+\text{c.c.}\right],
 \end{equation}
 \begin{equation}\label{heigenmodes}
     \mathbf{h}(\mathbf{r},t)=\sum_\beta \left[s_\beta\mathbf{h}_{\beta}(\mathbf{r})e^{-i\omega_\beta t}+\text{c.c.}\right],
 \end{equation}
  where $s_\beta$ is a complex amplitude, and $\mathbf{m}_\beta(\mathbf{r})$ and $\mathbf{h}_{\beta}(\mathbf{r})=-\nabla\psi_\beta(\mathbf{r})$ are the fields corresponding to an eigenmode, characterized by a set of mode indices $\beta$ and an oscillation frequency $\omega_\beta$.
 
 The calculation of these modes is summarized as follows, see Refs. \cite{Fletcher59,Roschmann} for more details. We start by considering the linearized Landau-Lifshitz equations for a given eigenmode, namely
 \begin{equation}\label{dipolarm1}
    i\omega m_x(\mathbf{r}) = \omega_M \partial_y\psi(\mathbf{r}) + \omega_0 m_y(\mathbf{r}),
\end{equation}
\begin{equation}\label{dipolarm2}
    i\omega m_y(\mathbf{r}) = -\omega_M \partial_x\psi(\mathbf{r}) - \omega_0 m_x(\mathbf{r}).
\end{equation}
By introducing these equations into Eq.~\eqref{eqforpsi} we obtain an equation for the magnetostatic potential only. Outside the micromagnet, we have
\begin{equation}
    \nabla^2\psi_{\rm out}(\mathbf{r})=0,
\end{equation}
whereas inside the micromagnet the potential fulfills
\begin{equation}\label{eqpsiint}
    (1+\chi_p)\left( \frac{\partial^2}{\partial x^2} + \frac{\partial^2}{\partial y^2}\right)\psi_{\rm in}(\mathbf{r}) +\frac{\partial^2}{\partial z^2}\psi_{\rm in}(\mathbf{r}) = 0,
\end{equation}
where $\chi_p(\omega)$ is the diagonal element of the Polder susceptibility tensor~\cite{StancilBook2009},
\begin{equation}
    \chi_p(\omega) \equiv\frac{\omega_M\omega_0}{\omega_0^2-\omega^2}.
\end{equation}

The general solution for the potential outside the sphere in spherical coordinates is given by
\begin{equation}
    \psi_\text{out}(\mathbf{r}) = \sum_{lm}\left[\frac{A_{lm}}{r^{l+1}}+B_{lm}r^{l}\right]Y_l^m(\theta,\phi),
\end{equation}
where $A_{lm}$ and $B_{lm}$ are expansion coefficients and $Y_l^m(\theta,\phi)$ are spherical harmonics. In order to solve for the potential inside the sphere, we express it in the set of non-orthogonal coordinates $\{\xi,\eta,\phi\}$ defined by
\begin{equation}\label{x0}
    x = \sqrt{\chi_p} R \sqrt{\xi^2-1}\sin\eta\cos\phi,
\end{equation}
\begin{equation}\label{y0}
    y = \sqrt{\chi_p} R \sqrt{\xi^2-1}\sin\eta\sin\phi,
\end{equation}
\begin{equation}\label{z0}
    z = \sqrt{\frac{\chi_p}{1+\chi_p}} R \xi\cos\eta.
\end{equation}
In these coordinates, Eq.~\eqref{eqpsiint} can be solved in terms of associated Legendre polynomials and spherical harmonics~\cite{Roschmann,Fletcher59} as
\begin{equation}\label{psiin}
    \psi_\text{in}(\mathbf{r}) = \sum_{lm}C_{lm}P_l^m(\xi)Y_l^m(\eta,\phi),
\end{equation}
with expansion coefficients $C_{lm}$.
Importantly, the new coordinates take a very simple form on the surface of the sphere, namely
\begin{equation}\label{surfaceDSC}
    \xi \to \xi_0 =\sqrt{\frac{1+\chi_p}{\chi_p}}\hspace{0.2cm} ; \hspace{0.2cm} \{\eta,\phi\} \to \{\theta,\phi\}.
\end{equation}
This allows to impose the boundary conditions in a relatively simple way. First, we require the potential $\psi$ to be regular at infinity, $B_{lm}=0$. Then,
we impose the continuity of the tangential component of $\mathbf{h}$, which is equivalent to imposing continuity of the potential $\psi$ across the surface,
\begin{equation}\label{AvsC}
    A_{lm} = C_{lm}P_l^m(\xi_0)R^{l+1}.
\end{equation}
The final boundary condition, namely the continuity of the normal component of the $\mathbf{b}$ field, can be expressed as~\cite{Fletcher59}
\begin{equation}
    \frac{\partial\psi_\text{out}}{\partial r}\Big\vert_{r=R} 
    =
    \frac{\xi_0}{R}\frac{\partial\psi_\text{in}}{\partial \xi}\Big\vert_{r=R}
    -i\frac{\kappa_p}{R} \frac{\partial\psi_\text{in}}{\partial \phi}\Big\vert_{r=R},
\end{equation}
where $i\kappa_p(\omega)$ is the off-diagonal element of the Polder susceptibility tensor~\cite{StancilBook2009},
\begin{equation}
    \kappa_p(\omega) =\frac{\omega_M\omega}{\omega_0^2-\omega^2}.
\end{equation}
By combining the above boundary condition with Eq.~\eqref{AvsC} we obtain the Walker mode eigenfrequency equation \cite{Fletcher59,Roschmann}
\begin{equation}
    \xi_0(\omega) \frac{P'{}_l^{m}(\xi_0(\omega))}{P_l^m(\xi_0(\omega))}+m\kappa_p(\omega) + l + 1 =0,
\end{equation}
some of whose solutions are displayed in Fig.~\figref{figmagnonfreqs}.
Note that this equation for $\omega$ does not depend on the radius $R$, and has no positive, i.e.\ physical, solutions for many pairs $\{l,m\}$, such as for $l=0$ or for $\{l,m\}=\{1,0\}$. In general, the eigenfrequency equations will have a set of discrete solutions for each $\{l,m\}$, which we label with the mode index $n$, starting at $n=0$ following historical conventions~\cite{Roschmann}.

For each mode $\{lmn\}$, the magnetization profile is obtained by expressing the potential $\psi_{\rm in}(\mathbf{r})$, Eq.~\eqref{psiin}, in Cartesian coordinates, introducing the result in the linearized Landau-Lifshitz equations, Eqs.~\eqref{dipolarm1} and \eqref{dipolarm2}, and solving the corresponding algebraic system of equations. To our knowledge, no general form of the magnetization field is known for all modes. However, the corresponding mode functions have been explicitly calculated for $l\le10$ in Ref.~\cite{Fletcher59}.

\subsection{Magnetostatic energy density}\label{appendixMagnetostaticEnergy}

The quantization of the Walker modes requires the definition of a magnetic energy functional reproducing the linearized Landau-Lifshitz equations Eq.~\eqref{LLElinearized}. Such micromagnetic energy functional, which, like the Landau-Lifshitz equation, is phenomenological, is given in the main text by Eq.~\eqref{micromagneticenergy}, i.e.,
\begin{equation}\label{micromagneticenergyapp}
\begin{split}
      E_m&(\{\mathbf{m}\},\{\mathbf{h}\}) =
      \\
      &=\frac{\mu_0}{2}\int dV \mathbf{m}(\mathbf{r},t)
      \cdot\left[\frac{H_I}{M_S}\mathbf{m}(\mathbf{r},t)-\mathbf{h}(\mathbf{r},t)\right].
\end{split}
\end{equation}
In this section, we prove that the above micromagnetic energy functional is adequate, namely that it reproduces the linearized Landau-Lifshitz equations Eq.~\eqref{LLElinearized}. 

The equations of motion for $\mathbf{m}$ and $\mathbf{h}$ are those which make the energy functional stationary, 
\begin{equation}
\begin{split}
    \dot{E}_m &= \mu_0\frac{H_I}{M_S}\int_{\mathcal{V}_a} dV \mathbf{m}(\mathbf{r},t)\cdot\dot{\mathbf{m}}(\mathbf{r},t)
      \\
      &-\frac{\mu_0}{2}\frac{d}{dt}\int_{\mathcal{V}_a} dV \mathbf{m}(\mathbf{r},t)\cdot\mathbf{h}(\mathbf{r},t)=0,
\end{split}
\end{equation}
where we extend the integration volume to an arbitrary volume $\mathcal{V}_a$ containing the magnet, by exploiting the fact that the magnetization field is zero outside the material.  We now use the identity \cite{brown1962book,Jackson1975classical}
$\int_{{\mathcal{V}_a}} dV \mathbf{b}\cdot\mathbf{h} = 0$,
 with $\mathbf{b} = \mu_0(\mathbf{h}+\mathbf{m})$, which holds for  fields satisfying the long-wavelength condition $\nabla\times\mathbf{h}=0$, and for a sufficiently large integration volume $\mathcal{V}_a$ containing all the sources, in this case the micromagnet and the free current density distribution $\mathbf{j}_f(\mathbf{r},t)$ responsible for generating the external fields. Using the above identity we readily find
 \begin{equation}\label{intermediateAD1}
 \begin{split}
    \frac{d}{dt}&\int_{\mathcal{V}_a} dV \mathbf{m}(\mathbf{r},t)\cdot \mathbf{h}(\mathbf{r},t)  = 
    \\
    &
    = 2\int_{\mathcal{V}_a}
    dV  \mathbf{h}(\mathbf{r},t)\cdot\left[\dot{\mathbf{m}}(\mathbf{r},t) -\frac{\dot{\mathbf{b}}(\mathbf{r},t)}{\mu_0}\right]
    . 
 \end{split}
 \end{equation}
 Finally, assuming no permanent electric polarization in the magnet and that the magnetic field $\mathbf{b}$ does not grow indefinitely with time, we can  use Maxwell equations for magnetostatic fields ($\nabla\times\mathbf{h}=0$) to express
\begin{equation}
     \dot{\mathbf{b}}(\mathbf{r},t) = -\frac{1}{\varepsilon_0 \varepsilon}\int^t dt'\nabla\times\mathbf{j}_f(\mathbf{r},t'),
 \end{equation}
where $\varepsilon_0$ is the vacuum permittivity and $\varepsilon$ the relative permeability of the medium. Using this expression, the vector identity $\nabla\cdot(\mathbf{A}\times\mathbf{B})=(\nabla\times\mathbf{A})\cdot\mathbf{B}-\mathbf{A} \cdot(\nabla\times\mathbf{B})$, the divergence theorem $\int_{\mathcal{V}_a}dV\nabla\cdot\mathbf{A} = \int_{\partial\mathcal{V}_a}d\mathbf{S}_a\cdot\mathbf{A}$, and choosing a volume $\mathcal{V}_a$ including all the sources, the last term in Eq.~\eqref{intermediateAD1} cancels out, and we arrive at
\begin{equation}
    \begin{split}
    \dot{E}_m =& \frac{1}{\vert\gamma\vert M_S}\int dV
    \\
    &\dot{\mathbf{m}}(\mathbf{r},t)\cdot \left[\omega_0\mathbf{m}(\mathbf{r},t) -\omega_M \mathbf{h}'(\mathbf{r},t) \right]=0.
    \end{split}
\end{equation}
This equality is satisfied if and only if the fields $\mathbf{m}$ and $\mathbf{h}$ fulfill the linearized Landau-Lifshitz equations Eq.~\eqref{LLElinearized}, thus demonstrating that the micromagnetic energy functional is correct.

 \subsection{Quantization of the magnetostatic dipolar spin wave modes}\label{appendixMagnonquantization}
 
  The magnetostatic dipolar spin wave modes are quantized following Ref.~\cite{Mills2006}. We commence by using Eq.~\eqref{LLElinearized} to cast the micromagnetic energy functional in the more convenient form
\begin{equation}\label{energyMills}
\begin{split}
    E_m&(\{\mathbf{m}\}) =
    \frac{1}{2M_S\vert\gamma\vert}\int dV 
    \\
    &
    \left(m_x(\mathbf{r},t)\frac{\partial m_y(\mathbf{r},t)}{\partial t}-m_y(\mathbf{r},t)\frac{\partial m_x(\mathbf{r},t)}{\partial t}\right).
    \end{split}
\end{equation}
We now introduce the magnetization expanded in terms of  eigenmodes, Eq.~\eqref{meigenmodes}, and make use of the orthogonality relations between eigenmodes, also called Walker identities \cite{WalkerPhysRev1957,brown1962book,Mills2006}, to finally reduce the energy to
\begin{equation}\label{energymagnonsHOS}
    E_m = \frac{1}{2M_S\hbar\vert\gamma\vert}\sum_{\beta} \hbar\omega_\beta\Lambda_\beta\left[s_\beta s^*_{\beta}+ s^*_\beta s_{\beta}\right],
\end{equation}
where $s_\beta$ are the expansion coefficients in Eq.~\eqref{meigenmodes} and we define
\begin{equation}
    \Lambda_\beta \equiv 2\text{Im}\int dV  m_{\beta y}(\mathbf{r}) m^*_{\beta x}(\mathbf{r}).
\end{equation}
Equation~\eqref{energymagnonsHOS} is equivalent to the energy of an ensemble of harmonic oscillators, which is expected, since the Walker modes describe small perturbations of the magnetization about a fully magnetized state. The quantization is then carried out, first, by promoting the expansion coefficients to bosonic magnon operators, $s_\beta\to\hat{s}_\beta$ and $s_\beta^*\to\hat{s}^\dagger_\beta$, 
and second, by defining the zero-point magnetization through an adequate choice of the eigenmode normalization, i.e.,  such that the factor $\Lambda_\beta/(  M_S\vert\gamma\vert \hbar)$ in Eq.~\eqref{energymagnonsHOS} cancels out. In other words, we perform the substitution
\begin{equation}
    \left[\begin{array}{c}
         \mathbf{m}_\beta(\mathbf{r})  \\
         \mathbf{h}_\beta(\mathbf{r}) 
    \end{array}\right]\to \mathcal{M}_{0\beta}\left[\begin{array}{c}
         \tilde{\mathbf{m}}_\beta(\mathbf{r})  \\
         \tilde{\mathbf{h}}_\beta(\mathbf{r}) 
    \end{array}\right].
\end{equation}
Here, we have defined new adimensional mode functions $\tilde{\mathbf{m}}_\beta(\mathbf{r})$ and $\tilde{\mathbf{h}}_\beta(\mathbf{r})$, as well as the zero-point magnetization Eq.~\eqref{zeropointM}. We can finally write the  magnetization and magnetic field operators in the Schr\"odinger picture as
\begin{equation}\label{mquantum}
    \hat{\mathbf{m}}(\mathbf{r})=\sum_\beta\mathcal{M}_{0\beta}\left[\tilde{\mathbf{m}}_\beta(\mathbf{r})\hat{s}_\beta + \text{H.c.}\right],
\end{equation}
\begin{equation}
    \hat{\mathbf{h}}(\mathbf{r})=\sum_\beta \mathcal{M}_{0\beta}\left[\tilde{\mathbf{h}}_\beta(\mathbf{r})\hat{s}_\beta + \text{H.c.}\right].
\end{equation}
As a final remark, let us emphasize that all the derivations in this section can be extended to cases where both magnetocrystalline anisotropy and exchange interactions are taken into account \cite{Mills2006,AriasPRB2005}.
 
 \section{Acoustic eigenmodes of a sphere}\label{appendixPHONONS}
 
 In this appendix, we summarize the derivation of the Lamb acoustic modes for a homogeneous spherical sample. At low energies, the dynamics of the continuous displacement field $\mathbf{u}(\mathbf{r},t)$ is described by the theory of linear elastodynamics, through the equation of motion
  \cite{Gurtinbook,Eringenbook,Achenbachbook}
\begin{equation}\label{EqsmotionU}
    \rho \ddot{u}_i = \sum_{jkl}C_{ijkl}\partial_j \partial_ku_l \hspace{0.5cm}(\{i,j,k,l\} = 1,2,3),
\end{equation}
 where $C_{ijkl}$ is the elasticity tensor. For an homogeneous and isotropic material, only two components of the tensor are independent, i.e.\ $C_{ijkl} = \rho(c_L^2-2c_T^2)\delta_{ij}\delta_{kl}+\rho c_T^2 (\delta_{ik}\delta_{jl}+\delta_{il}\delta_{jk})$ \cite{Eringenbook,Gurtinbook}, where $\delta_{ij}$ is the Kronecker delta. Equation \eqref{EqsmotionU} is simplified by decomposing the displacement field in terms of acoustic eigenmodes, 
 \begin{equation}
     \mathbf{u}(\mathbf{r},t) = \sum_\alpha \left[u_\alpha \mathbf{f}_\alpha(\mathbf{r})e^{-i\omega_\alpha t} + \text{c.c.}\right],
 \end{equation}
 where $u_\alpha$ is the complex amplitude of mode $\alpha$, which is characterized by a set of mode indices $\alpha$ and an oscillation frequency $\omega_\alpha$. The adimensional mode functions $\mathbf{f}_\alpha(\mathbf{r})$ are orthogonal, i.e.,
  \begin{equation}
     \int dV \mathbf{f}_\alpha(\mathbf{r})^*\cdot\mathbf{f}_{\alpha'}(\mathbf{r})= \mathcal{N}_\alpha\delta_{\alpha\alpha'}.
 \end{equation}
 where $\mathcal{N}_\alpha \equiv\int dV\vert\mathbf{f}_\alpha(\mathbf{r})\vert^2$ is the norm of mode $\alpha$. 
 Each of the eigenmodes obeys the simpler time-independent equation
 \begin{equation}\label{eqDISPeigenmodes}
     -\rho\omega_\alpha^2 f_{\alpha,i}(\mathbf{r}) = \sum_{jkl}C_{ijkl}\partial_j \partial_kf_{\alpha,l}(\mathbf{r}).
 \end{equation}
 The linear elastodynamics problem is thus reduced to computing the corresponding eigenmodes and eigenfrequencies using the above equation and an appropriate set of boundary conditions.

  Following Ref. \cite{Eringenbook}, we solve Eq.~\eqref{eqDISPeigenmodes} by decomposing the displacement field of a given mode into scalar and vector potentials,
	\begin{equation}\label{Helmholtzdecomp}
	\mathbf{f} = \nabla \varphi_p + \nabla\times \mathbf{L} + \nabla\times\nabla\times\mathbf{N}.
	\end{equation}
By construction, the three components above are orthogonal. 
This representation largely simplifies the equations since it is possible to demonstrate that, for any $\mathbf{f}$ satisfying Eq.~\eqref{eqDISPeigenmodes}, we can choose $\mathbf{L} = r\psi_p(\mathbf{r})\mathbf{e}_r$ and $\mathbf{N} = r\xi_p(\mathbf{r})\mathbf{e}_r$ so that, by choosing the convenient gauge \cite{Eringenbook} $\nabla\times[(c_T^2\nabla^2+\omega^2)\mathbf{L}(\mathbf{r},\omega)]=0$, the three unknown scalar functions $(\varphi_p,\psi_p,\xi_p)$ satisfy independent Helmholtz equations, 
	\begin{equation}
	\nabla^2\left(\begin{array}{c}
	\varphi_p
	\\
	\psi_p
	\\
	\xi_p
	\end{array}
	\right) = -\left(\begin{array}{c}
	\tilde{\alpha}^2\varphi_p
	\\
	\tilde{\beta}^2\psi_p
	\\
	\tilde{\beta}^2\xi_p
	\end{array}
	\right),
	\end{equation}
	where we define the two acoustic wavenumbers $\tilde{\alpha} = \omega/c_L$ and $\tilde{\beta} = \omega/c_T$. The solutions of the above equations in spherical coordinates read
	\begin{equation}
	\left(\begin{array}{c}
	\varphi_p
	\\
	\psi_p
	\\
	\xi_p
	\end{array}
	\right) =\left(\begin{array}{c}
	Aj_\lambda(\tilde{\alpha} r)
	\\
	Bj_\lambda(\tilde{\beta} r)
	\\
	Cj_\lambda(\tilde{\beta} r)
	\end{array}
	\right) P_\lambda^\mu(\cos\theta)e^{i\mu\phi},
	\end{equation}
	where $j_\lambda(z)$ and $P_\lambda^\mu(z)$ are  spherical Bessel functions of the first kind and associated Legendre polynomials, respectively.
	The general solution for an eigenmode is therefore given by Eq.~\eqref{Helmholtzdecomp} up to three arbitrary constants $A, B, C$ determining the contribution of the potentials $\varphi_p, \mathbf{L},$ and $\mathbf{M}$, respectively. In an infinite bulk with no boundary conditions, the three components are independent resulting in three families of modes, one longitudinally and two transversely polarized. Here, however, we impose stress-free boundary conditions at the surface of the sphere \cite{Eringenbook}:
	\begin{equation}
	\overline{\sigma}(R,\theta,\phi)\cdot \mathbf{e}_r = 0,
	\end{equation}
	where $\overline{\sigma}$ is the stress tensor [$\bar{\sigma}_{ij}\equiv\sum_{kl}C_{ijkl}\bar{\varepsilon}_{kl}$ with $\bar{\varepsilon}_{ij}$ given by Eq.~\eqref{straintensor}].
	These boundary conditions mix the amplitudes $A,B,C$ and the different polarizations couple to each other, resulting in two independent phonon families, the torsional and the spherioidal, respectively.
	
	The torsional modes correspond to $A=C=0$, and are therefore purely transverse, $\nabla\cdot\mathbf{u}=0$. Their dispersion relation is given by~\cite{Eringenbook}
	\begin{equation}\label{dispersionTORS}
		(\lambda-1)j_\lambda(\tilde{\beta}_{\lambda\nu} R) - (\tilde{\beta}_{\lambda\nu} R) j_{\lambda+1}(\tilde{\beta}_{\lambda\nu} R) = 0,
	\end{equation}
	where the index $\nu$ labels the discrete set of solutions and $\tilde{\beta}_{\lambda\nu} = \omega_{\lambda\nu}/c_T$. The corresponding mode functions read
	\begin{equation}\label{modefunctionTORS}
     \mathbf{f}_{t,\nu\lambda\mu}(\mathbf{r}) = e^{i\mu\phi}
 	\left(\begin{array}{c}
 	0
 	\\
 	\frac{i\mu}{\sin\theta}j_\lambda(\tilde{\beta}_{\nu\lambda} r)P_\lambda^\nu(\cos\theta)
 	\\
 	-j_\lambda(\tilde{\beta}_{\nu\lambda} r)\frac{d}{d\theta}P_\lambda^\mu(\cos\theta)
 	\end{array}\right),
 	\end{equation}
	where the vector components are ordered in the usual way, namely $(\mathbf{e}_r,\mathbf{e}_\theta,\mathbf{e}_\phi)$.
	The norm of these modes is
	\begin{equation}
	\begin{split}
	    \mathcal{N}_{t,\nu\lambda\mu}&=\frac{3 V}{4}\lambda(\lambda+1)\frac{2(\lambda+\mu)!}{(2\lambda+1)(\lambda-\mu)!}
	    \\
	    &
	    \times j_\lambda^2(\tilde{\beta}_{\lambda\nu} R)\frac{(\tilde{\beta}_{\lambda\nu} R)^2 +3\lambda(\lambda-1)}{(\tilde{\beta}_{\lambda\nu} R)^2}.
	\end{split}
	\end{equation}
	Note that no torsional mode with $\lambda=0$ exists, as the mode function vanishes, i.e.\ $\mathbf{f}_{t,\nu00}=0$.
	
	The spheroidal modes correspond to $B=0$, and $C$ being a given function of $A$. Their eigenfrequency equation can be written in compact form as~\cite{Eringenbook}
	\begin{equation}\label{dispersionSPH}
		T_{\lambda\nu}^{(a)}T_{\lambda\nu}^{(b)}-T_{\lambda\nu}^{(c)}T_{\lambda\nu}^{(d)} = 0,
	\end{equation}
	where the coefficients $T_{\lambda\nu}$ are given by	
	\begin{equation}
	\begin{split}
		T_{\lambda\nu}^{(a)} &= \left(\lambda(\lambda-1)-\frac{\tilde{\beta}_{\lambda\nu}^2R^2}{2}\right)j_\lambda(\tilde{\alpha}_{\lambda\nu} R) +
		\\
		&
		+2\tilde{\alpha}_{\lambda\nu} R j_{\lambda+1}(\tilde{\alpha}_{\lambda\nu} R),
	\end{split}
	\end{equation}
	\begin{equation}
	\begin{split}
	    T_{\lambda\nu}^{(b)}&=\left(\lambda^2-1-\frac{\tilde{\beta}_{\lambda\nu}^2R^2}{2}\right)j_\lambda(\tilde{\beta}_{\lambda\nu} R)+
	    \\
	    &
	    +\tilde{\beta}_{\lambda\nu} R j_{\lambda+1}(\tilde{\beta}_{\lambda\nu} R),
	\end{split}
	\end{equation}
	\begin{equation}
	\begin{split}
		T_{\lambda\nu}^{(c)} &= \lambda(\lambda+1)
		\\
		&
		\times\left[(\lambda-1)j_\lambda(\tilde{\beta}_{\lambda\nu} R)-\tilde{\beta}_{\lambda\nu} R j_{\lambda+1}(\tilde{\beta}_{\lambda\nu} R)\right],
	\end{split}
	\end{equation}
	\begin{equation}
	    T_{\lambda\nu}^{(d)} = (\lambda-1)j_\lambda(\tilde{\alpha}_{\lambda\nu} R)-\tilde{\alpha}_{\lambda\nu} R j_{\lambda+1}(\tilde{\alpha}_{\lambda\nu} R),
	\end{equation}
	and $\tilde{\alpha}_{\lambda\nu} = \omega_{\lambda\nu}/c_L$.
	The corresponding mode profile is given by
	\begin{equation}\label{modefunctionSPH}
     \mathbf{f}_{s,\nu\lambda\mu}(\mathbf{r}) = e^{i\mu\phi}
 	\left(\begin{array}{c}
 	\tilde{G}_{\nu\lambda}(r) P_\lambda^\mu(\cos\theta) 
 	\\
    \tilde{F}_{\nu \lambda}(r) \frac{d}{d\theta}P_\lambda^\mu(\cos\theta) 
    \\
 	\tilde{F}_{\nu\lambda}(r)\frac{i\mu}{\sin\theta}P_\lambda^\mu(\cos\theta)
 	\end{array}\right),
 \end{equation}
 where the two radial mode functions read
	\begin{equation}\label{Fnulambdadef}
	\begin{split}
		\frac{r}{R}\tilde{F}&_{\nu\lambda}(r) = j_\lambda(\tilde{\alpha}_{\lambda\nu} r) +
		\\
		&
		-\frac{T_{\lambda\nu}^{(d)}}{T_{\lambda\nu}^{(b)}}\left[(\lambda+1)j_\lambda(\tilde{\beta}_{\lambda\nu} r) - \tilde{\beta}_{\lambda\nu} r j_{\lambda+1}(\tilde{\beta}_{\lambda\nu} r)\right],
	\end{split}
	\end{equation}
	\begin{equation}\label{Gnulambdadef}
	\begin{split}
		\frac{r}{R}\tilde{G}_{\nu\lambda}(r) &=\lambda j_\lambda(\tilde{\alpha}_{\lambda\nu} r)-\tilde{\alpha}_{\lambda\nu} rj_{\lambda+1}(\tilde{\alpha}_{\lambda\nu} r) 
		\\
		&
		- \frac{T_{\lambda\nu}^{(d)}}{T_{\lambda\nu}^{(b)}}\lambda(\lambda+1)j_\lambda(\tilde{\beta}_{\lambda\nu} r).
	\end{split}
	\end{equation}
 Finally, the norm of the spheroidal modes is
 \begin{equation}
     \mathcal{N}_{s,\nu\lambda\mu}= 2\pi  \frac{2(\lambda+\mu)!}{(2\lambda+1)(\lambda-\mu)!} \mathcal{J}_{\nu\lambda},
 \end{equation}
 where the radial integral
 \begin{equation}
     \mathcal{J}_{\nu\lambda}=\int dr r^2\Big[\tilde{G}_{\nu\lambda}^2(r) + \lambda(\lambda+1) \tilde{F}_{\nu\lambda}^2(r)\Big],
 \end{equation}
 can be analytically expressed as a lengthy combination of hypergeometric functions. 

The quantization of the displacement field is straightforward given the quadratic nature of the elastodynamic Lagrangian, and will be omitted here since it can be found in the literature~\cite{AnghelJPAMT2007,HuemmerPRX2019}.

 \section{Calculation of single-magnon magnetoelastic couplings}\label{appendixCOUPLINGS}
 
 In this section, we compute the single-magnon magnetoelastic couplings given by Eq.~\eqref{singlemagnoncouplings}, for an arbitrary phonon and two magnon modes, namely the  Kittel ($\{110\}$) mode and the $\{210\}$ mode. Their corresponding adimensional mode functions are given by Eqs.~\eqref{mtildeKittel} and \eqref{mtilde210}, whereas their zero-point magnetizations are given in Eq.~\eqref{zeropointMmagnons}.
 Based on these expressions, the coupling rates for these modes can be written as
 \begin{equation}
     \left(g_{\alpha K}\right)^* = g_{\alpha K}^0\frac{1}{V}\int dV\left[ \tilde{\varepsilon}_{xz}^{(\alpha)}(\mathbf{r})-i\tilde{\varepsilon}_{yz}^{(\alpha)}(\mathbf{r})\right],
 \end{equation}
 \begin{equation}
     \left(g_{\alpha,210}\right)^* = g_{\alpha, 210}^0\frac{1}{V}\int dV \frac{z}{R}\left[ \tilde{\varepsilon}_{xz}^{(\alpha)}(\mathbf{r})-i\tilde{\varepsilon}_{yz}^{(\alpha)}(\mathbf{r})\right].
 \end{equation}
 The first step towards computing the above rates is to write the integrand explicitly. By taking the expressions for the strain tensor components in spherical coordinates \cite{Eringenbook} and contracting them with the Cartesian unit vectors, it is possible to show that
 \begin{equation}\label{straincylinder}
\begin{split}
    &\frac{2e^{i\phi}}{R}\!\big[\tilde\varepsilon_{xz}^{(\alpha)} - i \tilde\varepsilon_{yz}^{(\alpha)}\big]\! =   \sin(2\theta)\!\!\left[\partial_rf_{\alpha r}\!-\!\frac{f_{\alpha r}}{r}\!-\!\frac{\partial_\theta f_{\alpha \theta}}{r}\right] 
    \\
    &
    + \cos(2\theta)\left[\partial_rf_{\alpha \theta} + \frac{1}{r}\partial_\theta f_{\alpha r} -\frac{f_{\alpha \theta}}{r}\right]+
    \\
    &
    +\!i\!\left[\frac{\sin\theta}{r}\partial_\theta f_{\alpha \phi} \!-\! \cos\theta \partial_rf_{\alpha \phi} \!+\!\frac{i\mu(f_{\alpha \theta}\!-\!f_{\alpha r}\cot\theta)}{r}\right]\!\!,
\end{split}
\end{equation}
 where $\{f_{\alpha r},f_{\alpha \theta},f_{\alpha \phi}\}$ are the radial, polar, and azimuthal components of the mode function $\mathbf{f}_\alpha(\mathbf{r})$. It is evident from Eqs.  \eqref{modefunctionTORS} and \eqref{modefunctionSPH} that each of these components  is factorizable:
 \begin{equation}\label{factorizationf}
     f_{\alpha j}(\mathbf{r}) = R_{\alpha j}(r) \Theta_{\alpha j}(c_\theta) e^{i\mu\phi}\equiv f^{(0)}_{\alpha j}(r,c_\theta)e^{i\mu\phi},
 \end{equation}
 where we denote $c_\theta\equiv \cos \theta$, and the radial and polar functions are defined in the following way:
 \begin{enumerate}
     \item \underline{For torsional modes}:
     \begin{equation}\label{torsionalfactorization1}
    R_{\alpha r}^{(T)} = 0 \hspace{0.3cm};\hspace{0.3cm}R_{\alpha\theta}^{(T)} = R_{\alpha\phi}^{(T)} = j_\lambda(\tilde\beta_{\nu\lambda} r),
    \end{equation}
    \begin{equation}\label{torsionalfactorization2}
    \Theta_{\alpha\theta}^{(T)} = \frac{i\mu}{\sqrt{1-c_\theta^2}}P_\lambda^\mu(c_\theta) \hspace{0.2cm};\hspace{0.2cm}
    \Theta_{\alpha\phi}^{(T)} = \sqrt{1-c_\theta^2}\partial_{c_\theta} P_\lambda^\mu(c_\theta).
    \end{equation}
    \item \underline{For spheroidal modes}:
    \begin{equation}\label{spheroidalfactorization1}
    R_{\alpha r}^{(S)} = \tilde{G}_{\nu\lambda}(r) \hspace{0.3cm} ; \hspace{0.3cm}  R_{\alpha\theta}^{(S)} = R_{\alpha\phi}^{(S)} = \tilde{F}_{\nu\lambda}(r),
\end{equation}
\begin{equation}\label{spheroidalfactorization2}
    \Theta_{\alpha r}^{(S)} = \frac{\sqrt{1-c_\theta^2}}{i\mu} \Theta_{\alpha\phi}^{(S)} = P_\lambda^\mu(c_\theta) \hspace{0.3cm};\hspace{0.3cm} \Theta_{\alpha\theta}^{(S)} = - \Theta_{\alpha\phi}^{(T)}.
\end{equation}
  \end{enumerate}
 From Eq.~\eqref{straincylinder} and the decomposition \eqref{factorizationf} we conclude that
 \begin{equation}
     \tilde\varepsilon_{xz}^{(\alpha)}(\mathbf{r})  - i \tilde\varepsilon_{yz}^{(\alpha)}(\mathbf{r}) \propto R W_\alpha(r,c_\theta)e^{i\phi(\mu-1)},
 \end{equation}
 where the function $W_\alpha(r,c_\theta)$ depends on the functions $f_{\alpha j}^{(0)}(r,c_\theta)$ and their derivatives, and has an obvious definition (see Eq.~\eqref{straincylinder}). This factorization 
allows us to perform the integrals in the azimuthal angle $\phi$, thus finding the first selection rules:
 \begin{equation}\label{galphaK}
     \frac{g_{\alpha K}^*}{g_{\alpha K}^0}* = \frac{2\pi R}{V}\delta_{\mu,1}\int_0^R dr r^2\int _{-1}^{1}\!\!dc_\theta W_\alpha(r,c_\theta),
 \end{equation}\label{galpha210}
 \begin{equation}\label{couplinggeneral210}
     \frac{g_{\alpha,210}^*}{g^0_{\alpha,210}} = \frac{2\pi R}{V}\delta_{\mu,1}\int_0^R dr \frac{r^3}{R}\int_{-1}^1\!\!dc_\theta c_\theta W_\alpha(r,c_\theta).
 \end{equation}
 From here on we have to compute each of these coupling rates separately.
 
 \subsection{Single-magnon couplings for the Kittel mode}
 
 We commence with the Kittel mode, for which we must compute the integral
 \begin{equation}
  I_\alpha^K \equiv \int_{-1}^1 dc_\theta W_\alpha(r,c_\theta).
\end{equation}
 We start by explicitly writing $W_\alpha(r,c_\theta)$ inside the integral. Then, we group together the terms containing $f_{\alpha r}^{(0)}$, $f_{\alpha\theta}^{(0)}$, and $f_{\alpha \phi}^{(0)}$ respectively, and integrate by parts to eliminate the derivatives with respect to $c_\theta$. This results in the simplified expression
\begin{equation}
 \begin{split}
    I_\alpha^K =& \int_{-1}^1dc_\theta\bigg\lbrace
    t\sqrt{1-c_\theta^2}\left(\partial_rf_{\alpha r}^{(0)}+\frac{2f_{\alpha r}^{(0)}}{r}\right)+
    \\
    &
    +\left(\frac{2c_\theta^2-1}{2}\partial_rf_{\alpha \theta}^{(0)}
   + (2c_\theta^2-1)\frac{f_{\alpha \theta}^{(0)}}{r}\right)+
   \\
   &
   +\frac{1}{2i}c_\theta\left(\partial_r f_{\alpha \phi}^{(0)} +\frac{2}{r}f_{\alpha \phi}^{(0)}\right)
    \bigg\rbrace.
\end{split}
\end{equation}
  Our next step is to explicitly write the mode functions in their decomposed form using Eq.~\eqref{factorizationf} and rearrange the above equation into
\begin{equation}\label{ialphaKexpression1}
\begin{split}
    I_\alpha^K &= I_{\alpha1}\left[\partial_r R_{\alpha\theta} +  \frac{2R_{\alpha\theta}}{r}\right]+
    \\
    &
    +\left[\partial_rR_{\alpha r}+\frac{2R_{\alpha r}}{r}
    \right]I_{\alpha2},
\end{split}
\end{equation}
 where we used the fact that $R_{\alpha\theta}(r)=R_{\alpha\phi}(r)$, and defined the integrals
 \begin{equation}
    I_{\alpha 1} \equiv \int_{-1}^1dc_\theta \frac{2c_\theta^2-1}{2}\Theta_{\alpha\theta}(c_\theta) -\frac{i}{2}c_\theta\Theta_{\alpha\phi}(c_\theta),
\end{equation}
 \begin{equation}
     I_{\alpha 2}\equiv \int_{-1}^1dc_\theta
    c_\theta\sqrt{1-c_\theta^2}\Theta_{\alpha r}(c_\theta).
 \end{equation}
 With this representation and the explicit expressions given by Eq.~\eqref{torsionalfactorization2}, it is straightforward to show, using integration by parts, that $I_{\alpha 1}=0$ for torsional modes and, since $R_{\alpha r}^{(T)}(r)=0$, we conclude from Eq.~\eqref{ialphaKexpression1} that torsional modes do not couple to the Kittel mode, i.e.,
 \begin{equation}
     I_\alpha^K \propto \delta_{\sigma,s}.
 \end{equation}
 We thus focus on the spheroidal modes in the following. Using Eq.~\eqref{spheroidalfactorization2} and integration by parts, we reduce the above integrals to
 \begin{equation}
     I_{\alpha 1} = 3I_{\alpha 2} =\!\int_{-1}^1\!dc_\theta3c_\theta\sqrt{1-c_\theta^2}P_\lambda^\mu(c_\theta)=\frac{12}{5}\delta_{\lambda 2},
 \end{equation}
 where in the last step we have used the orthogonality relations of the associated Legendre polynomials, assumed $\mu=1$ due to the selection rule $\delta_{\mu1}$ (see Eq. \eqref{galphaK}), and taken into account that no mode with $\lambda=0$ and $\mu=1$ exists. The angular integral is thus
 \begin{equation}
     I_{\alpha}^K = \frac{4}{5}\delta_{\sigma,s}\delta_{\lambda 2}\Big[3\partial_r R_{\alpha \theta} + \frac{6 R_{\alpha \theta}}{r} + \partial_r R_{\alpha r} + \frac{2 R_{\alpha r}}{r}\Big].
 \end{equation}
and the coupling rate is
 \begin{equation}
 \begin{split}
     \left(g_{\alpha K}\right)^* &= g_{\alpha K}^0\frac{2\pi R}{V}\frac{4}{5}\delta_{\sigma,s}\delta_{\lambda 2}\delta_{\mu,1}\int_0^R dr r^2
     \\
     &
     \times  \Big[3\partial_r R_{\alpha \theta}^{(S)} + \frac{6 R_{\alpha \theta}^{(S)}}{r} + \partial_r R_{\alpha r}^{(S)} + \frac{2 R_{\alpha r}^{(S)}}{r}\Big],
\end{split}
 \end{equation}
 which contains all the selection rules stated in the main text, i.e.\ the Kittel mode can couple only to spheroidal modes $S_{\nu 21}$. We finally notice that, for $\lambda=2$, the integrand of the radial integral can be expressed as
 \begin{equation}
    \begin{split}
        &r^2\Big[3\partial_r R_{\alpha \theta}^{(S)} + \frac{6 R_{\alpha \theta}^{(S)}}{r} + \partial_r R_{\alpha r}^{(S)} + \frac{2 R_{\alpha r}^{(S)}}{r}\Big]=
        \\
        &
        =
        R(\tilde{\alpha}_{\nu2} r)^2j_0(\tilde{\alpha}_{\nu2} r) - 3R\frac{T_{2\nu}^{(d)}}{T_{2\nu}^{(b)}}(\tilde{\beta}_{\nu2} r)^2j_0(\tilde{\beta}_{\nu2} r),
    \end{split}
 \end{equation}
 where we used Eq.~\eqref{spheroidalfactorization1}, equations \eqref{modefunctionSPH}-\eqref{Gnulambdadef}, and the recurrence relations of the spherical Bessel functions. In this form, the radial integral is straightforward, and we obtain the coupling rates given in Table~\ref{TableCouplings} in the main text.
 
 \subsection{Single magnon couplings for the $\{210\}$ mode}
 
To compute the coupling rate for the $\{210\}$ mode, we need to solve the integral
 \begin{equation}
  I_\alpha^{210} \equiv \int_{-1}^1 dc_\theta c_\theta W_\alpha(r,c_\theta),
\end{equation}
which we do by following similar steps as above. First, we explicitly write the integrand, group the terms containing $f_{\alpha r}^{(0)}$, $f_{\alpha\theta}^{(0)}$, and $f_{\alpha \phi}^{(0)}$ respectively, and integrate each term by parts to eliminate the derivatives with respect to $c_\theta$. This leads to
\begin{equation}
 \begin{split}
  I_\alpha^{210} =\!&\int_{-1}^1\!dc_\theta \Bigg\lbrace \!\!\!\left(c_\theta\frac{2c_\theta^2-1}{2}\partial_r
   + (3c_\theta^2-2)\frac{c_\theta}{r}\right) f_{\alpha \theta}^{(0)}
   \\
   &
   +
  \!\!\left( c_\theta^2\sqrt{1-c_\theta^2}\partial_r-\frac{6c_\theta^4-7c_\theta^2+1}{2 r\sqrt{1-c_\theta^2}}\!\right)\!\!f_{\alpha r}^{(0)}
   \\
   &
   -\frac{i}{2}\left(c_\theta^2\partial_r +\frac{3c_\theta^2-1}{r}\right)f_{\alpha \phi}^{(0)}
    \bigg\rbrace.
\end{split}
\end{equation}
We now introduce the explicit decomposition of the functions $f_{\alpha j}^{(0)}$, namely Eq.~\eqref{factorizationf} and, using that $R_{\alpha \theta} = R_{\alpha \phi}$, we rearrange the above integral into
\begin{equation}\label{Ialpha210}
\begin{split}
    I_\alpha^{210} &= \Big(\tilde I_{\alpha 1}\partial_r R_{\alpha\theta} + \tilde I_{\alpha 2}\frac{R_{\alpha\theta}}{r} +
    \\
    &
    +\tilde I_{\alpha 3}\partial_rR_{\alpha r} + \tilde I_{\alpha 4}\frac{R_{\alpha r}}{r}\Big),
\end{split}
\end{equation}
with integrals
\begin{equation}\label{Ialpha1tilde}
\begin{split}
    \tilde I_{\alpha 1} &\equiv\int_{-1}^1 dc_\theta
    \\
    &
    \left[c_\theta\frac{2c_\theta^2-1}{2}\Theta_{\alpha\theta}(c_\theta)-\frac{i}{2}c_\theta^2\Theta_{\alpha\phi}(c_\theta)\right],
\end{split}
\end{equation}
\begin{equation}\label{Ialpha2tilde}
\begin{split}
    \tilde I_{\alpha 2} &\equiv \int_{-1}^1 dc_\theta
    \\
    &
    \left[c_\theta(3c_\theta^2-2)\Theta_{\alpha\theta}(c_\theta)-\frac{i}{2}(3c_\theta^2-1)\Theta_{\alpha\phi}(c_\theta)\right],
\end{split}
\end{equation}
\begin{equation}\label{Ialpha3tilde}
    \tilde I_{\alpha 3} \equiv \int_{-1}^1 dc_\theta c_\theta^2\sqrt{1-c_\theta^2}\Theta_{\alpha r}(c_\theta),
\end{equation}
\begin{equation}\label{Ialpha4tilde}
    \tilde I_{\alpha 4}\equiv\int_{-1}^1 dc_\theta \frac{-6c_\theta^4+7c_\theta^2-1}{2\sqrt{1-c_\theta^2}}\Theta_{\alpha r}(c_\theta).
\end{equation}
Let us compute the above integrals for the two different phonon families.

\subsubsection{Torsional modes}

For torsional modes, we introduce Eqs.~\eqref{torsionalfactorization2} into Eqs.~\eqref{Ialpha1tilde} and \eqref{Ialpha2tilde} and integrate by parts to eliminate the derivatives, thereby obtaining
\begin{equation}
\begin{split}
    \tilde I_{\alpha 1}^{(T)}&=\frac{1}{3}\tilde I_{\alpha 2}^{(T)} = 
    \\
    &=\frac{i}{2} \int_{-1}^1 dc_\theta P_\lambda^1(c_\theta)c_\theta\sqrt{1-c_\theta^2}=-\frac{2i}{5}\delta_{\lambda 2}.
\end{split}
\end{equation}
 where we have used the fact that $\mu=1$ and used the orthogonality relations of the associated Legendre polynomials. Because $R_{\alpha r}^{(T)}=0$, we can write the desired integral for torsional modes as
 \begin{equation}
     I_{T,\nu\lambda 1}^{210}=-\frac{2i}{5}\delta_{\lambda 2}\left[\partial_r R_{\alpha\theta}^{(T)} + 3\frac{R_{\alpha\theta}^{(T)}}{r}\right].
 \end{equation}
 Alternatively, introducing Eq.~\eqref{torsionalfactorization1} and using the recurrence relations of the spherical Bessel functions, on can show that
  \begin{equation}
     I_{T,\nu\lambda 1}^{210}=-\frac{2i}{5}\delta_{\lambda 2}\tilde{\beta}_{2\nu}j_1(\tilde{\beta}_{2\nu}r).
 \end{equation}
 The coupling rate between the $\{210\}$ magnon and torsional phonons thus reads
 \begin{equation}
 \begin{split}
    \frac{g_{t\nu\lambda\mu,210}}{g_{t\nu\lambda\mu,210}^0} &= \frac{4\pi i }{5V}\delta_{\lambda 2}\delta_{\mu 1}\tilde{\beta}_{2\nu}\int_0^R dr r^3 j_1(\tilde{\beta}_{2\nu}r)
     \\
     &
     =
     \frac{3 i}{5 }\delta_{\lambda 2}\delta_{\mu 1}j_2(\tilde{\beta}_{2\nu}R),
 \end{split}
 \end{equation}
 which is the expression given in Table~\ref{TableCouplings} in the main text.
 
 \subsubsection{Spheroidal modes}
 
 Let us now compute the coupling for the spheroidal modes. We begin with introducing Eq.~\eqref{spheroidalfactorization2} into Eqs.~\eqref{Ialpha1tilde}-\eqref{Ialpha4tilde}, and eliminating all the derivatives with respect to $c_\theta$ through integration by parts. We thus obtain the following identities:
 \begin{equation}
\begin{split}
    \tilde I_{\alpha 1}^{(S)}= \int_{-1}^1 dc_\theta\frac{P_\lambda^1(c_\theta)}{2\sqrt{1-c_\theta^2}}\left(
    -8c_\theta^4+9c_\theta^2-1
    \right),
\end{split}
 \end{equation}
 \begin{equation}
     \tilde{I}_{\alpha 2}^{(S)}=\int_{-1}^1 dc_\theta\frac{P_\lambda^1(c_\theta)}{2\sqrt{1-c_\theta^2}}(-24c_\theta^4+29c_\theta^2-5),
 \end{equation}
 \begin{equation}
     \tilde I_{\alpha 3}^{(S)}=\int_{-1}^1 dc_\theta c_\theta^2\sqrt{1-c_\theta^2}P_\lambda^1(c_\theta),
 \end{equation}
 \begin{equation}
     \tilde I_{\alpha 4}^{(S)}=\int_{-1}^1 dc_\theta \frac{-6c_\theta^4+7c_\theta^2-1}{2\sqrt{1-c_\theta^2}}P_\lambda^1(c_\theta),
 \end{equation}
 where we have particularized to the case $\mu=1$. The above integrals are solved as follows: for $\tilde I_{\alpha 3}^{(S)}$, we write $c_\theta^2\sqrt{1-c_\theta^2}=(-1/5)[P_1^1(c_\theta)+(2/3)P_3^1(c_\theta)]$ and use the orthogonality relations for the associated Legendre polynomials. For the remaining three integrals, we use the following relation between the associated Legendre polynomials and the Legendre polynomials, 
 \begin{equation}
     \frac{P_{\lambda}^1(c_\theta)}{\sqrt{1-c_\theta^2}}=-\frac{d}{dc_\theta}P_\lambda(c_\theta).
 \end{equation}
We substitute the above into the integrals, eliminate the derivative through integration by parts, and express the resulting polynomials accompanying $P_\lambda(c_\theta)$ in terms of Legendre polynomials themselves. Then, the integrals are solved by using the orthogonality relation of the Legendre polynomials and we find
\begin{equation}
\begin{split}
    \tilde I_{\alpha 1}^{(S)}= -\frac{1}{5}\left[\frac{64}{7}\delta_{\lambda3}+2\delta_{\lambda1}\right],
\end{split}
 \end{equation}
 \begin{equation}
     \tilde{I}_{\alpha 2}^{(S)}=-\frac{1}{5}\left[96\frac{2}{7}\delta_{\lambda3}-\frac{2}{3}\delta_{\lambda1}\right],
 \end{equation}
 \begin{equation}
     \tilde I_{\alpha 3}^{(S)}=-\frac{1}{5}\left[\delta_{\lambda3}\frac{16}{7}+\delta_{\lambda1}\frac{4}{3}\right],
 \end{equation}
 \begin{equation}
     \tilde I_{\alpha 4}^{(S)}=-\frac{1}{5}\left[\delta_{\lambda3}\frac{48}{7}+\delta_{\lambda1}\frac{2}{3}\right].
 \end{equation}
 We now introduce the above expressions into Eq.~\eqref{Ialpha210}, substitute the explicit expressions of the radial functions Eq.~\eqref{spheroidalfactorization1} and, using the recurrence relations for the spherical Bessel functions and the results in Appendix~\ref{appendixPHONONS}, we cast the integral into the simplified form
 \begin{equation}
 \begin{split}
     I_{S,\nu\lambda 1}^{210}&=\left(\delta_{\lambda 1}-\delta_{\lambda3}\right)\frac{2R}{5}\bigg[\kappa_{\lambda\nu}\tilde{\beta}_{\lambda\nu}^2j_1(\tilde{\beta}_{\lambda\nu} r)
     \\
     &+\frac{2^\lambda}{2\lambda+1}\tilde{\alpha}_{\lambda\nu}^2j_1(\tilde{\alpha}_{\lambda\nu} r)\bigg],
 \end{split}
 \end{equation}
 where
 \begin{equation}
     \kappa_{1\nu}\equiv-\frac{1}{2}\frac{j_1(\tilde{\alpha}_{1\nu}R)}{j_1(\tilde{\beta}_{1\nu}R)},
 \end{equation}
 and
 \begin{equation}
 \begin{split}
     \kappa_{3\nu}\!\equiv\!\frac{(64/7)\left[\tilde{\alpha}_{3\nu}Rj_4(\tilde{\alpha}_{3\nu}R)-2j_3(\tilde{\alpha}_{3\nu}R)\right]}{\left[16-(\tilde{\beta}_{3\nu}R)^2\right]j_3(\tilde{\beta}_{3\nu}R)+2\tilde{\beta}_{3\nu}Rj_4(\tilde{\beta}_{3\nu}R)}.
\end{split}
 \end{equation}
 In the above form, the coupling rate Eq.~\eqref{couplinggeneral210} can be computed analytically by integrating the radial coordinate $r$, which results in the functions given in Table~\ref{TableCouplings} in the main text.
 
 \section{Further Derivations}\label{appendixFurtherDerivations}
 
 In this appendix we add two useful derivations. First, in Sec.~\ref{appendixSELECTION}, we prove the key selection rule $\int dV\tilde{\mathbf{m}}_\beta(\mathbf{r})\propto \delta_{\beta,\rm Kittel}$ for the coupling between the Kittel mode and the center-of-mass motion. Then, in Sec.~\ref{appendixHEATING}, we compute the power absorbed by the micromagnet due to the inhomogeneous magnetic driving field $\mathbf{H}_d(\mathbf{r},t)$.
 
 \subsection{Kittel-to-Center-of-Mass Selection rule}\label{appendixSELECTION}
 
 Here we demonstrate the identity $\int dV\tilde{\mathbf{m}}_\beta(\mathbf{r})\propto\delta_{\beta,\rm Kittel}$, instrumental in the derivation of Eq.\eqref{Vsecondquantization}.
 We consider the volume integral of a given Cartesian component of the adimensional magnetization mode function, namely
 \begin{equation}\label{Fbetaj}
     F_{\beta j} \equiv \int dV \mathbf{e}_j\cdot\tilde{\mathbf{m}}_\beta(\mathbf{r})\equiv \int dV \tilde{m}_{\beta j}(\mathbf{r}).
 \end{equation}
 Each component is obtained by solving the system of equations \eqref{dipolarm1} and \eqref{dipolarm2}, and takes the general form
 \begin{equation}\label{mjab}
     \mathbf{e}_j\cdot\tilde{\mathbf{m}}_\beta(\mathbf{r}) = a_{j}\partial_x\psi_\beta + b_j\partial_y\psi_\beta,
 \end{equation}
 where the particular expression of the mode-dependent coefficients $a_j$ and $b_j$ is not relevant for the following derivation. The magnetostatic potential corresponding to mode $\beta\equiv\{nlm\}$ inside the micromagnet has already been calculated in Appendix~\ref{appendixMAGNONS},
 \begin{equation}
     \psi_\beta \propto P_l^m(\xi)P_l^m(\cos\eta)e^{im\phi},
 \end{equation}
and is naturally expressed in the coordinate system $\{\xi,\eta,\phi\}$ given by Eqs.~\eqref{x0}-\eqref{z0}. Our aim is to express the integral Eq.~\eqref{Fbetaj} in these coordinates. To this end, we remark that some quantities derived in this coordinate system, such as e.g. the Jacobian of the transformation to Cartesian coordinates, or the inverse transformation $\{x(\xi,\eta,\phi),y(\xi,\eta,\phi),z(\xi,\eta,\phi)\}$, have different qualitative forms depending on the sign of $\chi_p$ or $\chi_p+1$. However, one can verify that the final result $F_{\beta j}$ does not dependent on the sign and we will hereafter consider only the case $\chi_p>0$.

Applying the chain rule, we express the spatial derivatives of the Cartesian coordinates in terms of derivatives with respect to $\{\xi,\eta,\phi\}$ and write Eq.~\eqref{mjab} as
 \begin{equation}\label{mbDSCcoords}
 \begin{split}
    \tilde{m}_{\beta j}&(\mathbf{r}) \propto \Bigg\{\frac{im (a_j\sin\phi - b_j\cos\phi)}{\sqrt{1-c_\eta^2}\sqrt{\xi^2-1}}+
    \\
    &
    +\frac{\sqrt{\xi^2-1}\sqrt{1-c_\eta^2}}{c_\eta^2-\xi^2}(a_j\cos\phi + b_j\sin\phi)
    \\
    &
    \times\left[\xi\frac{\partial}{\partial \xi}-c_\eta\frac{\partial}{\partial c_\eta}\right]\Bigg\}P_l^m(\xi)P_l^m(c_\eta)e^{im\phi},
\end{split}
\end{equation}
 where we used the shorthand $c_\eta\equiv\cos\eta$ to keep the notation simple. We also express the integral over the micromagnet volume as an integral over the coordinates $\{\xi,\eta,\phi\}$, which requires computing the full Jacobian of the coordinate transformation. After lengthy but straightforward algebra one can show that
 \begin{equation}
 \begin{split}
     \int_{\rm magnet}&dV f(\mathbf{r}) =\frac{\chi_p R^3}{\xi_0}
     \\
     &
     \times\!\!\int_0^{\xi_0} \! d\xi\int_{-1}^1\!\! dc_\eta\!\int_0^{2\pi}\!\! d\phi (\xi^2-c_\eta^2)f(\xi,c_\eta,\phi),
 \end{split}
 \end{equation}
 for any function $f(\mathbf{r})$, where $\xi_0$ is defined in Eq.~\eqref{surfaceDSC}. Combining the above equation with Eq.~\eqref{mbDSCcoords} we write
 \begin{equation}
 \begin{split}
     &F_{\beta j} \propto \!\int_0^{\xi_0} \!d\xi\!\int_{-1}^1\!dc_\eta\!\int_0^{2\pi}\!d\phi e^{im\phi}\sqrt{(\xi^2-1)(1-c_\eta^2)}
     \\
     &
     \times
     \bigg\{-(a_j\cos\phi + b_j\sin\phi)\left[\xi\frac{\partial}{\partial \xi}-c_\eta\frac{\partial}{\partial c_\eta}\right]
     \\
     &
     -\!\frac{m (\xi^2-c_\eta^2)(a_j\sin\phi - b_j\cos\phi)}{i(1-c_\eta^2)(\xi^2-1)}\!\Bigg\}
     P_l^m(\xi)P_l^m(c_\eta).
 \end{split}
 \end{equation}
 The azimuthal integrals in $\phi$ are straightforward and provide the first selection rule for $F_{\beta j}$\footnote{Note that, in order to derive this expression, we must fix a sign convention for the imaginary unit, due to the sign arbitrariness in ratios of the form $\sqrt{1-\xi^2}/\sqrt{\xi^2-1}=\pm i$. The choice of convention is arbitrary, but must be consistently followed throughout the entire derivation. In this work we choose the positive sign of the above equality.}:
 \begin{equation}
 \begin{split}
    &F_{\beta j} \propto \delta_{\vert m\vert,1} \int_0^{\xi_0} d\xi\int_{-1}^1dc_\eta\sqrt{(1-\xi^2)(1-c_\eta^2)}
    \\
    &
    \times\!
    \bigg\{\!\xi\frac{\partial}{\partial \xi}\!-\!c_\eta\frac{\partial}{\partial c_\eta} 
    \!-\!\frac{\xi^2-c_\eta^2}{(1\!-\!c_\eta^2)(1\!-\!\xi^2)}\!\bigg\}P_l^m(\xi)P_l^m(c_\eta). 
\end{split}
 \end{equation}
 Since the associated Legendre polynomials fulfill $P_l^{-m}(z)\propto P_l^m(z)$, we will focus on the case $m=+1$ from now on, without loss of generality. Using the properties of the associated Legendre polynomials and their relation to the usual Legendre polynomials $P_l(z)$, we recast the integral in the form
 \begin{equation}
    \begin{split}
        &F_{\beta j}\! \propto \delta_{\vert m\vert,1}\! \int_0^{\xi_0}\! d\xi\int_{-1}^1\!dc_\eta \bigg\{\!
        (\xi^2-c_\eta^2)P_l'(\xi)P_l'(c_\eta)
        \\
        &
        +\frac{\xi}{2}\!\left[l(l\!+\!1)P_l(\xi)-(1\!-\!\xi^2)P_l''(\xi)\right]\!(1\!-\!c_\eta^2)P_l'(c_\eta)
        \\
        &-\!
        \frac{c_\eta}{2}(1\!-\!\xi^2)P_l'(\xi)\!\!\left[l(l\!+\!1)P_l(c_\eta)\!-\!(1\!-\!c_\eta^2)P_l''\!(c_\eta)\right]
        \!\!\bigg\}. 
    \end{split}
 \end{equation}
 We now carry out all the integrals in the variable $c_\eta$ by combining integration by parts, the orthogonality relations of the Legendre polynomials, and the parity properties $P_l(1)=1$ and $P_l(-1) = (-1)^l$. This integration leads to the second and final selection rule:
 \begin{equation}
     F_{\beta j} = \int dV \mathbf{e}_j\cdot\tilde{\mathbf{m}}_{\{lmn\}}(\mathbf{r}) \propto \delta_{l1}\delta_{\vert m\vert,1}.
 \end{equation}
Note that there is no $\{1,-1,n\}$ magnon mode, and that the only solution for $l=m=1$ is the $\{110\}$ mode, namely the Kittel mode. We thus conclude that the integral of the spin wave magnetization mode function, $\tilde{\mathbf{m}}_\beta(\mathbf{r},t)$, across the volume of the micromagnet is exactly zero for all magnon modes except for the Kittel mode.

\subsection{Driving-induced heating of the micromagnet.}\label{appendixHEATING}

In this appendix we address the unavoidable internal heating of the micromagnet induced by the time-dependent magnetic drive $\mathbf{H}_d(\mathbf{r},t)$. Such heating could be detrimental, especially for a levitated micromagnet where the absence of a physical thermal contact prevents it from rapidly equilibrating with its surroundings. 
The differential equation obeyed by the internal temperature of the micromagnet, $T_{\rm MM}$, is \cite{MessinaPRB2013}
\begin{equation}
    c_v V \dot{T}_{\rm MM} = \mathcal{P}_{\rm abs} - \mathcal{P}_{\rm em} (T_{\rm MM}),
\end{equation}
where $c_v$ is the specific heat of YIG per unit volume, $\mathcal{P}_{\rm abs}$ is the total power absorbed from the drive, and $\mathcal{P}_{\rm em}(T_{\rm MM})$ is the total power emitted into the electromagnetic field modes by radiative thermal emission. In the steady state, the micromagnet temperature does not evolve in time, $\dot{T}_{\rm MM} = 0$, allowing to compute the steady-state temperature by solving the implicit equation 
\begin{equation}\label{SteadyStateT}
    \mathcal{P}_{\rm abs} = \mathcal{P}_{\rm em}(T_{\rm MM, ss}).
\end{equation}

\begin{figure} 
	\centering
	\includegraphics[width=\linewidth]{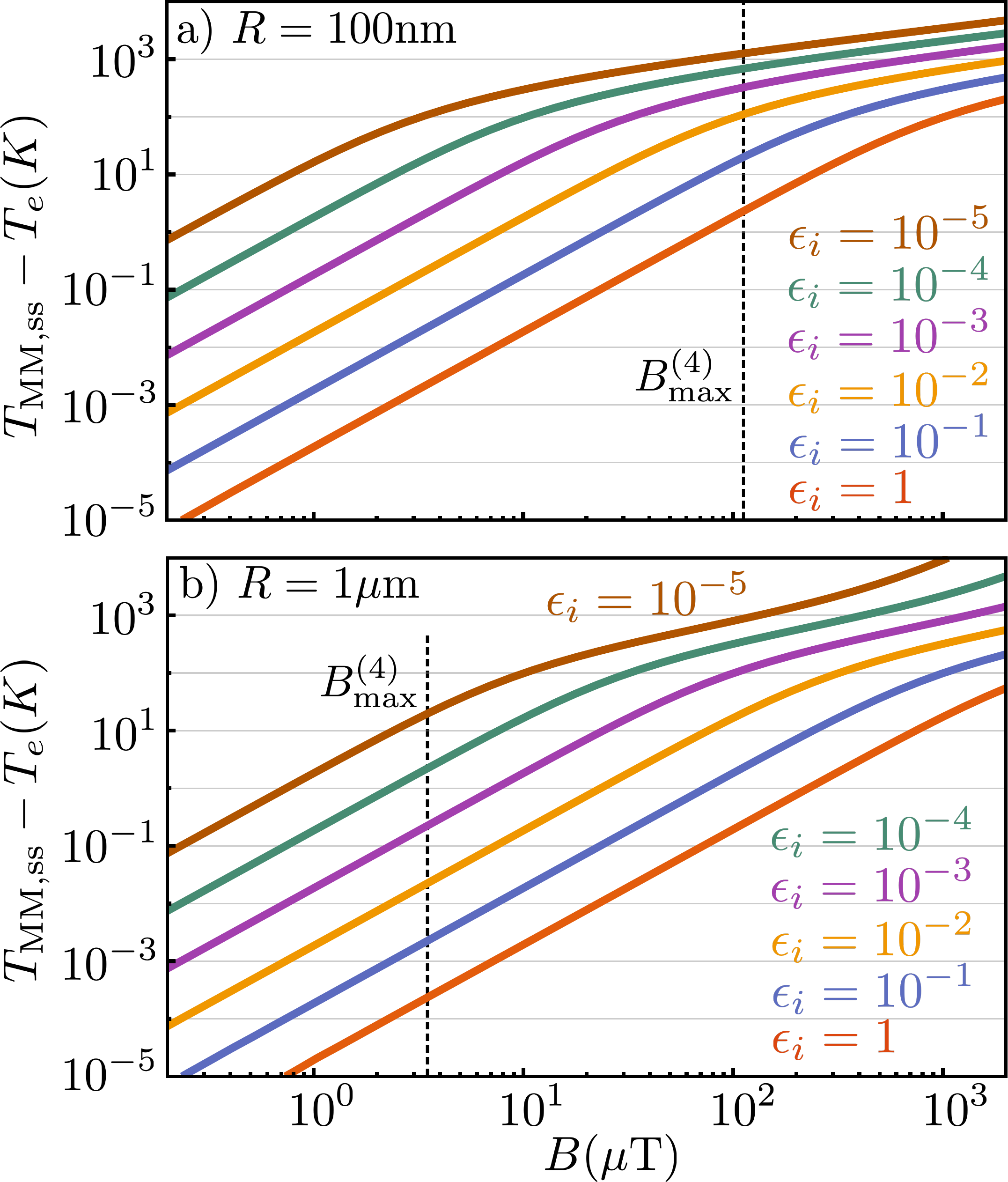}
	\caption{Increase in the internal temperature of the micromagnet computed from Eq.~\eqref{SteadyStateT} versus homogeneous driving field $B$, at $T_e=300$K and for radius $R=100$nm (panel a) and $R=1\mu m$ (panel b). Different colors correspond to different values of the imaginary part of the permittivity, $\epsilon_i$. The dashed line in each panel marks the maximum field felt by the micromagnet under the inhomogeneous driving Eq.~\ref{Hdparticular}, on its thermal motion at a temperature $T_e$ and a field gradient $b=10^4$ T/m (Eq.~\eqref{Bmax4}). Unspecified parameters are taken as in the main text.
	}\label{figheatingroomT}
\end{figure}

Assuming the external field is purely magnetic and monochromatic, i.e., $\mathbf{H}_d(\mathbf{r},t) = (1/2)\text{Re}[\mathbf{H}_{d0}(\mathbf{r})\exp(-i\omega_dt)]$, the total absorbed power is given by \cite{EvlyukhinPRB2010,DongPRB2017,MessinaPRB2013}
\begin{equation}
    \mathcal{P}_{\rm abs} = \frac{\omega_d\mu_0}{2}\text{Im}[\chi_M(\omega_d)]\vert \mathbf{H}_{d0}(\mathbf{r})\vert^2,
\end{equation}
where $\chi_M(\omega_d) \equiv \alpha_M(\omega_d) - i(12\pi c^3/\omega_d^3)^{-1}\vert\alpha_M(\omega_d)\vert^2$, and
$\alpha_M(\omega)=3V[\mu(\omega)-1]/[\mu(\omega)+2]$ is the magnetic polarizability of the micromagnet~\cite{Jackson1975classical,DongPRB2017}, 
with $\mu(\omega)$ the relative permeability of YIG. 
Note that the absorbed power is approximately proportional to the driving frequency $\omega_d$, and thus depends on the frequency of the acoustic phonon coupled to the center-of-mass motion. Conversely, the thermally emitted power is dominated by the fluctuations of the thermally induced electric dipole moment of the micromagnet, and can be computed as~\cite{MessinaPRB2013,RubioLopezPRB2018}
\begin{equation}\label{Pemitted}
\begin{split}
    \mathcal{P}_{\rm em} (T_{\rm MM}) &= \int_0^\infty d\omega\left[n(\omega,T_{\rm MM})-n(\omega,T_{e})\right]
    \\
    &
    \times\frac{\hbar\omega^4}{\pi^2\varepsilon_0c^3}\frac{\text{Im}[\alpha_E(\omega)]}{\vert1-i\alpha_E(\omega)\omega^3/(6\pi\varepsilon_0 c^3))\vert^2},
\end{split}
\end{equation}
where $n(\omega,T)=(\exp[\hbar\omega/k_B T]-1)^{-1}$ is the Bose-Einstein distribution, and
$\alpha_E(\omega) = \varepsilon_0 V (\varepsilon(\omega)-1)/(\varepsilon(\omega) + 2)$ is the electric polarizability of the micromagnet, with $\varepsilon(\omega)$ the relative permittivity of YIG.
In deriving the above expression, we assumed that the surrounding electromagnetic modes are at the temperature of the environment, namely $T_e$, and took the long wavelength approximation $k_B T_e/\hbar \gg 2\pi c/R$. Note that the emitted power $\mathcal{P}_{\rm em}$ is zero if $T_{\rm MM} = T_e$, and decreases at low temperatures $T_{\rm MM}$, following the decrease of the integrand $\omega^4 n(\omega,T_{\rm MM})$.

\begin{figure} 
	\centering
	\includegraphics[width=\linewidth]{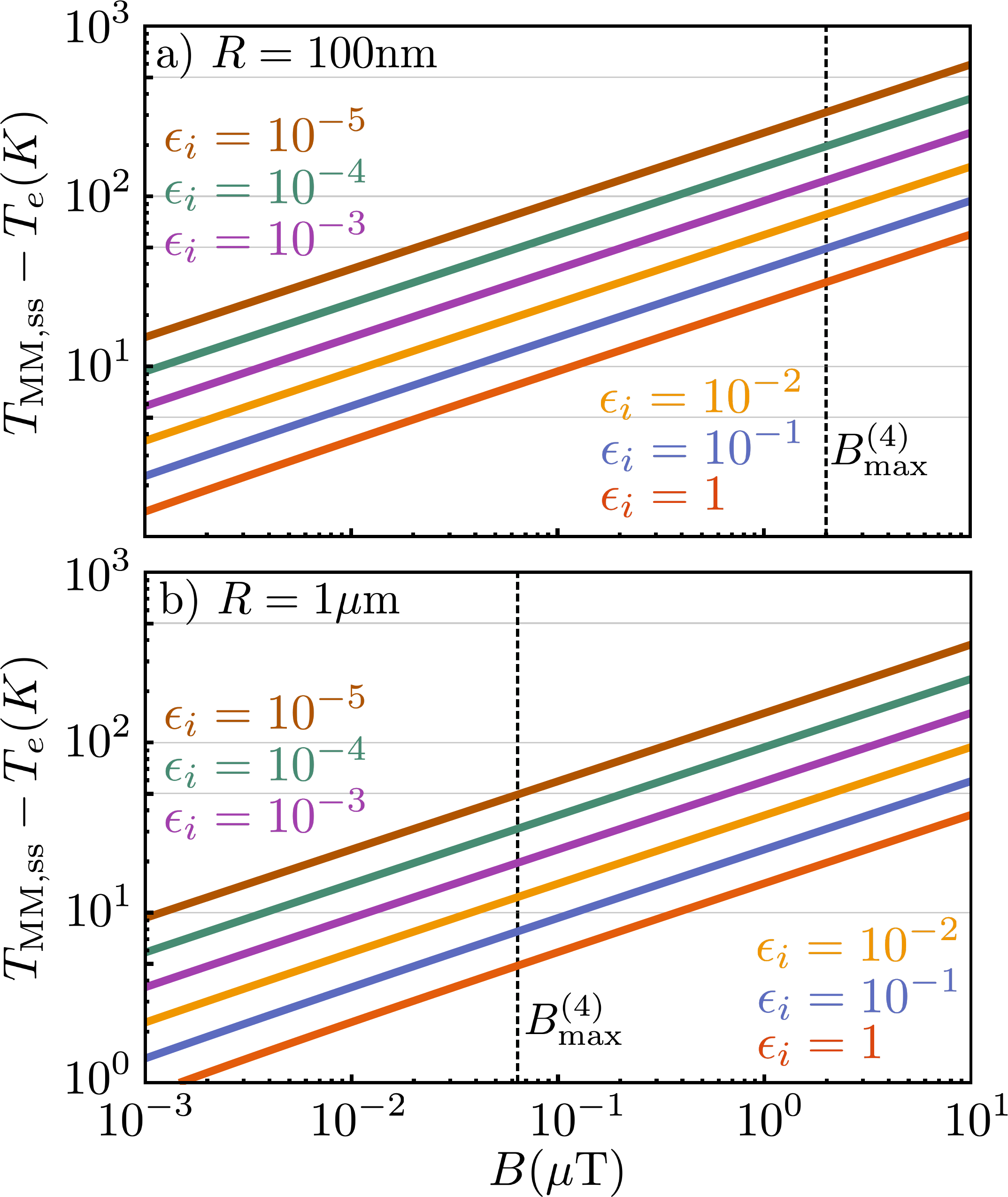}
	\caption{Same as \figref{figheatingroomT} with $T_e=100$\,mK.
	}\label{figheatingcryo}
\end{figure}

We solve Eq.~\eqref{SteadyStateT} numerically in the simple scenario of a homogeneous driving field $\vert\mathbf{H}_{d0}(\mathbf{r})\vert\approx B/\mu_0$, and assuming a constant dielectric permittivity for YIG across the thermal wavelength range, $\varepsilon(\omega)=\varepsilon \equiv\epsilon_r + i\epsilon_i$. The total increase in the internal temperature of the micromagnet, $T_{\rm MM,ss}-T_e$, is shown as a function of the driving field $B$ in Figs.~\ref{figheatingroomT} and \ref{figheatingcryo} for $T_e=300$\,K and $T_e=100$\,mK respectively.  In both plots we assume the driving frequency $\omega_d$ is tuned to couple the center-of-mass motion to the $S_{121}$ acoustic mode as described in the main text, and take typical values for YIG, namely~\cite{HasanIEEE2018} $\epsilon_r=4.4$ and $\mu(\omega_d)=1+i0.005$. The latter is consistent with drivings far away from the magnon resonance~\cite{ChenSciRep2016,Derzakowski2010}, which is a good assumption because $\omega_d = \omega_m-\omega_{tx}+g/\chi_0 \approx \omega_m+100 g$.
As evidenced by Figs.~\ref{figheatingroomT} and \ref{figheatingcryo}, larger values of the dielectric loss $\epsilon_i$ result in lower heating, as the radiative emission is dominated by the intrinsic thermal fluctuations of the electric dipole moment associated with larger value of $\epsilon_i$. The heating also decreases for higher values of $T_e$, as the power dissipated via thermal emission is larger. Finally, the aforementioned dependence of the absorbed power on the driving frequency $\omega_d$, and thus with the frequency of the chosen acoustic phonon, is responsible for the approximately linear increase of $T_{\rm MM,ss}$ both with $R^{-1}$ and with the acoustic mode index $\nu$ (the latter not shown in the figures).

In the main text, we consider the inhomogeneous driving field given by Eq.~\eqref{Hdparticular}. Using this field, we estimate the maximum field experienced by the micromagnet, at the maximum field gradient considered in this work, namely $b_{\rm max}=10^4$T/m, as
\begin{equation}\label{Bmax4}
    B_{\rm max}^{(4)}\equiv b_{\rm max}\langle\hat{X}^2 \rangle_{\rm{ss}}^{1/2}\Big\vert_{T=T_e}.
\end{equation}
This field is marked by the dashed lines in each panel of Figs. \ref{figheatingroomT} and \ref{figheatingcryo}. At this values of the magnetic field, and assuming a dielectric loss comparable to that of a very low-loss material such as silica at thermal frequencies  ($\epsilon_i\approx 10^{-3}$), the temperature of a $R=1\mu$m micromagnet increases by around $20$\,K at cryogenic temperatures and by less than $1$\,K at room temperature. 
These final temperatures are not only low, but also largely overestimate the temperature increase experienced by the micromagnet under the inhomogeneous field Eq.~\eqref{Hdparticular}, for various reasons. First, at low frequencies, the dielectric loss of YIG usually takes on larger values than the one for silica~\cite{HasanIEEE2018}. 
Second, the average field experienced by the micromagnet is in general much smaller than $B_{\rm max}^{(4)}$, because, on the one hand, the motional amplitude of the center of mass is significantly reduced when it is cooled, that is $\langle\hat{X}^2 \rangle_{\rm{ss}} \ll \langle\hat{X}^2 \rangle_{\rm{ss}}\Big\vert_{T=T_e}$. On the other hand, the required field gradients for motional cooling are usually at least one order of magnitude smaller than $b=10^4$T/m (see \figref{figCMcoolingS121}). From these results we  safely conclude that the heating of the nanomagnet is not critical and allows for efficient cooling of the center-of-mass motion.



 
\bibliographystyle{apsrev4-1}
\bibliography{bibliography}

\end{document}